\documentclass[AMA,STIX1COL]{WileyNJD-v2}

\articletype{RESEARCH ARTICLE}%

\received{15 February 2023}
\revised{}
\accepted{}

\usepackage{amssymb}
\usepackage{multicol}
\usepackage{amsmath}

\makeatletter
\renewcommand*\env@matrix[1][*\c@MaxMatrixCols c]{%
  \hskip -\arraycolsep
  \let\@ifnextchar\new@ifnextchar
  \array{#1}}
\makeatother

\usepackage{amssymb,mleftright}
\usepackage{amsmath}
\usepackage{setspace}
\usepackage{amsfonts}
\usepackage{amsthm}
\usepackage{bbm}
\usepackage{mathtools}
\usepackage{subfig}
\usepackage{lineno}
\usepackage{hyperref}
\usepackage{float}
\usepackage{array,multirow}
\usepackage{color}
\usepackage{stmaryrd}
\usepackage{tikz}
\usepackage{graphicx}
\usepackage{rotating}

\usepackage{bm}
\usepackage{caption}
\usepackage{booktabs}
\usepackage{lscape}

\definecolor{lightgray}{gray}{0.80}

\newcommand{\mat}[1]{\mathbf{#1}} 
\newcommand{\vect}[1]{\boldsymbol{#1}} 							
\newcommand{\List}[1]{\left\{ \right. #1 \left.\right\} }				

\newcommand{\Pe}{\mathrm{Pe}}		
\renewcommand{\div}{\nabla \cdot}		
\newcommand{\grad}{\nabla}				

\newcommand{\jump}[1]{\llbracket #1 \rrbracket}		
\newcommand{\bform}[3]{#1\left(#2 , #3 \right)}		
\newcommand{\lform}[2]{#1\left(#2 \right)}		

\newcommand{\domain}{\Omega}		
\newcommand{\boundary}{\partial}		

\newcommand{\mesh}{\mathcal{T}_h}			
\newcommand{\skeleton}{\mathcal{E}_h}		
\newcommand{\border}{\Gamma}	
\newcommand{\element}{\mathrm{e}}	
\newcommand{\face}{\mathrm{f}}	

\newcommand{\Pair}[3]{\left(#1 , #2 \right)_{#3}}	
\newcommand{\pair}[3]{\langle#1 , #2 \rangle_{#3}}	

\modulolinenumbers[5]

\begin{document}

\title{A matrix-free macro-element variant of the hybridized discontinuous Galerkin method}

\author[1]{Vahid Badrkhani*}
\author[1]{Ren\'e R. Hiemstra}
\author[1]{Micha{\l} Mika}
\author[1]{Dominik Schillinger}

\authormark{Badrkhani V,  Hiemstra R, Mika M AND Schillinger D}

\address[1]{Institute for Mechanics, Computational Mechanics Group, Technical University of Darmstadt, Germany}

\corres{*Vahid Badrkhani, Institute for Mechanics, Computational Mechanics Group, Technical University of Darmstadt, Germany. \email{ vahid.badrkhani@tu-darmstadt.de}}
\fundingInfo{ German Research Foundation (Deutsche Forschungsgemeinschaft), Grant/Award Numbers: 1249/2-1}

\abstract[Summary]{ We investigate a macro-element variant of the hybridized discontinuous Galerkin (HDG) method, using patches of standard simplicial elements that can have non-matching interfaces. Coupled via the HDG technique, our method enables local refinement by uniform simplicial subdivision of each macro-element. By enforcing one spatial discretization for all macro-elements, we arrive at local problems per macro-element that are embarrassingly parallel, yet well balanced. Therefore, our macro-element variant scales efficiently to n-node clusters and can be tailored to available hardware by adjusting the local problem size to the capacity of a single node, while still using moderate polynomial orders such as quadratics or cubics. Increasing the local problem size means simultaneously decreasing, in relative terms, the global problem size, hence effectively limiting the proliferation of degrees of freedom. The global problem is solved via a matrix-free iterative technique that also heavily relies on macro-element local operations.  We investigate and discuss the advantages and limitations of the macro-element HDG method via an advection-diffusion model problem.}


\maketitle

\section{Introduction \label{sec:introduction}}

Discontinuous Galerkin (DG) formulations \cite{arnold2002unified} exhibit desirable properties when applied to solve conservation problems, such as a rigorous mathematical foundation, the ability to use arbitrary orders of basis functions on general unstructured meshes, and a natural stability property for convective operators \cite{bassi1997high,cockburn2018discontinuous,hesthaven2007nodal, peraire2008compact}. About one decade ago, the hybridized discontinuous Galerkin (HDG) method was introduced in \cite{cockburn2009unified}, which distinguishes itself by several unique features from other DG methods. First, it provides, for smooth and diffusion dominated problems, approximations of all variables that converge with the optimal order of $ p+1 $ in the $ L^{2}$ norm \cite{nguyen2009implicit}. In addition, it enables the computation of new approximate velocity fields that converge with order $ p+2 $ for $p \geq 1$ by means of an inexpensive element by-element postprocessing procedure \cite{cockburn2009superconvergent,nguyen2012hybridizable}. Second, it provides a novel and systematic way of imposing different types of boundary conditions \cite{nguyen2010hybridizable,peraire2010hybridizable}. Third, the linear systems that arise from the HDG method are equivalent to two different linear systems: a first one that couples globally the numerical trace of the solution on element boundaries, thereby leading to a significant reduction in the degrees of freedom; and a second one that couples at the element level (locally) the conserved quantities and their gradients and, therefore, can be solved in an element-by-element manner \cite{cockburn2016static,cockburn2009hybridizable}. In light of these benefits, a number of research efforts have extended the HDG method to a wide variety of initial boundary value problems and PDE systems, see e.g.\
 \cite{cockburn2009unified,nguyen2009implicit,cockburn2009superconvergent, nguyen2012hybridizable,peraire2010hybridizable,  cockburn2009hybridizable,Nguyen_2009, nguyen2011implicit, HDGLES, vila2021hybridisable, nguyen2011high,la2020weakly}.

A disadvantage of DG methods in general is the proliferation and increased global coupling of degrees of freedom with respect to standard continuous finite element methods. One objective of the HDG method has been to improve the computational efficiency of DG schemes and the associated solutions in discontinuous finite element spaces. In particular, the concept of hybridization based on additional unknowns on element interfaces reduces global coupling and facilitates static condensation strategies. For large-scale computations \cite{foucart2018distributed,pazner2017stage}, however, hybridization alone has often turned out insufficient to overcome memory and time-to-solution limitations, resulting in ongoing algorithms-centered research \cite{fabien2019manycore, kronbichler2018performance, roca2013scalable, roca2011gpu}.
 
Compared to its discontinuous counterpart, the continuous Galerkin (CG) method, that is, the classical finite element formulation \cite{hughes2012finite,paipuri2019coupling}, requires a smaller number of unknowns for the same mesh \cite{huerta2013efficiency}. The CG solution can also be accelerated by means of static condensation, which produces a globally coupled system involving only those degrees of freedom on the mesh skeleton.
Unstructured mesh generators often produce meshes with high vertex valency (number of elements touching a given vertex). Therefore, CG methods involve rather complex communication patterns in parallel runs on distributed memory systems, which has a negative impact on scalability \cite{kirby2012cg,yakovlev2016cg}. The HDG method generates a global system for the approximation on the boundaries of the elements that although larger in rank than the traditional static condensation system in CG, has significantly smaller bandwidth at moderate polynomial degrees \cite{cockburn2009hybridizable,kirby2012cg}.

In this paper, we investigate a discretization strategy that combines elements of the continuous and hybridized discontinuous approach. Its main idea is to apply the HDG concept not between individual elements, but between larger macro-elements that can contain a flexible number of finite elements themselves. While in the standard HDG approach, the structure and properties of the global and local problems are rigidly determined by a given mesh and the polynomial degree on each element, the macro-element HDG concept enables us to flexibly influence this structure. Due to this flexibility, the proposed macro-element variant offers additional algorithmic possibilities that standard HDG methods do not offer in the same way. In this paper, we will explore how these possibilities can be leveraged to arrive at a practical and effective approach to adaptive refinement, domain decomposition and load balancing on modern heterogeneous compute systems.

It is worthwhile to note that the macro-element HDG concept shares similarities with domain decomposition methods \cite{arbogast2000mixed,belgacem1999mortar,klawonn2006dual,bertsekas2014constrained}, targeted at decomposing a given finite element discretization \cite{toselli2004domain,toselli2001feti} into subdomain patches. The distinguishing feature, however, is its algebraic structure in terms of one globally coupled system based on trace variables between macro-elements and many decoupled systems, one for each macro-element, which gives rise to the unique advantages of the HDG approach in terms of parallel computing.

Our article is organized as follows: in Section 2, we briefly describe our model problem and fix basic concepts and notation fundamental for the remainder of the paper. In Section~3, we juxtapose our variant of HDG on macro-elements to standard HDG discretization and illustrate basic properties in terms of accuracy and computing cost. In Section~4, we discuss and highlight practical advantages of our discretization strategy in terms of accuracy, adaptive refinement, domain decomposition and load balancing. In Sections~5 and 6, we thoroughly investigate the computational efficiency and scalability of our macro-element HDG variant in comparison with the standard HDG method, first in terms of theoretical estimates and then via numerical tests with a parallel implementation.

\section{Background and notation}
In the following, we will briefly describe our model problem. We then review some relevant background and fix our notation in a general setting, which applies both to the standard HDG method on fully discontinous elements as well as to our variant on macro-elements.

\subsection{Model problem}
In this paper, the methods and their numerical properties are illustrated via the stationary linear advection-diffusion equation in $d=2,3$ space dimensions.
Let $\domain \subset \mathbb{R}^{d}$ be an open set with piecewise smooth boundary $\boundary \domain = \overline{\border_D \cup \border_N}$, $\emptyset = \border_D  \cap \border_N$, and outward unit vector $\vect{n}$. Consider a concentration $u$ of a species in an incompressible fluid, modeled by

\begin{subequations}\label{C21}
\begin{align}
		\nabla \cdot\left( \vect{a} u\right) -\nabla \cdot\left( \kappa \nabla u\right)  & = f   \quad\quad &\text{in} \; &\domain, \\
		u & = g_D    \quad\quad &\text{on} \; &\border_D, \\
		(\vect{a} u - \kappa \nabla u) \cdot \vect{n} & = g_N    \quad\quad &\text{on} \; &\border_N.
\end{align}
\end{subequations}

Here, $\vect{a} \in \left(L^\infty(\domain) \right)^d$ denotes a divergence-free advective velocity field, $\kappa > 0$ is a constant coefficient of diffusion, $f \in L^2(\domain)$ is a scalar source term and $g_D$ and $g_N$ are the prescribed Dirichlet and Neumann boundary data, respectively.

Considering the auxiliary variable $\vect{q} = -\nabla u$, we may rewrite \eqref{C21} as the following set of first order equations

\begin{subequations}\label{C22}
\begin{align}
		\vect{q} + \nabla u &= 0 \quad\quad &\text{in} \; &\domain,\\
		\div \left( \vect{a} u + \kappa \vect{q} \right)  &=f \quad\quad &\text{in} \; &\domain,\\
		u &= g_D \quad\quad &\text{on} \; &\border_D, \\
		(\vect{a} u + \kappa \vect{q}) \cdot \vect{n} & = g_N \quad\quad &\text{on} \; &\border_{N} .
\end{align}
\end{subequations}

\subsection{Mesh, mesh skeleton, and trace variables}

We briefly fix the notation to describe the broken computational domain. Let $\mesh$ denote a partitioning of domain $\domain$ in a collection of disjoint subdomains $\List{K \; : \; \bigcup K = \Omega, \; K_i \cap K_j = \emptyset \; \text{for } K_i \neq K_j}$. The collection of sub-domain boundaries will be denoted by $\boundary \mesh = \List{\boundary K  \; : \;  K \in \mesh}$. The common interface $F = \boundary K^+ \cap \boundary K^-$ shared by adjacent subdomains  $K^+$ and $K^-$ in $\mesh$ is called an interior face. The collection of all interior faces is the set $\skeleton^\circ$, called the ``mesh skeleton''. Similarly, a boundary face is the non-empty intersection $\boundary K \cap \boundary \domain$ of a subdomain $K \in \mesh$ with the boundary $\boundary \domain$ and their collection is denoted by $\skeleton^{\boundary}$. The collection of all faces $\skeleton^{\circ} \cup \skeleton^{\boundary}$ will be denoted by $\skeleton$.

We recall the definition of the jump operator. Let $K^+$ and $K^-$ denote adjacent subdomains and $\vect{n}^+$ and $\vect{n}^-$ the outward unit normals of $\boundary K^+$ and $\boundary K^-$, respectively. Let $u^\pm$ and $\vect{q}^\pm$ denote the traces of $u$ and $\vect{q}$ on $F$ from the interior of $K^\pm$. We define the jump $\llbracket \cdot \rrbracket$ of a scalar and vector valued function as

\begin{subequations}
\begin{align}
	\llbracket u \vect{n} \rrbracket &:= u^+ \vect{n}^+ + u^- \vect{n}^- 	&& \text{on } F \in \skeleton^{\circ}, &
	\llbracket u \vect{n} \rrbracket &:= u \vect{n}	&& \text{on } F \in \skeleton^{\boundary} ,\\
	\llbracket \vect{q}  \cdot \vect{n} \rrbracket &:= \vect{q}^+ \cdot \vect{n}^+ + \vect{q}^- \cdot \vect{n}^- && \text{on } F \in \skeleton^{\circ}, &
	\llbracket \vect{q} \cdot \vect{n} \rrbracket &:= \vect{q} \cdot \vect{n} && \text{on } F \in \skeleton^{\boundary}.
\end{align}
\end{subequations}
Note that the jump of a scalar function is vector valued, while the jump of a vector valued function is scalar valued.

\subsection{Functional setting and interpolation spaces}

We briefly fix the notation to describe the relevant discontinuous test and trial spaces.
Let $L^2(D)$ denote the space of square integrable functions on a generic subdomain $D \subset \mathbb{R}^d$. 
For functions $\vect{q}, \vect{v} \in \left(L^2(D) \right)^d$, we denote $\Pair{\vect{q}}{\vect{v}}{D} := \int_D \vect{q} \cdot \vect{v}$. For scalar functions, $u, w \in L^2(D)$, we denote $\Pair{u}{w}{D} := \int_D u w$ if $D$ is a $d$-dimensional subdomain, and $\pair{u}{w}{D} := \int_D u w$ if $D$ is of dimension $d-1$. We also introduce

\begin{align}
	&\Pair{u}{w}{\mesh} = \sum_{K \in \mesh} \Pair{u}{w}{K}, &
	&\pair{\rho}{\eta}{\boundary \mesh} = \sum_{K \in \mesh} \pair{\rho}{\eta}{\boundary K}, &
	&\pair{\lambda}{\mu}{\skeleton} = \sum_{F \in \skeleton} \pair{\lambda}{\mu}{F}, &
\end{align}
for functions $u,w$ defined on $\mesh$, $\rho, \eta$ on $\boundary \mesh$ and $\lambda, \mu$ on $\skeleton$.

We now introduce the finite dimensional spaces defined on the broken computational domain. Let $W(K)$, $\vect{V}(K)$ and $M(F)$ denote spaces of $C^0$-continuous piecewise polynomials of order $p \geq 1$, defined on subdomains $K \in  \mesh$ and $F \in  \skeleton$, respectively. We define the following discontinuous finite element spaces on $\mesh$
\begin{subequations}
\begin{align}
	W^h &:= \left\{ w \in L^2(\domain) \; : \; \left. w \right|_{K} \in W(K) \; \forall K \in \mesh   \right\},	\\
	\vect{V}^h &:= \left\{ \vect{v} \in \left(L^2(\domain)\right)^d \; : \; \left. \vect{v} \right|_{K} \in \vect{V}(K) \; \forall K \in \mesh   \right\}.
\end{align}
On the mesh skeleton, $\skeleton$, we define a discontinuous finite element space for the trace variable
\begin{align}
	M^h &:= \left\{ \mu \in L^2(\skeleton) \; : \; \left. \mu \right|_{F} \in M(F) \;  \; \forall F \in \skeleton \right\}.
\end{align}
\end{subequations}
We will also need spaces with prescribed Dirichlet and Neumann conditions, $M^h(g_\alpha) = \List{\mu \in M^h \; : \; \mu = P (g_\alpha) \; \text{on } \border_\alpha}$, $(\alpha = D, N)$, which are applied by means of an $L^2$-projection: $P \; : \; L^2(\border_\alpha) \mapsto L^2(\border_\alpha)$, $P \circ P = \mathrm{Id}$.
\section{The hybridized DG method on macro-elements}
In the following, we will discuss our variant of the HDG method on macro-elements, highlighting its differences to the standard HDG method that uses fully discontinuous elements. To this end, we will first focus on theory and formulation aspects and then detail a matrix-free solution methodology for the global problem. 

\subsection{The hybridized DG method}
We start by reviewing the hybridyzed DG formulation and discuss some of its mathematical properties. Let $\hat{u}_h = \hat{v}_h + \hat{g}_h$, where $\hat{v}_h \in M^h(0)$, and $\hat{g}_h \in M^h(g_D)$ is zero for all $F \in \skeleton^\circ$. The weak formulation reads: find $(\vect{q}_h, \, u_h, \, \hat{v}_h) \in \vect{V}^h \times W^h \times M^h(0)$ such that

\begin{subequations}
\begin{align}
	\Pair{\vect{q}_h}{\vect{v}_h}{\mesh} - \Pair{u_{h}}{\nabla \cdot \vect{v}_{h}}{\mesh}  + \pair{\hat{v}_{h}}{\vect{v}_{h} \cdot \vect{n}}{\boundary \mesh} &= -\pair{g_D}{\vect{v}_{h} \cdot \vect{n}}{\border_D} && \forall \vect{v}_{h} \in \vect{V}^{h},    \label{eq:mixed} \\
	-\Pair{\vect{a} u_{h} + \kappa \vect{q}_{h}}{\nabla w_{h}}{\mesh} + \pair{\left(\widehat{\vect{a} u_{h}} + \widehat{\kappa \vect{q}_{h}}\right) \cdot \vect{n}}{w_{h}}{\boundary \mesh} &= \Pair{f}{w_{h}}{\mesh}	&& \forall w_{h} \in W^{h},   \label{eq:pde} \\
	\pair{\jump{\left(\widehat{\vect{a} u_{h}} + \widehat{\kappa \vect{q}_{h}}\right) \cdot \vect{n}}}{\hat{\mu}_{h}}{\skeleton} &= \pair{g_N}{\hat{\mu}_{h}}{\border_N} && \forall \hat{\mu}_{h} \in M^{h}(0). \label{eq:transmission}
\end{align}  
Here, ``the numerical flux'' $\widehat{\vect{a} u_{h}} + \widehat{\kappa \vect{q}_{h}}$ and the trace $\hat{u}_{h}$ over $\boundary K$ are suitable approximations of $\vect{a} u_{h} +\kappa \vect{q}_{h}$ and $u_{h}$, respectively, that satisfy

\begin{align}
	\pair{(\widehat{\vect{a} u_{h}} + \widehat{\kappa \vect{q}_{h}}) \cdot \vect{n}}{w_{h}}{\boundary K} &= \pair{(\vect{a} u_{h} + \kappa \vect{q}_{h}) \cdot \vect{n}}{w_{h}}{\boundary K}, \label{eq:numerical_flux_approx} \\
	\pair{\hat{u}_{h}}{\vect{v}_{h} \cdot \vect{n}}{\boundary K} &= \pair{u_{h}}{\vect{v}_{h} \cdot \vect{n}}{\boundary K}. \label{eq:trace_approx}
\end{align}

Following \cite{nguyen2009implicit,cockburn2009hybridizable}, we choose the following relationship for the numerical flux

\begin{align}
	\widehat{\vect{a} u_{h}} + \widehat{\kappa \vect{q}_{h}} &= 
	\vect{a} \hat{u}_{h} + \kappa \vect{q}_{h} + \tau_K \left(u_{h} - \hat{u}_{h} \right) \vect{n} && \text{on } \boundary K.
\end{align}
Here $\tau_K$ is the ``local stabilization parameter'', which provides a mechanism to control both the accuracy and stability of the method. This choice for the numerical flux corresponds to the so-called local DG method. Existence and uniqueness of the stated weak form is shown in \cite{nguyen2009implicit,cockburn2009superconvergent} and proper choices for the stabilization parameter are discussed.
\end{subequations}

To understand strong versus weak imposition of the constraints, it is useful to transfer the above weak statement into its residual form. Using equations \ref{eq:numerical_flux_approx} and \ref{eq:trace_approx} and integration by parts, we may restate equations \eqref{eq:mixed}, \eqref{eq:pde}, and \eqref{eq:transmission} in residual form for each subdomain $K$:

\begin{subequations}
\begin{align}
	\Pair{\vect{q}_h + \grad u_h}{\vect{v}_{h}}{K} &= \vect{0}	&& \forall \vect{v}_{h} \in \vect{V}(K), \; K \in \mesh ,			\\
	\Pair{\div \left( \vect{a} u_h + \kappa \vect{q}_h  \right)}{w_{h}}{K} &= \Pair{f}{w_{h}}{K}  	&& \forall w_{h} \in W(K), \; K \in \mesh, 			\\
	\pair{\jump{(\vect{a} u_h + \kappa \vect{q}_h) \cdot \vect{n}}}{\hat{\mu}_{h}}{F} &=0			&& \forall \hat{\mu}_{h} \in M(F),  \; F \in \skeleton^\circ ,	\label{eq:transmission2}	\\
	\pair{u_h}{\hat{\mu}_{h}}{F} &= \pair{g_D}{\hat{\mu}_{h}}{F}												&& \forall \hat{\mu}_{h} \in M(F),  \; F \in \border_D ,		\\
	\pair{(\vect{a} u_h + \kappa \vect{q}_h) \cdot \vect{n}}{\hat{\mu}_{h}}{F} &= \pair{g_N}{\hat{\mu}_{h}}{F}							&& \forall \hat{\mu}_{h} \in M(F),  \; F \in \border_N .
\end{align}
\end{subequations}

If $\jump{\left(\vect{a} u_{h} + \kappa \vect{q}_{h}\right) \cdot \vect{n}}$ is an element of $M(F)$, then \eqref{eq:transmission2} implies that $\jump{\left(\vect{a} u_{h} + \kappa \vect{q}_{h}\right) \cdot \vect{n}} = 0$ pointwise in all faces $F \in \skeleton^\circ$, which means that the normal component of $\vect{a} u_{h} + \kappa \vect{q}_{h}$ is continuous across interior faces. Otherwise, continuity accross the normal is satisfied in a weak sense.

\subsection{Interpolation spaces}

The standard HDG method uses fully discontinuous elements, where the spaces $W(K), \vect{V}(K)$ and $M(F)$ are polynomials. There are several choices that lead to well posed discrete systems, see \cite{cockburn2009unified}. We use equal order interpolation spaces on simplicial elements, where $W(K)$ and $\vect{V}(K)$ consist of scalar and vector valued polynomials of total degree $p$ on element $K$, respectively, and $M(F)$ is a scalar valued space of polynomials of total degree $p$ on face $F$, i.e.

\begin{subequations}
\begin{align}
	W(K) &:= \mathcal{P}_p(K),	\\
	\vect{V}(K) &:= \left(\mathcal{P}_p(K) \right)^d, \\
	M(F) &:= \mathcal{P}_p(F).
\end{align}
\end{subequations}
Our macro-element variant of the HDG method uses patches of standard $C^0$ continuous elements that are discontinuous only across patch boundaries. Hence, on macro-elements, we use continuous piece-wise polynomials, that is,

\begin{subequations}
\begin{align}
	W(K) 			&:= \List{w \in C^0(K) \; : \; \left. w \right|_{\element} \in \mathcal{P}_p(\element), \; \forall \element \in K},	\\
	\vect{V}(K) 	&:= \List{v \in \left(C^0(K)\right)^d \; : \; \left. w \right|_{\element} \in \left(\mathcal{P}_p(\element)\right)^d, \; \forall \element \in K},	\\
	M(F) 				&:= \List{m \in C^0(F) \; : \; \left. m \right|_{\face} \in \mathcal{P}_p(\face), \; \forall \face \in F}.
\end{align}
\end{subequations}

\begin{figure}
    \centering
     \subfloat[Standard HDG]{{\includegraphics[width=0.45\textwidth]{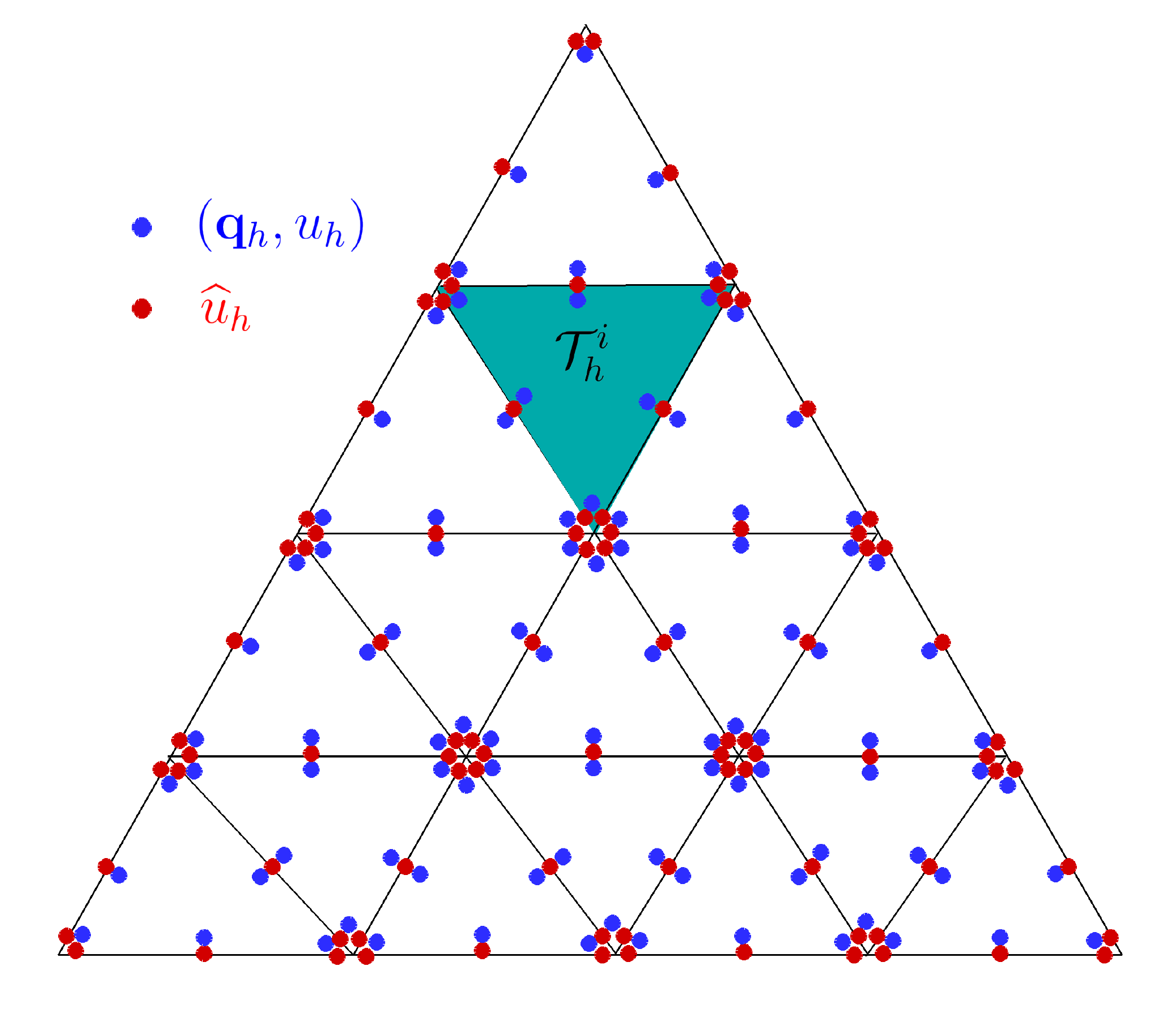} }}%
     \subfloat[Zoom-in local degrees of freedom]{{\includegraphics[width=0.45\textwidth]{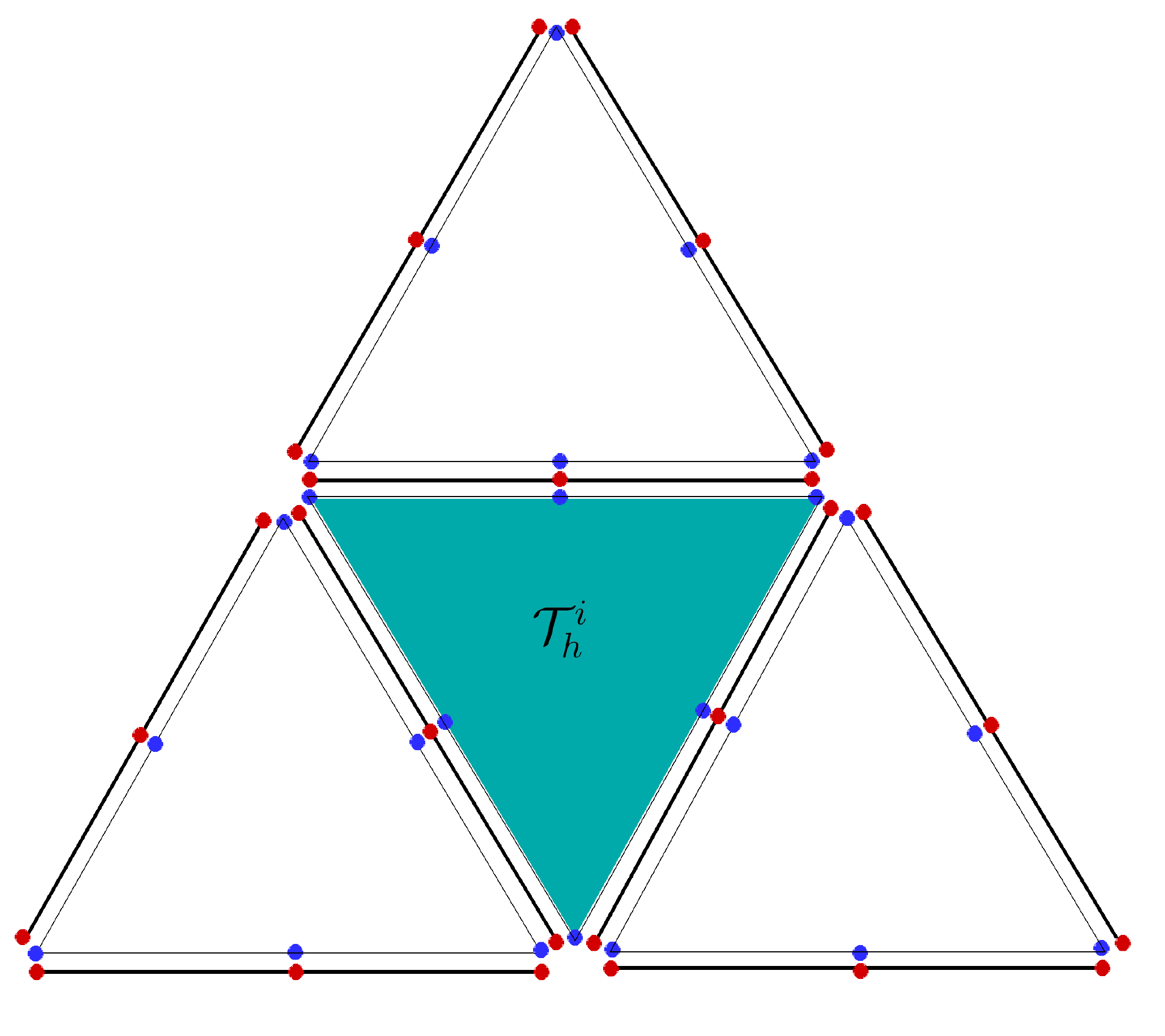} }}
    \caption{Illustration of the degrees of freedom for the standard HDG method with $p=2$. The black lines represent the boundaries of the standard discontinuous elements.}
    \label{fig:shdg}%
\end{figure}

\begin{figure}
    \centering
     \subfloat[ Macro-element HDG]{{\includegraphics[width=0.45\textwidth]{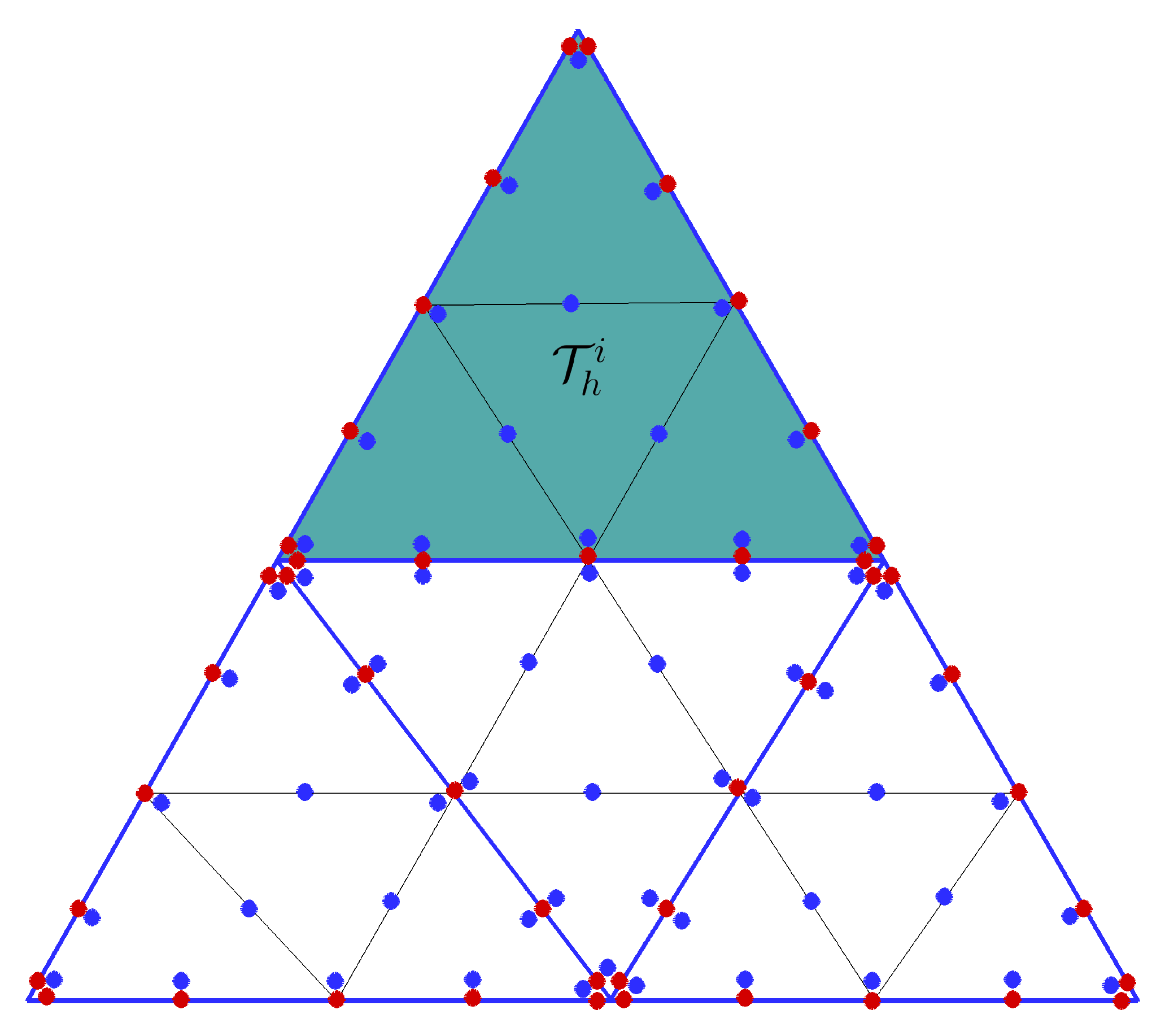} }}%
     \subfloat[Zoom-in local degrees of freedom]{{\includegraphics[width=0.45\textwidth]{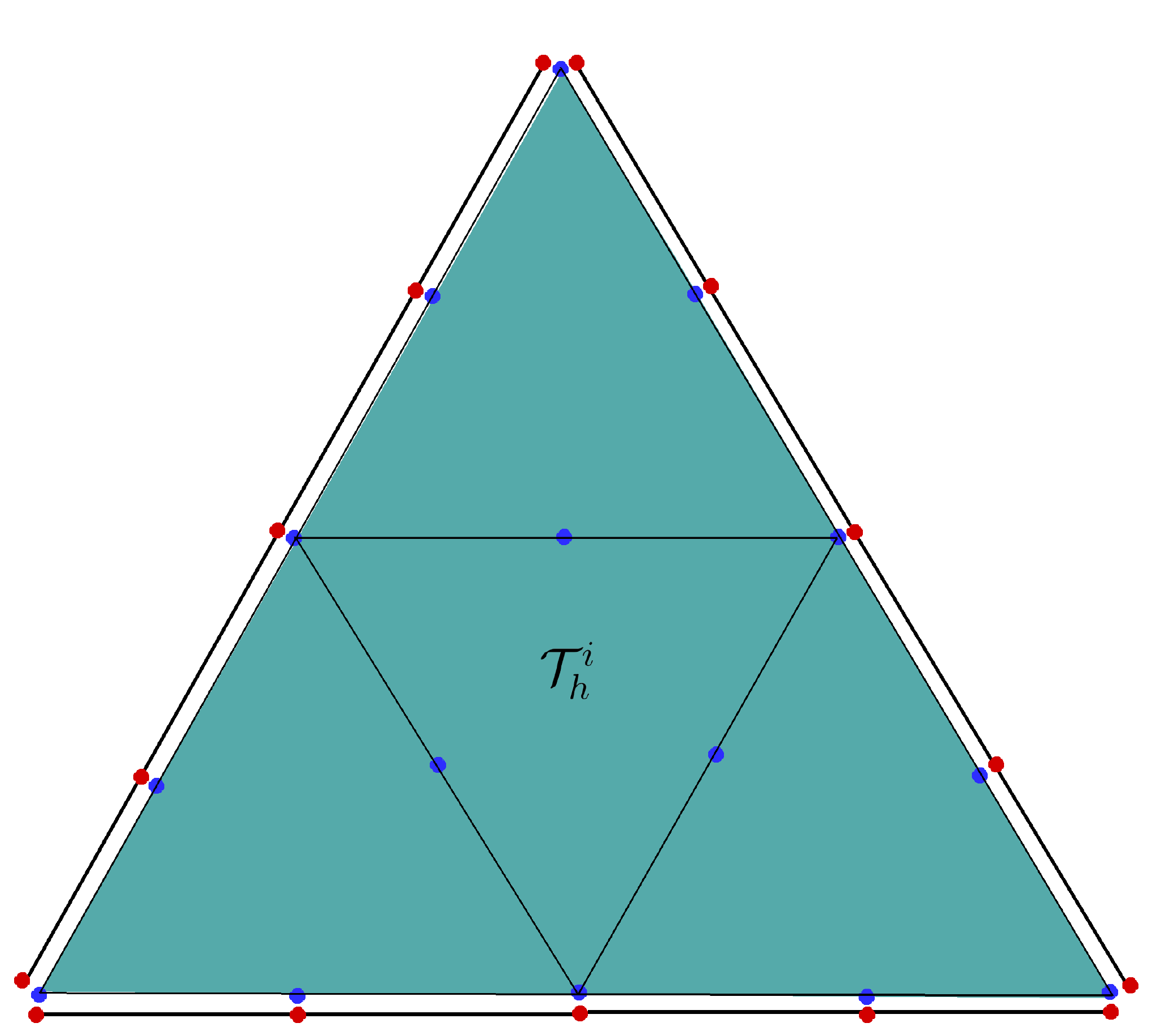} }}%
    \caption{Illustration of the degrees of freedom for the macro-element HDG method with $p=2$. The blue lines represent the boundary of the macro-element patches, while the black lines represent the boundaries of the $C^0$-continuous elements within each macro-element.}%
    \label{fig:mhdg}%
\end{figure}

Figures \ref{fig:shdg} and \ref{fig:mhdg} illustrate the degree-of-freedom structure of standard HDG versus macro-element HDG, respectively. It is easy to see that the macro-element HDG method contains the standard HDG method as a special case. 

The concept of using macro-elements in an HDG context is driven by the idea to arrive at a HDG variant that can be easily tailored to the available computational resources, both in terms of shared and distributed memory systems. In addition, the macro-element variant of the HDG method implicitly contains a simple approach to domain decomposition. 

Using the relationship for the numerical flux, we may summarize the weak formulation as follows: find $U_{h} = (\vect{q}_{h}, u_{h}) \in \vect{V}^{h} \times W^{h} $ and $\hat{v}_{h} \in M^{h}(0)$ such that

\begin{subequations}
\label{eq:weak_form}
\begin{align}
	\bform{A}{U_{h}}{V_{h}} + \bform{B}{\hat{v}_{h}}{V_{h}} &= \lform{l}{V_{h}} && \forall V_{h} = (\vect{v}_{h}, w_{h}) \in \vect{V}^{h} \times W^{h} ,			\\
	\bform{C}{U_{h}}{\hat{\mu}_{h}}  + \bform{D}{\hat{v}_{h}}{\hat{\mu}_{h}} &= \lform{m}{\hat{\mu}_{h}} && \forall \hat{\mu}_{h} \in M^{h}(0),
\end{align}
where, for any $U_{h}, V_{h} \in \vect{V}^{h} \times W^{h}$ and $\hat{v}_{h}, \hat{\mu}_{h} \in M^{h}(0)$, we have
\begin{align}
\bform{A}{U_{h}}{V_{h}} &= 
	\Pair{\vect{q}_{h}}{\vect{v}_{h}}{\mesh} 
	- \Pair{u_{h}}{\nabla \cdot \vect{v}_{h}}{\mesh}
	-\Pair{\vect{a} u_{h} + \kappa \vect{q}_{h}}{\nabla w_{h}}{\mesh}
	+\pair{\kappa \vect{q}_{h} \cdot \vect{n} + \tau u_{h}}{w_{h}}{\boundary \mesh}, \\
\bform{B}{\hat{v}_{h}}{V_{h}} &=
	\pair{\hat{v}_{h}}{\vect{v}_{h} \cdot \vect{n}}{\boundary \mesh}
	+ \pair{\left(\vect{a}\cdot \vect{n} - \tau \right) \hat{v}_{h}}{w_{h}}{\boundary \mesh}, \\
\lform{l}{V_{h}} &= \Pair{f}{w_{h}}{\mesh} - \pair{g_D}{\vect{v}_{h} \cdot \vect{n}}{\border_D},  	\\
\bform{C}{U_{h}}{\hat{\mu}_{h}} &= \pair{\jump{\kappa \vect{q}_{h} \cdot \vect{n} + \tau u_{h} }}{\hat{\mu}_{h}}{\skeleton},\\
\bform{D}{\hat{v}_{h}}{\hat{\mu}_{h}} &= \pair{\jump{(\vect{a} \cdot \vect{n} - \tau) \hat{v}_{h} }}{\hat{\mu}_{h}}{\skeleton} ,		\\
\lform{m}{\hat{\mu}_{h}} &= \pair{g_N}{\hat{\mu}_{h}}{\border_N}.
\end{align}
\end{subequations}

The variational statements in \eqref{eq:weak_form} only contain first-order differential operators. Hence, the weak form \eqref{eq:weak_form} remains well defined when $C^0$-continuous interpolation spaces defined on each macro-element are applied. We therefore conclude that the weak form \eqref{eq:weak_form} holds for both the standard as well as the macro-element HDG variants.

\subsection{Matrix-free solution of the global part of the linear system of equations}

\subsubsection{Matrix equations and solution approach}
The weak form in \eqref{eq:weak_form} translates naturally to the following system of matrix equations

\begin{align}
\label{eq:D}
	\begin{bmatrix}
		\mat{A} & \mat{B} \\
		\mat{C} & \mat{D}
	\end{bmatrix}
	\begin{bmatrix}
		\mat{u}  \\
		\mat{\hat{u}} 
	\end{bmatrix} = 
	\begin{bmatrix}
		\mat{R}_u  \\
		\mat{R}_{\hat{u}} 
	\end{bmatrix},
\end{align}
where $\mat{u} \in \mathbb{R}^z$ denote degrees of freedom associated with the discretization of the domain into discontinuous sub-domains (standard elements or macro-elements) and $\mat{\hat{u}} \in \mathbb{R}^{\hat{z}}$ denotes a vector with degrees of freedom associated with the discretization of trace variables associated with the mesh skeleton. The matrices have a special block-structure due to the discontinuous nature of the approximation spaces and the coupling of patches by means of interface variables. In particular, matrix $\mat{A}$ is a block-diagonal matrix, where each block is associated with one macro-element. All block-matrices of $\mat{A}$ are invertible.

Using the Schur complement, we can solve the above matrix problem for $\mat{\hat{u}}$ by condensing out the degrees of freedom $\mat{u}$, which leads to

\begin{subequations}
\begin{align}
	\left( \mat{D} - \mat{C} \mat{A}^{-1} \mat{B} \right) \mat{\hat{u}} &= \mat{R}_{\hat{u}} - \mat{C} \mat{A}^{-1} \mat{R}_{u} && (\text{Solve for interface dofs - global problem}),	\label{eq:global_problem}	\\
	\mat{A} \, \mat{u} &= \mat{R}_{u} - \mat{B} \mat{\hat{u}} && (\text{Solve for interior dofs - local problem}).
	\label{eq:local_problem}
\end{align}
\end{subequations}
It is common to solve the global problem with an iterative solver. In a parallel computing environment, this can be particularly efficiently with a matrix-free solution approach.

\subsubsection{Matrix-free solution methodology}
We consider the following iterative solution approach to the global problem \eqref{eq:global_problem}. We first compute $\mat{f} = \mat{R}_{\hat{u}} - \mat{C} \mat{A}^{-1} \mat{R}_{u}$. We then solve $\left( \mat{D} - \mat{C} \mat{A}^{-1} \mat{B} \right) \mat{\hat{u}} = \mat{f}$ by an iterative solver. Here we utilize the block-structure of the matrices $\mat{A}, \mat{B}, \mat{C}, \mat{D}$ to perform all operations locally and in parallel. Let the subscript $(\element)$ denote the localization of operations to the (macro) element level and $(\face)$ denote localization of operations to the (macro) element faces. Then, each iteration of the matrix-free algorithm can be performed for each (macro) element, $\element$,  in parallel, in the following four steps

\begin{enumerate}
	\item $\mat{x}_{(\element)} = \mat{B}_{(\element)} \mat{\hat{u}}_{(\element)},$
	\item $\mat{y}_{(\element)} = \mat{A}_{(\element)}^{-1} \mat{x}_{(\element)},$ 
	\item $\mat{\hat{v}}_{(\element)} = \mat{C}_{(\element)} \mat{y}_{(\element)}.$
\end{enumerate}
Step four involves a reduction that requires data associated with the mesh skeleton faces and data from (macro) elements, and thus requires communication among processors. 

\begin{enumerate}
	\setcounter{enumi}{3}
	\item $\mat{\hat{w}}_{(\face)} = \mat{D}_{(\face)} \mat{\hat{u}}_{(\face)} - \mat{\hat{v}}_{(\face1)} -  \mat{\hat{v}}_{(\face2)}.$
\end{enumerate}

\begin{remark} The inverse $\mat{A}_{(\element)}$ is applied by computing and storing its LU factorization in-place.
\end{remark}

\begin{remark} A simple pre-conditioner of the matrix-free method is to apply $\mat{D}^{-1}_{(\face)}$ to step four of the matrix-free algorithm. 
\end{remark}

\begin{figure}[h!]
	\centering
	\includegraphics[trim = 0cm  0cm 0cm 0cm,clip,width=0.28\textwidth]{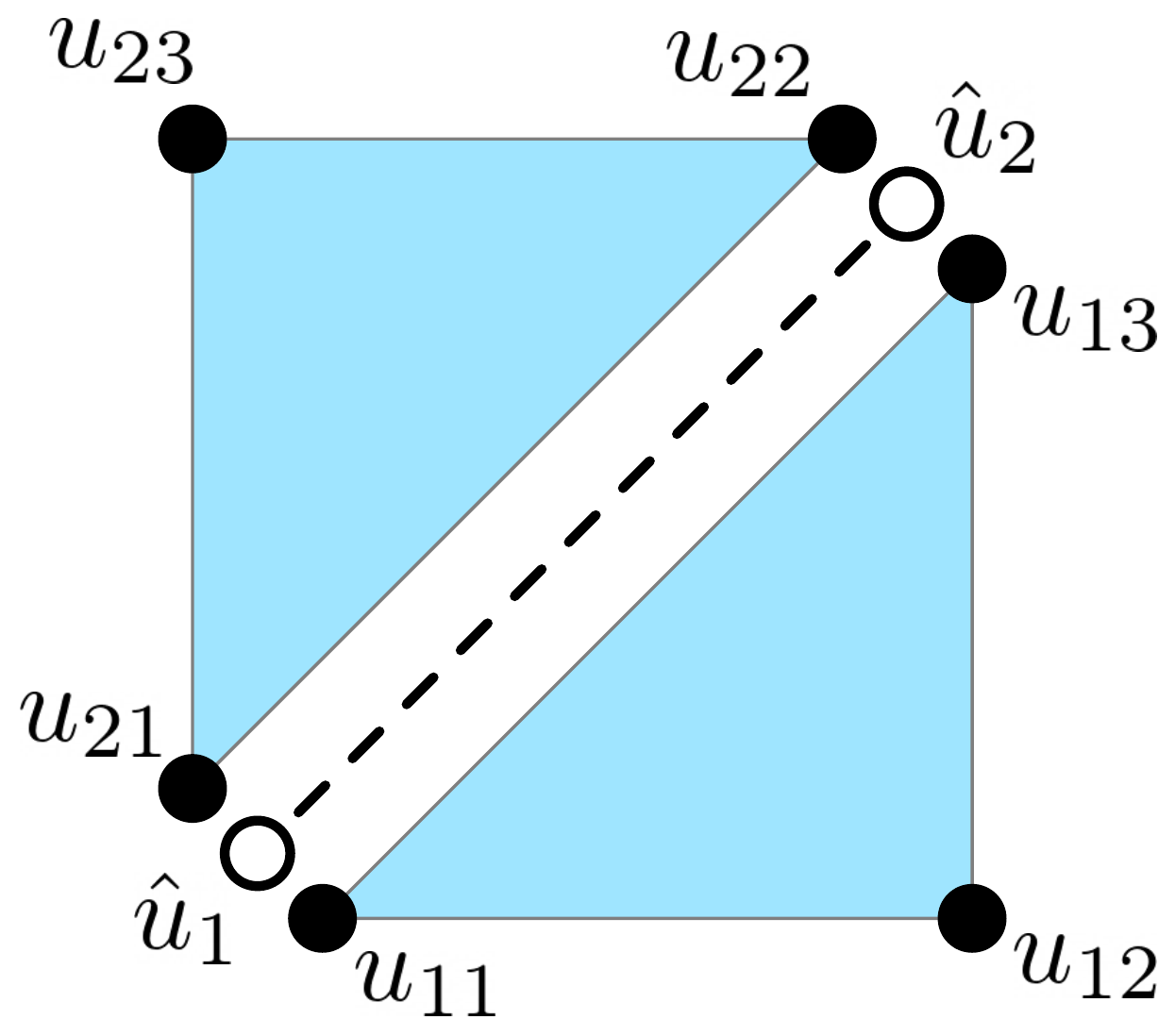}
	\caption{Example discretization for a two-macro-element mesh with homogeneous boundary conditions and one macro-element interface.}
\label{fig:hdg}
\end{figure}

\begin{eexample}
Consider the two-macro-element mesh depicted in Figure \ref{fig:hdg}. We are interested in performing one matrix-vector multiplication in the iterative solution of \eqref{eq:global_problem}. The first three steps can be performed as follows

\begin{align*}
	\begin{bmatrix}
		x_{11} \\ x_{12} \\ x_{13} \\ \hline  x_{21} \\ x_{22} \\ x_{23}
	\end{bmatrix} = \underbrace{ 
	\begin{bmatrix}
		 * &  *  \\ * & * \\ * & * \\ \hline * & * \\ * & * \\ * & *
	\end{bmatrix}}_{\mat{B}}
	\begin{bmatrix}
		\hat{u}_1  \\
		\hat{u}_2 
	\end{bmatrix},
&&
	\begin{bmatrix}
		y_{11} \\ y_{12} \\ y_{13} \\ \hline  y_{21} \\ y_{22} \\ y_{23}
	\end{bmatrix} = \underbrace{ 
	\begin{bmatrix}[c c c | c c c]
		 * & * & * &    &    &  \\ * & * & * &   &   &   \\ * & * & * &   &   &   \\ \hline 
			&		&		&		* &	*	& * \\	&		&		&		* &	*	& * \\ &		&		&		* &	*	& *
	\end{bmatrix}}_{\mat{A}^{-1}}
	\begin{bmatrix}
		x_{11} \\ x_{12} \\ x_{13} \\ \hline  x_{21} \\ x_{22} \\ x_{23}
	\end{bmatrix},
&&
	\begin{bmatrix}
		\hat{v}_{11}  \\ \hat{v}_{12}  \\ \hline  \hat{v}_{21}  \\ \hat{v}_{22}
	\end{bmatrix} = \underbrace{ 
	\begin{bmatrix}[c c c | c c c]
		 * & * & * &    &    &  \\ * & * & * &   &   &   \\ \hline 
			&		&		&		* &	*	& * \\	&		&		&		* &	*	& *
	\end{bmatrix}}_{\mat{C}}
	\begin{bmatrix}
		y_{11} \\ y_{12} \\ y_{13} \\ \hline  y_{21} \\ y_{22} \\ y_{23}
	\end{bmatrix}.
\end{align*}

Here the horizontal and vertical lines depict the partitioning of the matrices and vectors, possibly on different processors. Hence, all computations can be performed locally for each macro-element in the mesh.

The final step, step four, can be performed for each macro-element face in the mesh skeleton. It involves a reduction and thus requires communication among processors.

\begin{align*}
	\begin{bmatrix}
		\hat{w}_1  \\ \hat{w}_2
	\end{bmatrix} =   \underbrace{ 
	\begin{bmatrix}
		 * & * \\ * & *
	\end{bmatrix}}_{\mat{D}}
	\begin{bmatrix}
		\hat{u}_1  \\ \hat{u}_2
	\end{bmatrix} - 
	\begin{bmatrix}
		\hat{v}_{11}  \\ \hat{v}_{12}
	\end{bmatrix} - 
	\begin{bmatrix}
		\hat{v}_{21}  \\ \hat{v}_{22}
	\end{bmatrix}.
\end{align*}
\end{eexample}

\subsection{Additional stabilization for advection-dominated problems}
We must expect that for advection-dominated problems, oscillations in the solution that are typical for standard $C^0$-continuous finite elements, will appear locally within each macro-element, in particular for patches with a large number of $C^0$-continuous elements. To mitigate these oscillations within macro-elements, we add standard residual-based stabilization within each macro-element, using the streamline-upwind-Petrov Galerkin (SUPG) method \cite{brooks1980streamline,brooks1982streamline,christie1976finite,donea2003finite,heinrich1977upwind}. The modified bi-linear and linear forms are,

\begin{subequations}
\begin{align}
\bform{B_{supg}}{U_{h}}{V_{h}} &= \bform{B}{U_{h}}{V_{h}}  + \sum_{\element \in K} \int_{\element} \mathcal{P}(w_{h}) \tau_{supg} \div \left(\vect{a} u_{h} - \kappa \grad u_{h}\right), 	\\
\lform{l_{supg}}{V_{h}} &= \lform{l}{V_{h}} + \sum_{\element \in K} \int_{\element} \mathcal{P}(w_{h}) \tau_{supg} f
\end{align}
with

\begin{align}
	\mathcal{P}(w_{h}) &= \vect{a} \cdot \nabla w_{h} 	,						\\
	\tau_{supg} &= \frac{h}{2 |\vect{a}|} \left( \coth{\Pe_\element} + 1 / \Pe_\element  \right),
\end{align}
\end{subequations}
where $\Pe_\element$ is the element Peclet number.

\section{Properties of the macro-element hybridized DG method}
In the following, we first demonstrate accuracy and convergence of our macro-element variant of the HDG method. We then discuss and highlight practical advantages of our discretization strategy in terms of adaptive refinement, domain decomposition and load balancing

\subsection{Accuracy and convergence}

\subsubsection{A benchmark with a customizable internal layer skew to the mesh}

Based on our model problem \eqref{C21}, we adopt a benchmark with the following exact solution, 

\begin{align}
\label{C24a}
		{u}_{exact}  &=0.5 \left(1+\tanh{\left(\frac{y-2x+0.4}{\kappa}\right)} \right)       \quad   &\text{for} \; & d  =  2, \\
		\label{C24b}
		{u}_{exact}	 &=0.5 \left(1+\tanh{\left(\frac{x+y+z-0.5}{\kappa}\right)} \right) 	  \quad   &\text{for} \; & d  = 3,
\end{align}
inspired by the one-dimensional steady-state phase-field solution of the one-dimensional Allen-Cahn problem that features a diffuse interface, see e.g.\ \cite{nguyen2017phase} and the references therein. The field solutions \eqref{C24a} and \eqref{C24b} represent spatially translated and rotated variants, such that an internal layer of varying sharpness develops skew to the (structured) mesh. In our case, we choose the line $2x -y = 0.4$ and the surface $x+y+z = 0.5$ as the center of the diffuse internal layer for the two-dimensional case and the three-dimensional case, respectively. The corresponding boundary conditions can be derived directly from \eqref{C24a} and \eqref{C24b}, and the corresponding source terms can be determined by inserting the solutions \eqref{C24a} and \eqref{C24b} in the advection-diffusion equation \eqref{C21} and solving for $f$. We emphasize that, provided we choose the advective velocity field $\bm{a}$ constant and in the same direction as the internal layer, $f$ is independent of $\bm{a}$, as $\bm{a} \cdot \nabla u = 0$.

\subsubsection{Diffusion vs.\ advection dominated settings}

\begin{figure}
    \centering
    \subfloat[\centering 2D domain]{{\includegraphics[width=0.5\textwidth]{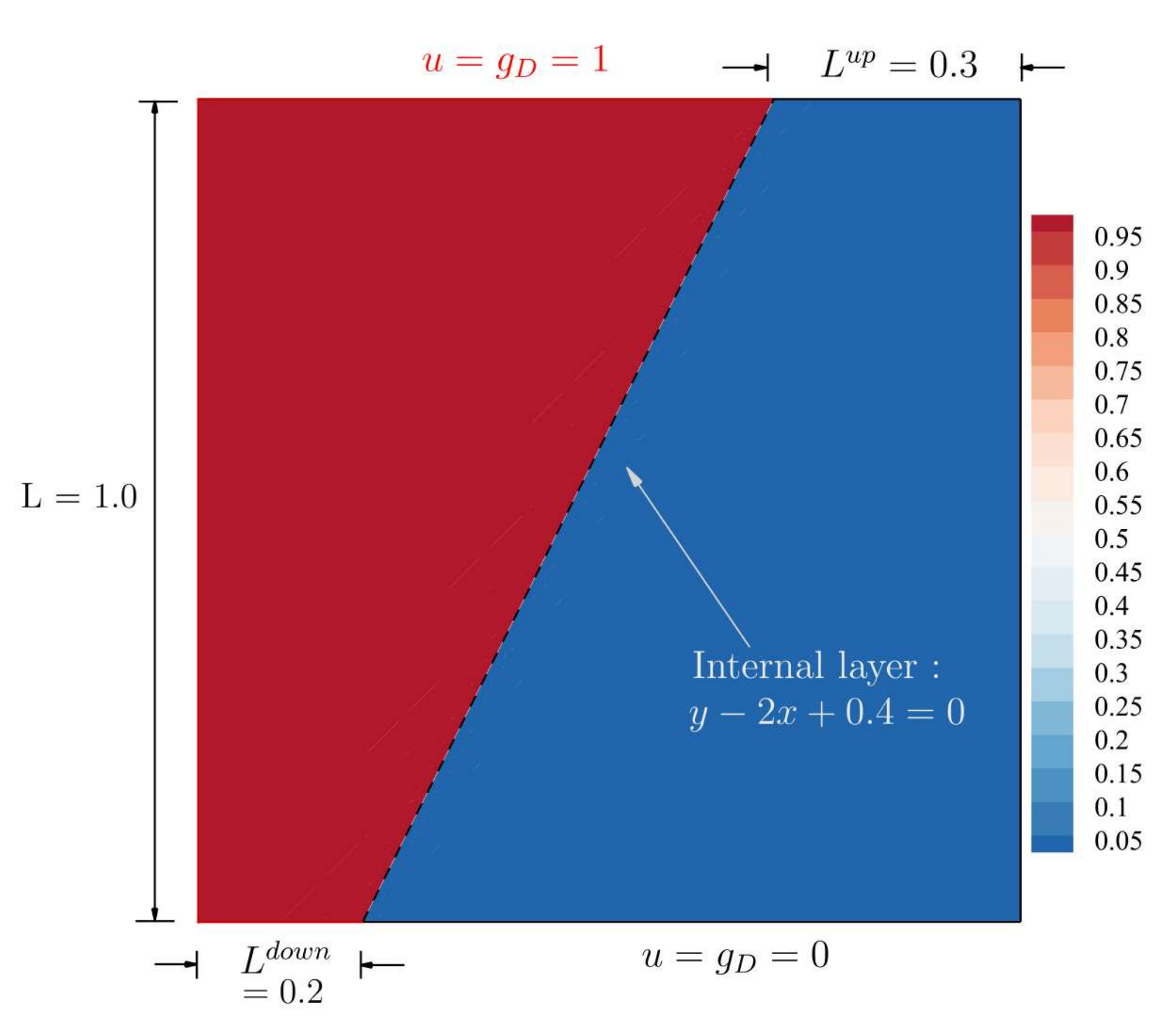} }}%
    \subfloat[\centering 3D domain]{{\includegraphics[width=0.5\textwidth]{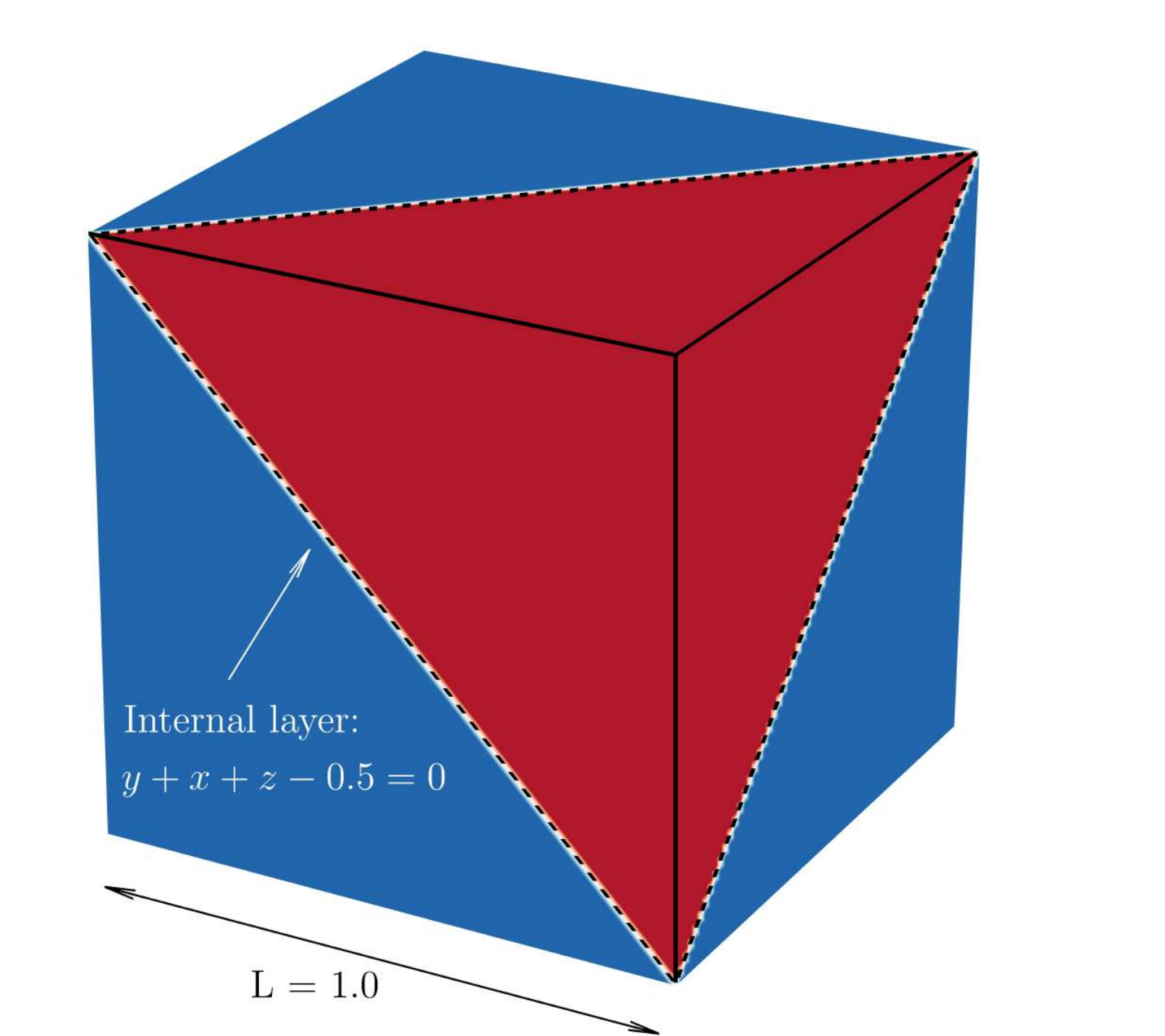} }}%
    \caption{Exact solutions of the advection-diffusion benchmark with a customizable internal layer skew to the mesh ($\text{Pe} \gg 1$) on a unit square and a unit cube.}%
    \label{fig:test case}%
\end{figure}

The solution regime is determined by the Peclet number, $\text{Pe} = \left| \vect{a} \right| L / \kappa$, where $L$ denotes the characteristic length scale of the domain $\Omega$. In our benchmark, the Peclet number can be modified by choosing different diffusion coefficients $\kappa$ at a given velocity and length scale. Figures \ref{fig:test case}a and \ref{fig:test case}b plot the exact solutions \eqref{C24a} and \eqref{C24b} on a unit square and a unit cube, showing a sharp internal layer in the advection dominated regime ($\text{Pe} \gg 1$).

\begin{figure}
    \centering
    \subfloat[2D structured mesh.]{{\includegraphics[width=0.48\textwidth]{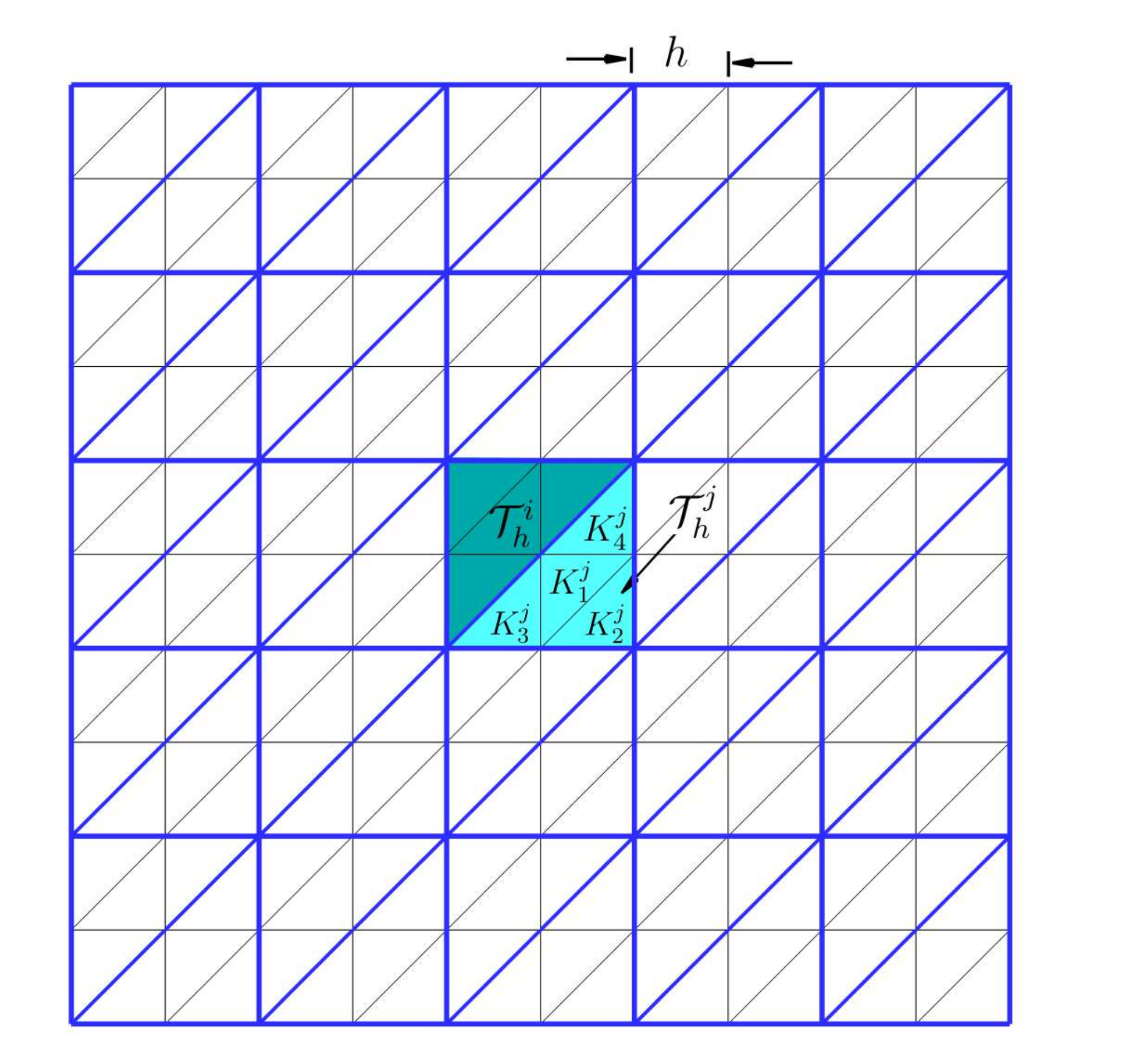} }}%
    \subfloat[3D structured mesh.]{{\includegraphics[width=0.51\textwidth]{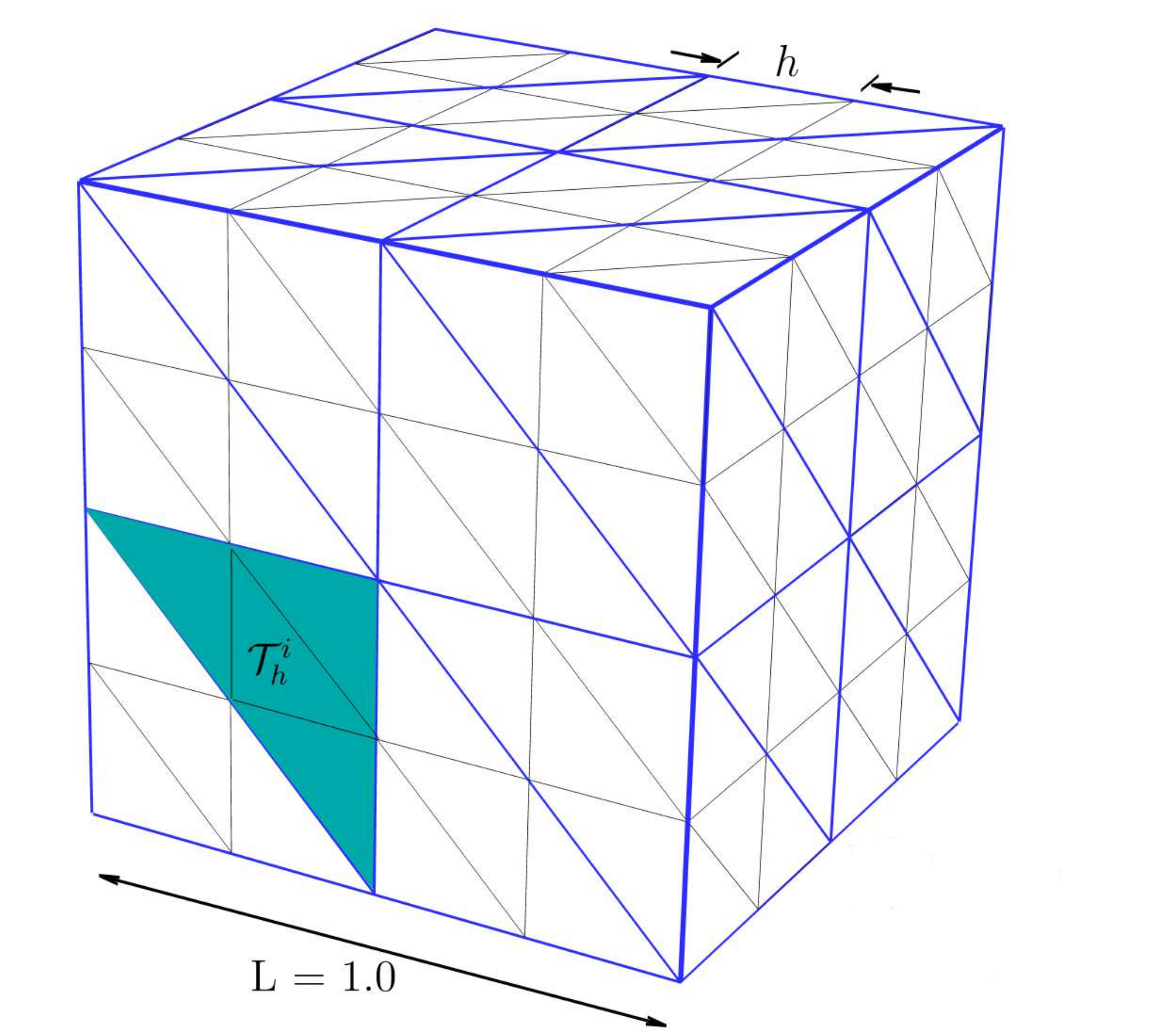} }} %
    \caption{Plot of standard elements (black) and macro-elements (blue).}%
    \label{fig:mesh}%
\end{figure}

We first assess our macro-element HDG variant in terms of its ability to achieve optimal rates of convergence. To this end, we fix $\kappa = 0.4 $ at an advective field of $\vect{a}=(1,1)$ in 2D and $\vect{a}=(1,1,1)$ in 3D, both parallel to the internal layer, arriving at a diffusion-dominated solution. We consider polynomial orders $p=1$ to $p=4$ and consider meshes as shown in Figures \ref{fig:mesh}a and \ref{fig:mesh}b, where we choose the number $m$ of elements along a macro-element edge in each direction to be two.

\begin{figure}
    \centering
    \subfloat[Macro-element HDG ($m=2$).]{{\includegraphics[width=0.5\textwidth]{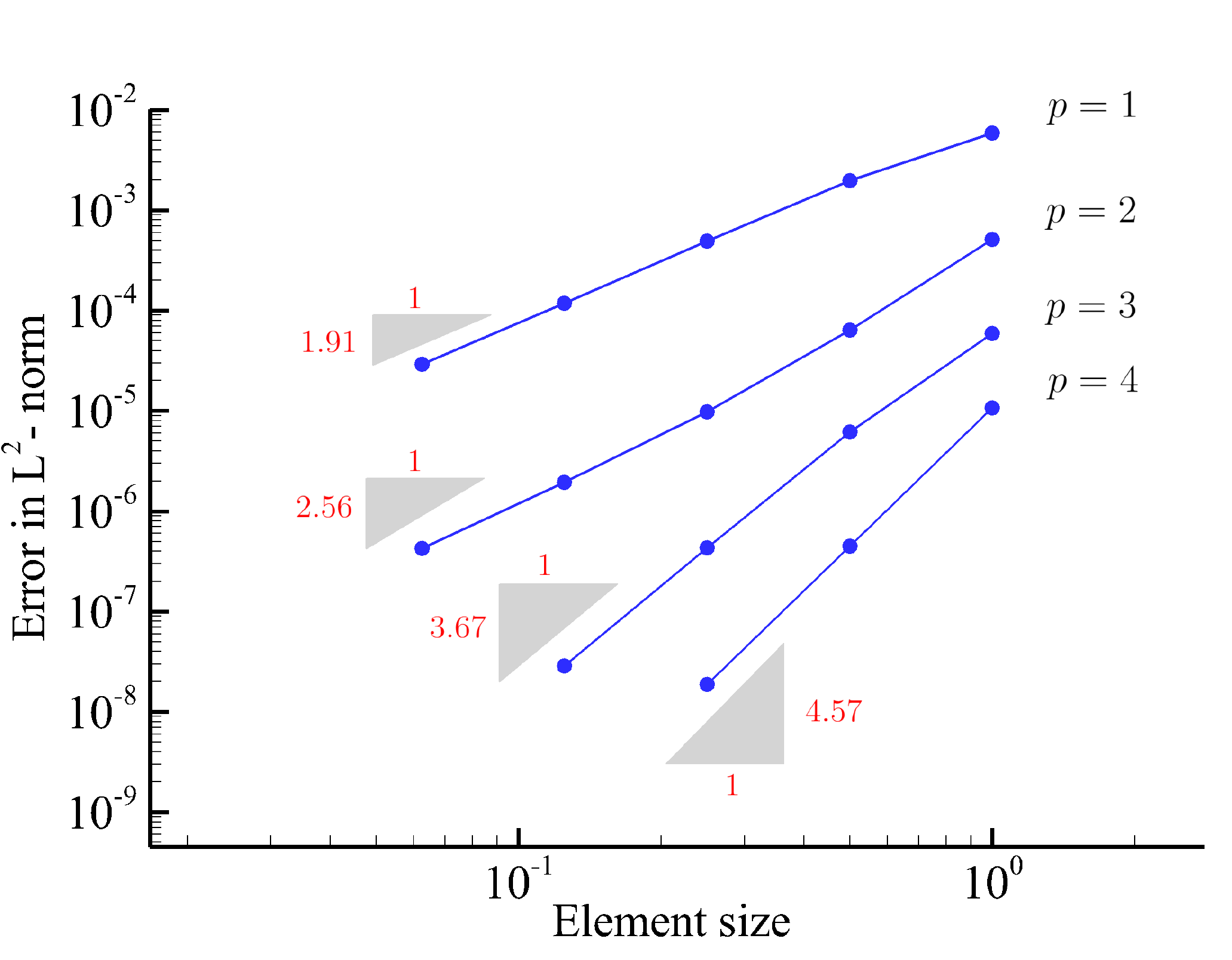} }}%
    \subfloat[Standard HDG.]{{\includegraphics[width=0.5\textwidth]{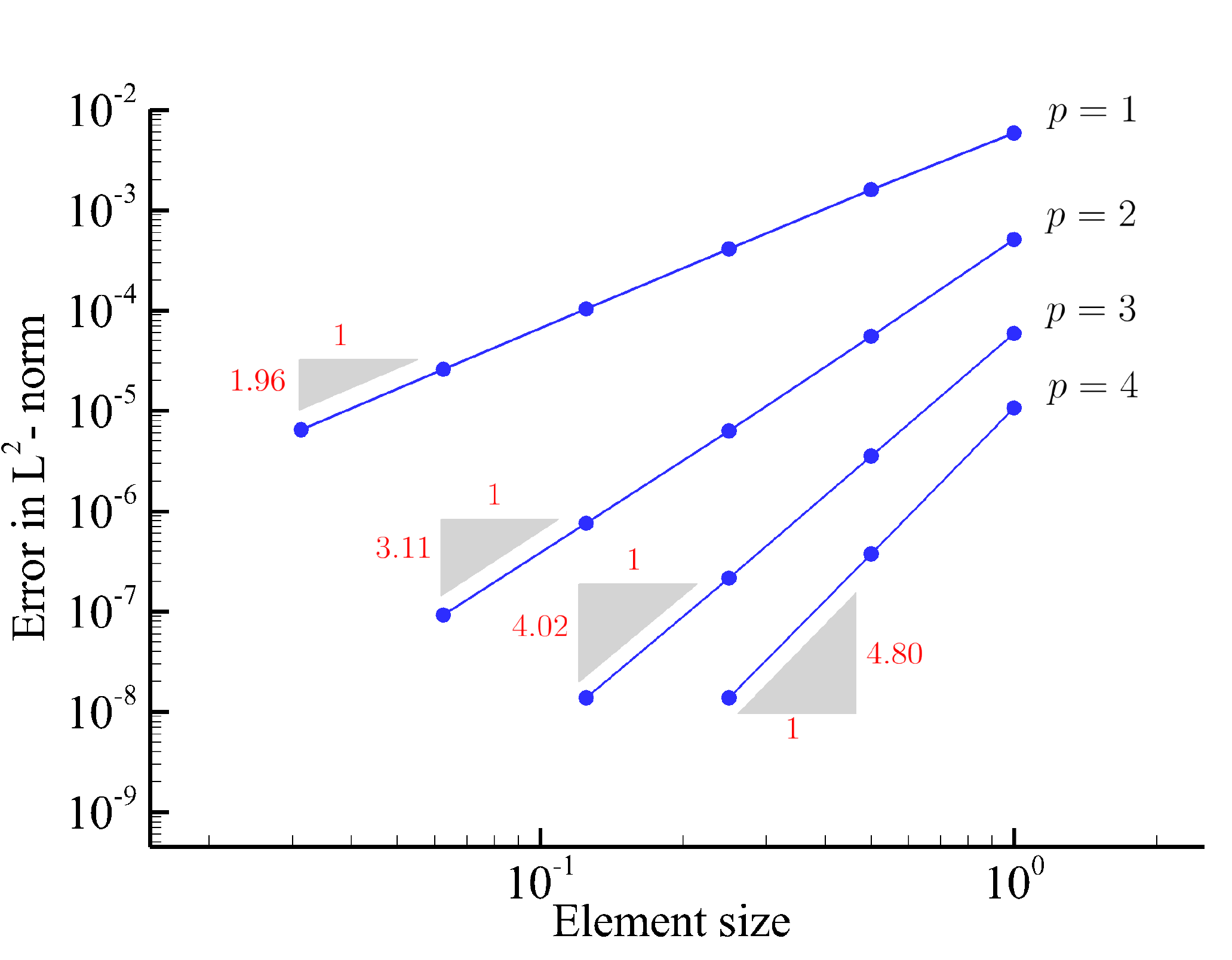} }}
    \caption{$L^2$ error obtained for the 2D benchmark with $\kappa=0.4$ (Pe = 3.5) under uniform refinement.}%
    \label{fig:rate2d}%
\end{figure}

\begin{figure}
    \centering
    \subfloat[Macro-element HDG ($m=2$).]{{\includegraphics[width=0.5\textwidth]{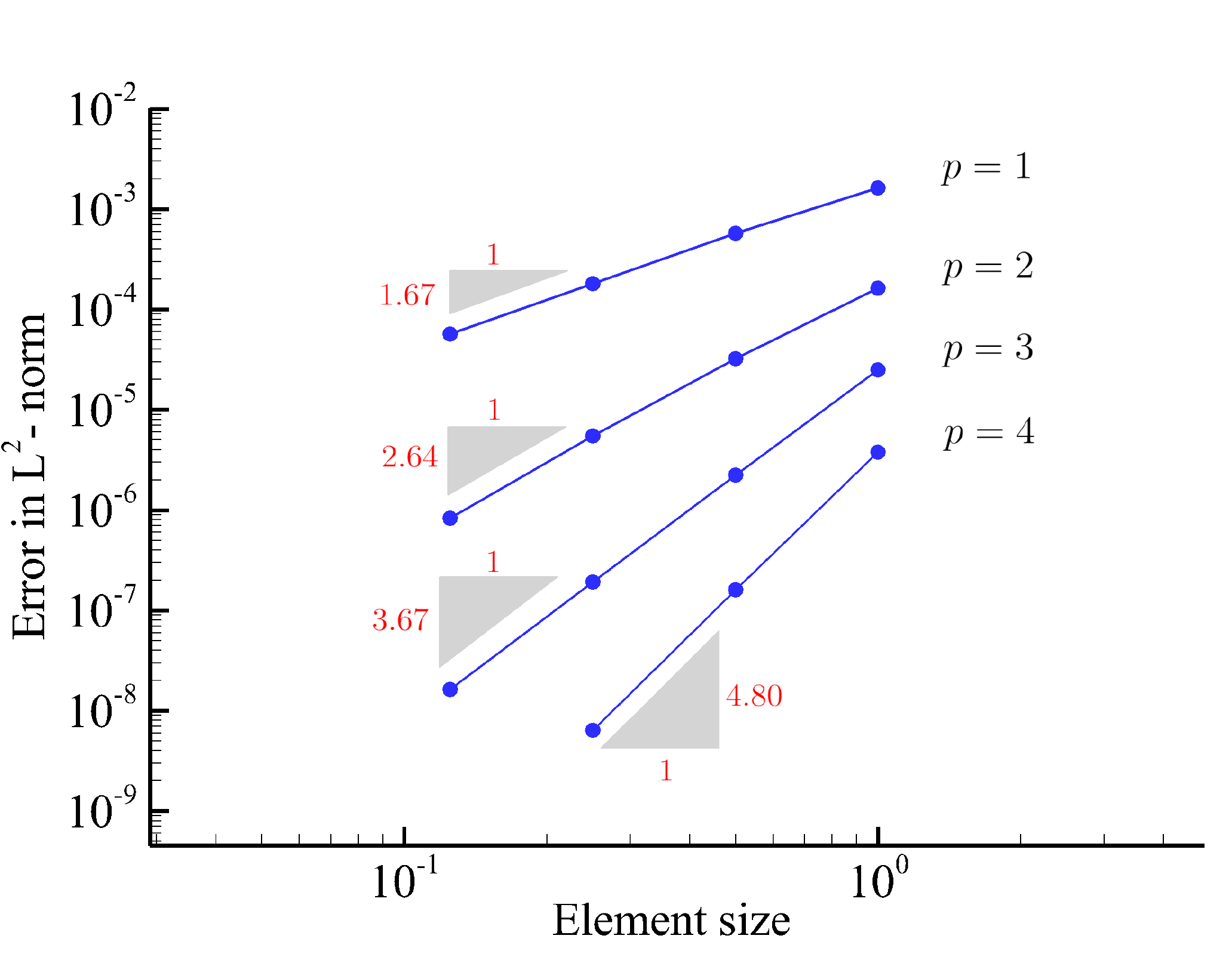} }}%
    \subfloat[Standard HDG.]{{\includegraphics[width=0.5\textwidth]{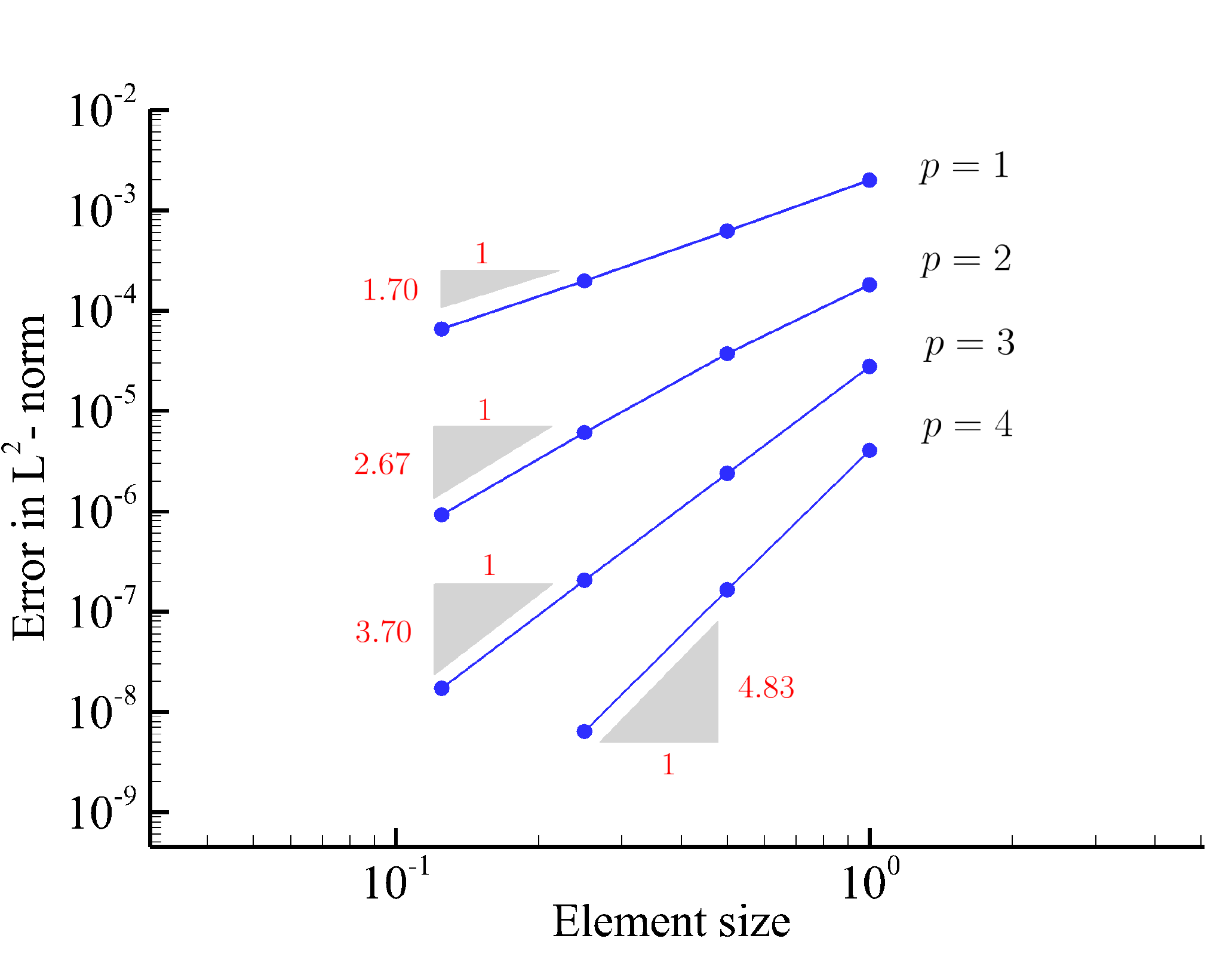} }}%
    \caption{$L^2$ error obtained for the 3D benchmark with $\kappa=0.4$ (Pe = 4.3) under uniform refinement.}%
    \label{fig:rate3d}%
\end{figure}

In Figures \ref{fig:rate2d}a and \ref{fig:rate3d}a, we plot the error in the $L^2$ norm versus the element size under uniform refinement of the macro-elements for the two-dimensional benchmark and the three-dimensional benchmark, respectively. We observe that we achieve optimal rates of convergence $p+1$ in all cases. We also compute the same examples with the standard HDG method and plot the resulting convergence curves in Figures \ref{fig:rate2d}b and \ref{fig:rate3d}b. We observe that there is practically no difference both in terms of error level and convergence rate between the two HDG variants.

\begin{figure}[t]
    \centering
    \subfloat[\centering Without SUPG stabilization.]{{\includegraphics[width=0.5\textwidth]{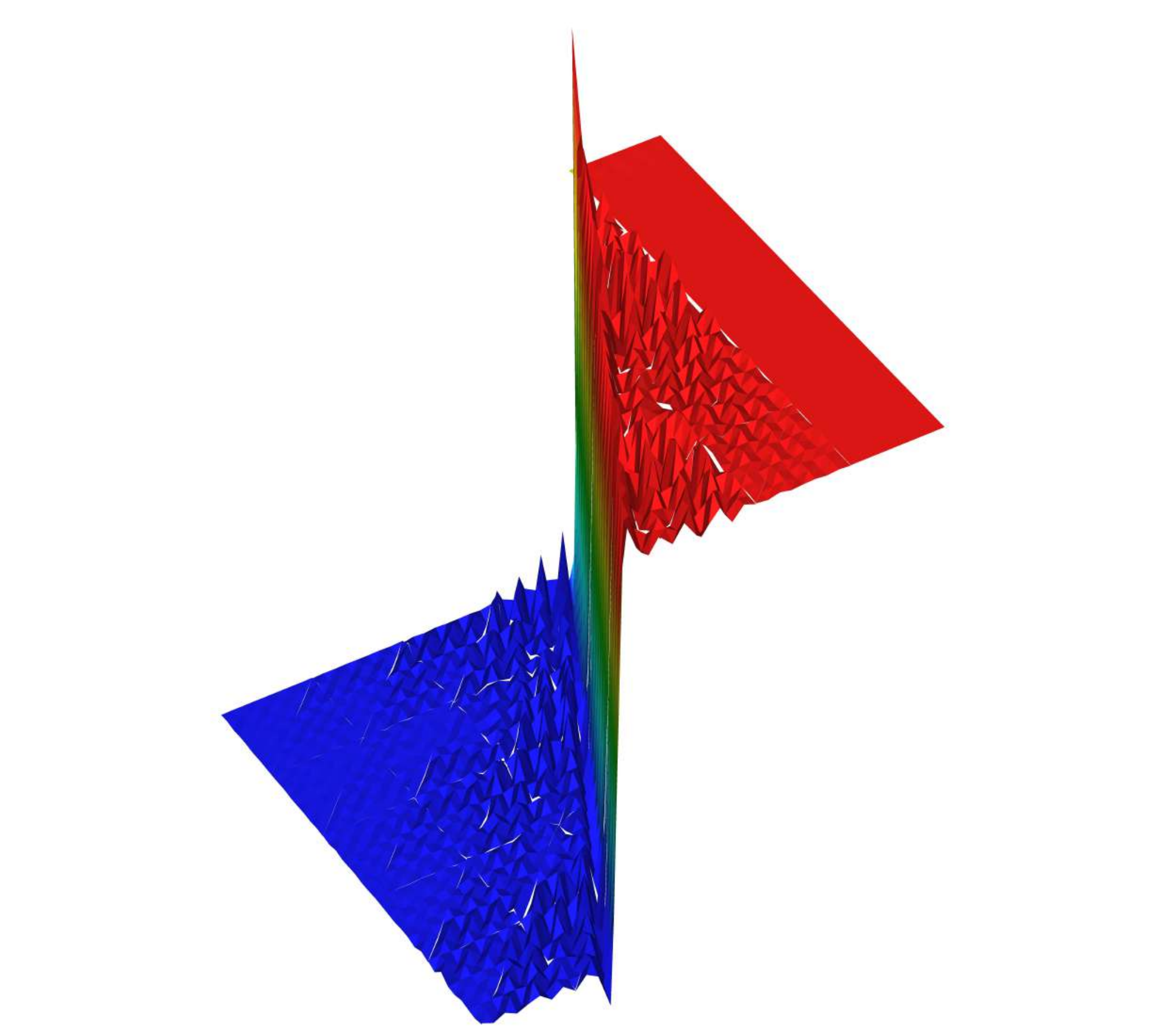} }}%
    \subfloat[\centering With SUPG stabilization.]{{\includegraphics[width=0.5\textwidth]{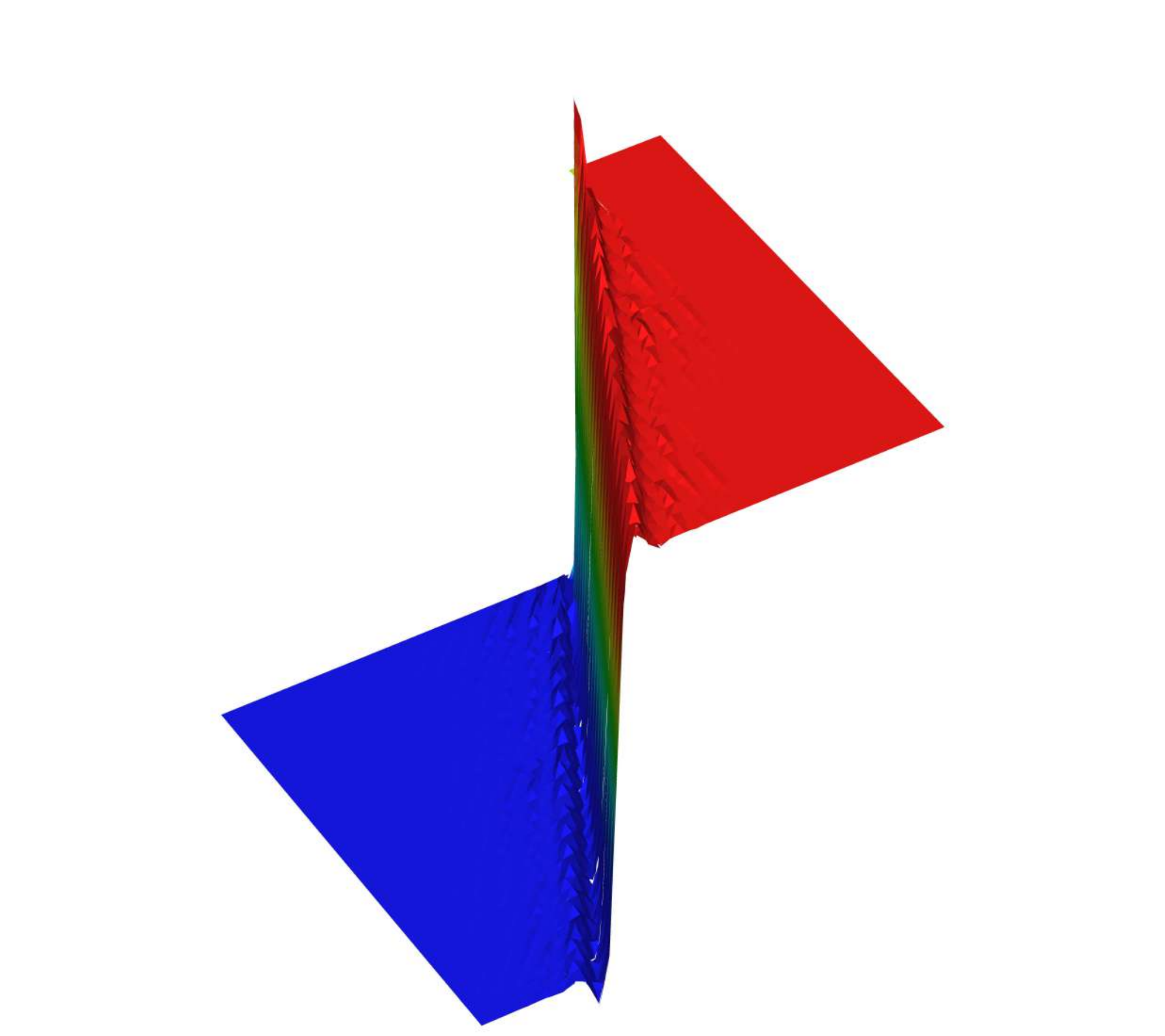} }}%
    \caption{The macro-element HDG method with and without SUPG stabilization applied in the advection dominated setting ($ \kappa=10^{-5} $, Pe = $\sqrt{2}\cdot 10^5$).}%
    \label{fig:supg}%
\end{figure}

In the next step, we illustrate the importance of adding additional stabilization to achieve accurate solutions with the macro-element HDG method in the advection-dominated setting. To this end, we decrease the diffusion coefficient to $\kappa = 10^{-5}$. We first plot the solution in Figure\ref{fig:supg}a, where we use the mesh shown in Figure \ref{fig:mesh}a without additional stabilization. We observe that overall, the solution is stable, but within each macro-element, the solution exhibits strong oscillatory behavior. These oscillations largely vanish in Figure \ref{fig:supg}b after adding SUPG stabilization. Over- and under-shoots are mitigated and poor behavior is limited to a band of macro-elements near the sharp layer. In the remainder of this paper, SUPG stabilization is used in all results obtained with the macro-element HDG method.

\subsection{Local adaptive refinement}
The standard HDG method provides a simple mechanism for local refinement made possible by its discontinuous bases and the presence of a trace variable. Elements can be recursively subdivided and a proper choice of trace space is key in maintaining weak continuity between elements  \cite{samii2016parallel}. This ability of local refinement naturally generalizes to the macro-element HDG method. Again, the choice of trace space is key in maintaining weak continuity accross macro-element boundaries.

\begin{figure}[t]
    \centering
     \subfloat[Before local subdivision]{\includegraphics[width=0.49\textwidth]{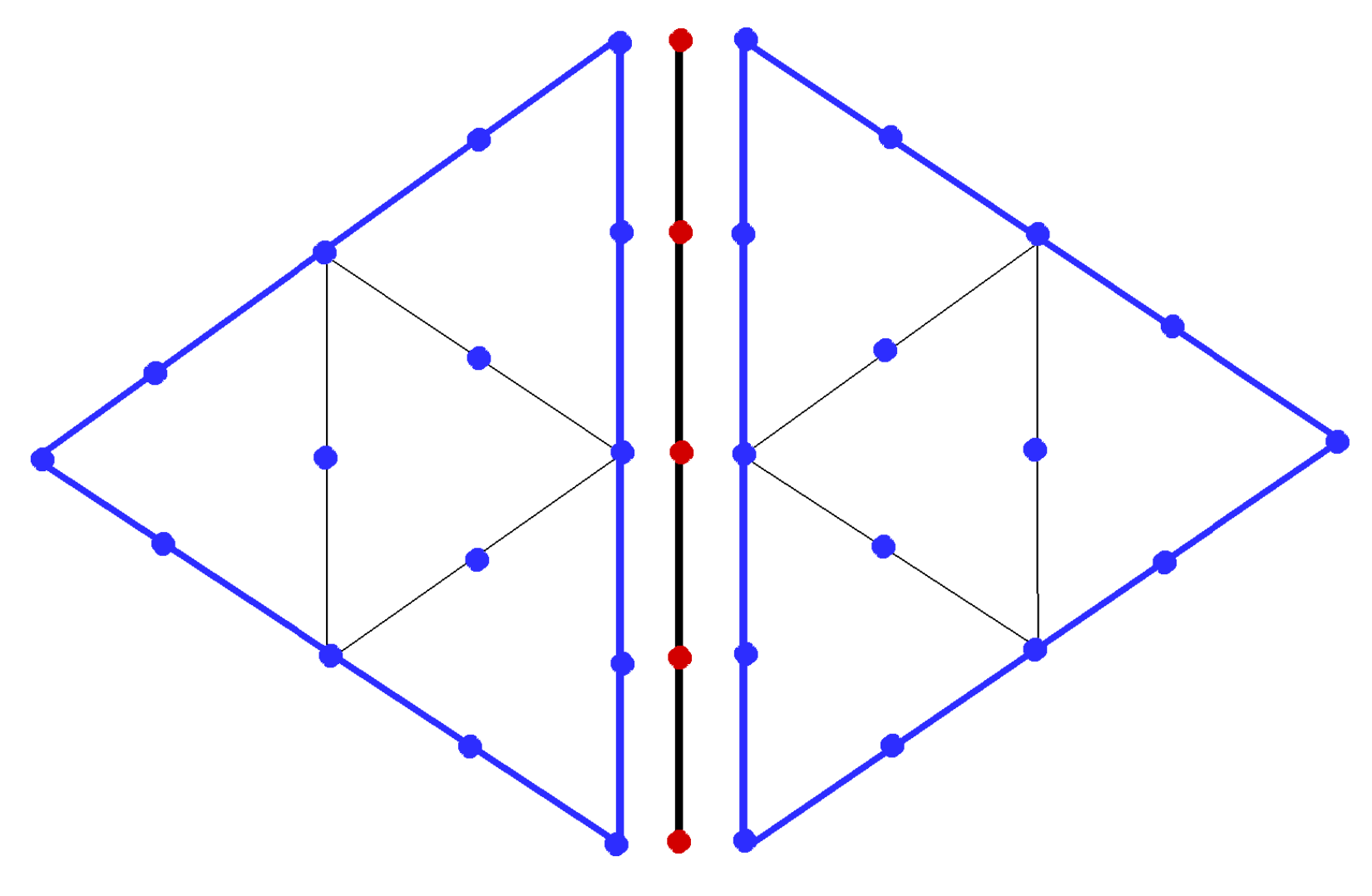} }%
     \subfloat[After local subdivision]{\includegraphics[width=0.46\textwidth]{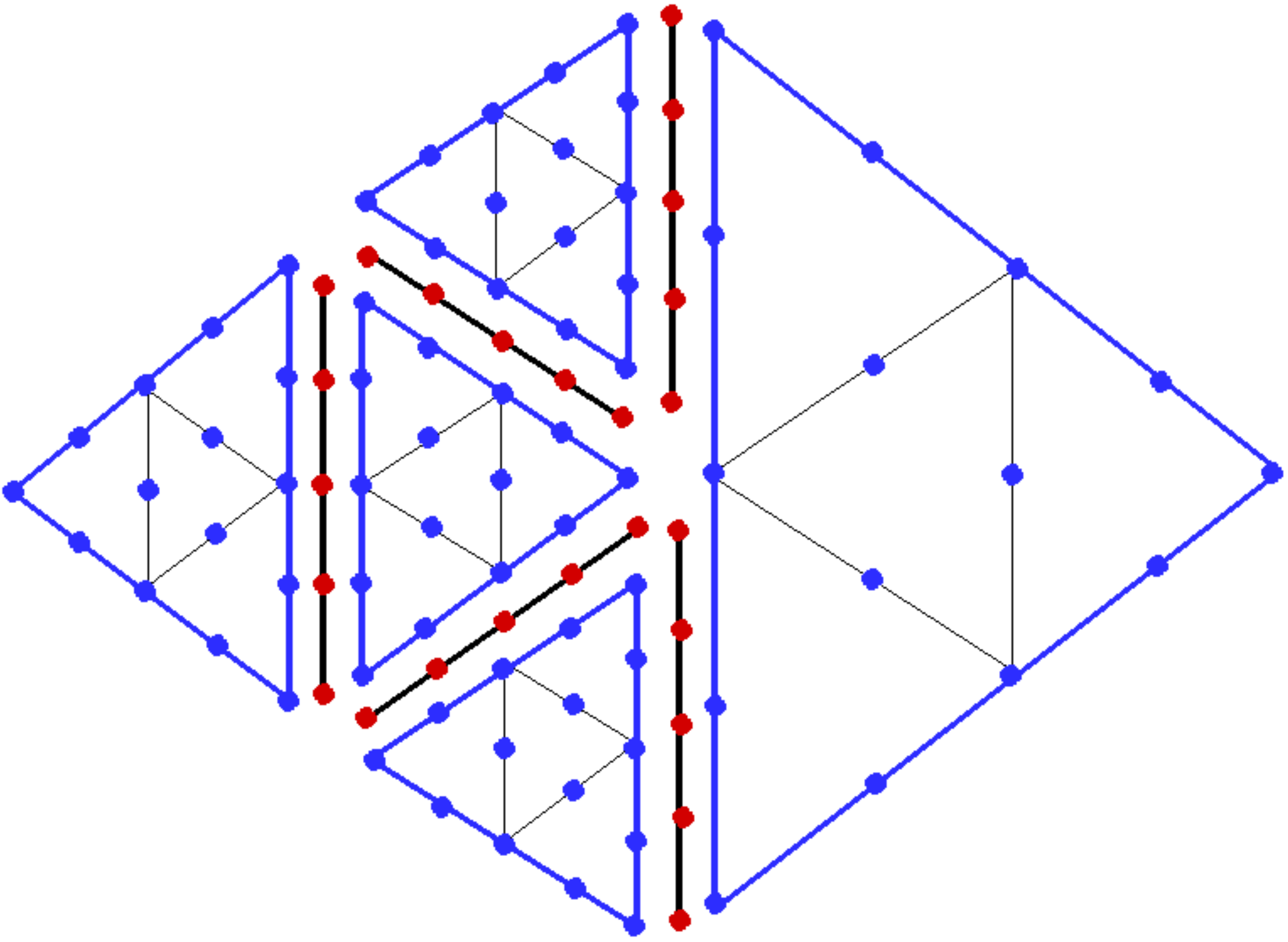} }%
    \caption{Local refinement of a macro-element and its corresponding trace variable for $p=2$ and $m=2$.}%
    \label{fig:subdivision}%
\end{figure}

An example of our approach to refinement is depicted in Figure \ref{fig:subdivision}. Macro-elements are dyadically refined. The corresponding trace variable is refined such that it can locally represent the trace of the refined macro-element variable. This means that continuity across the normal is maintained exactly.  Other options are available, but according to our computational tests, this choice is in general most accurate and works well in practice.

\begin{figure}
    \centering
     \subfloat[Level 1.]{{\includegraphics[width=0.3\textwidth]{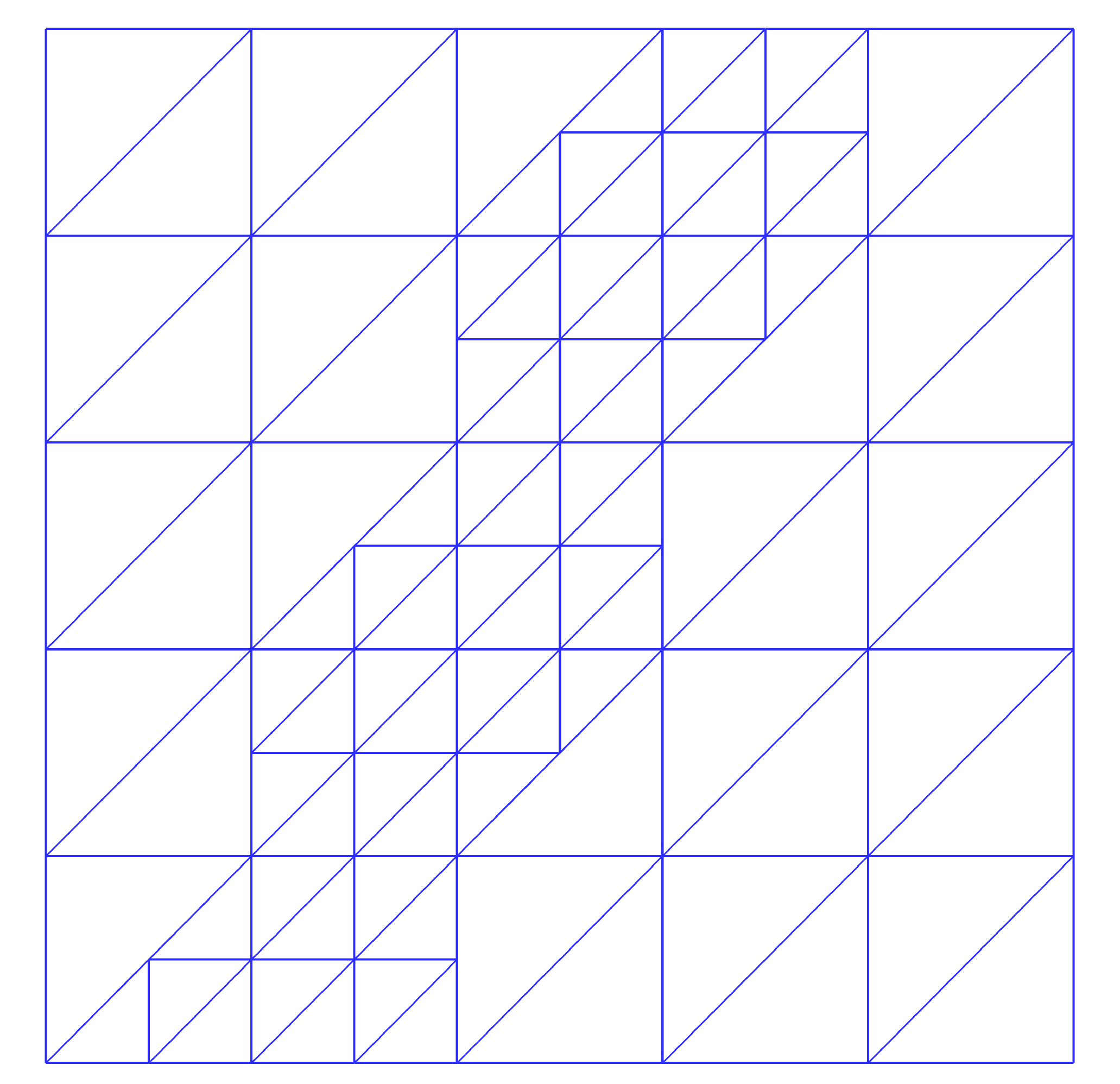} }}%
     \subfloat[Level 3.]{{\includegraphics[width=0.3\textwidth]{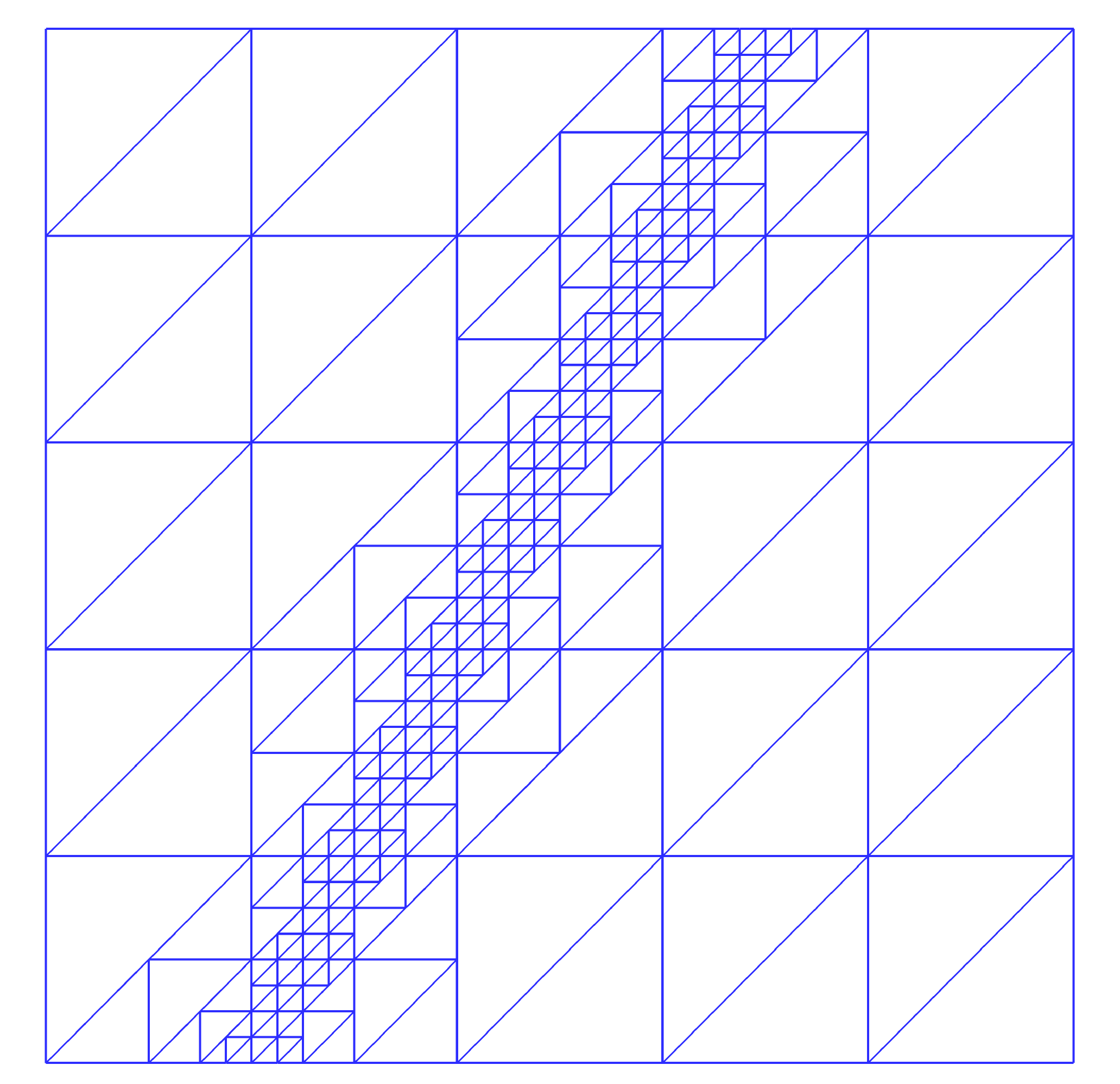} }} 
     \subfloat[Level 5.]{{\includegraphics[width=0.3\textwidth]{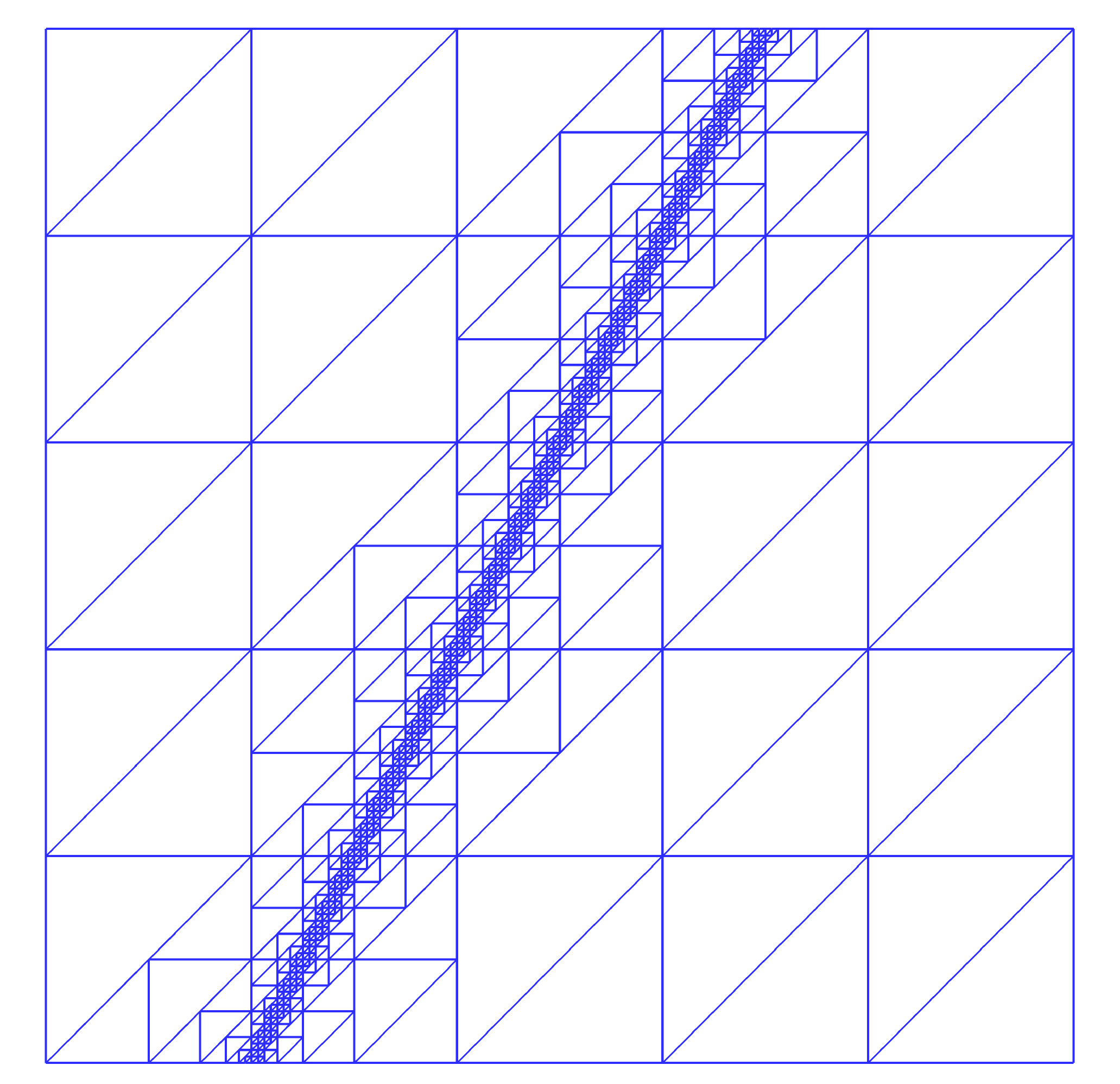} }}
    \caption{Sequence of local macro-element refinements for the 2D benchmark in the advection dominated setting ($\Pe=10^{10}$).}
    \label{fig:Mesh refine}%

    \subfloat[Convergence under uniform and adaptive refinement.]{{\includegraphics[width=0.5\textwidth]{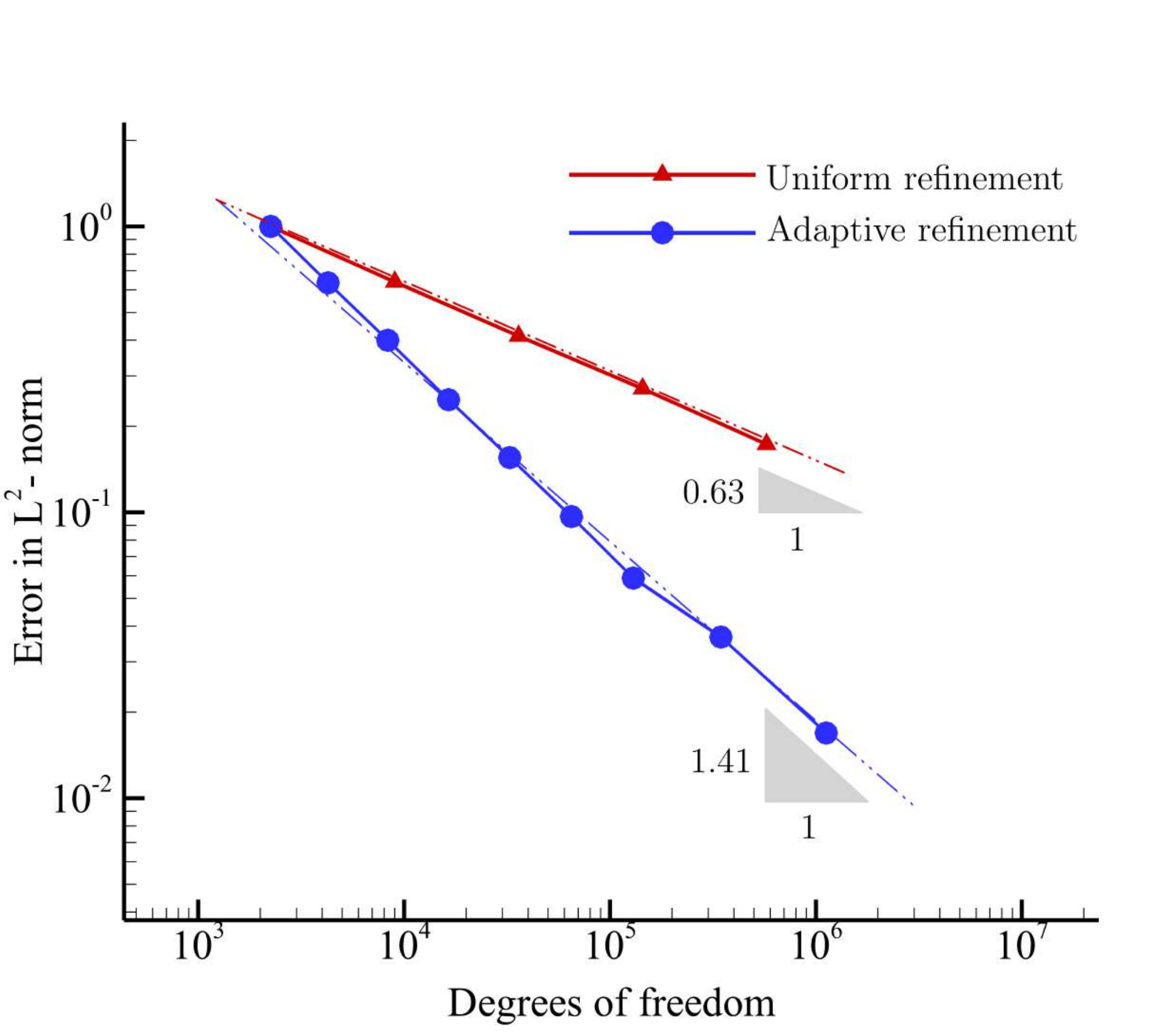} }}%
    \subfloat[Solution obtained refinement level 9.]{{\includegraphics[width=0.4\textwidth]{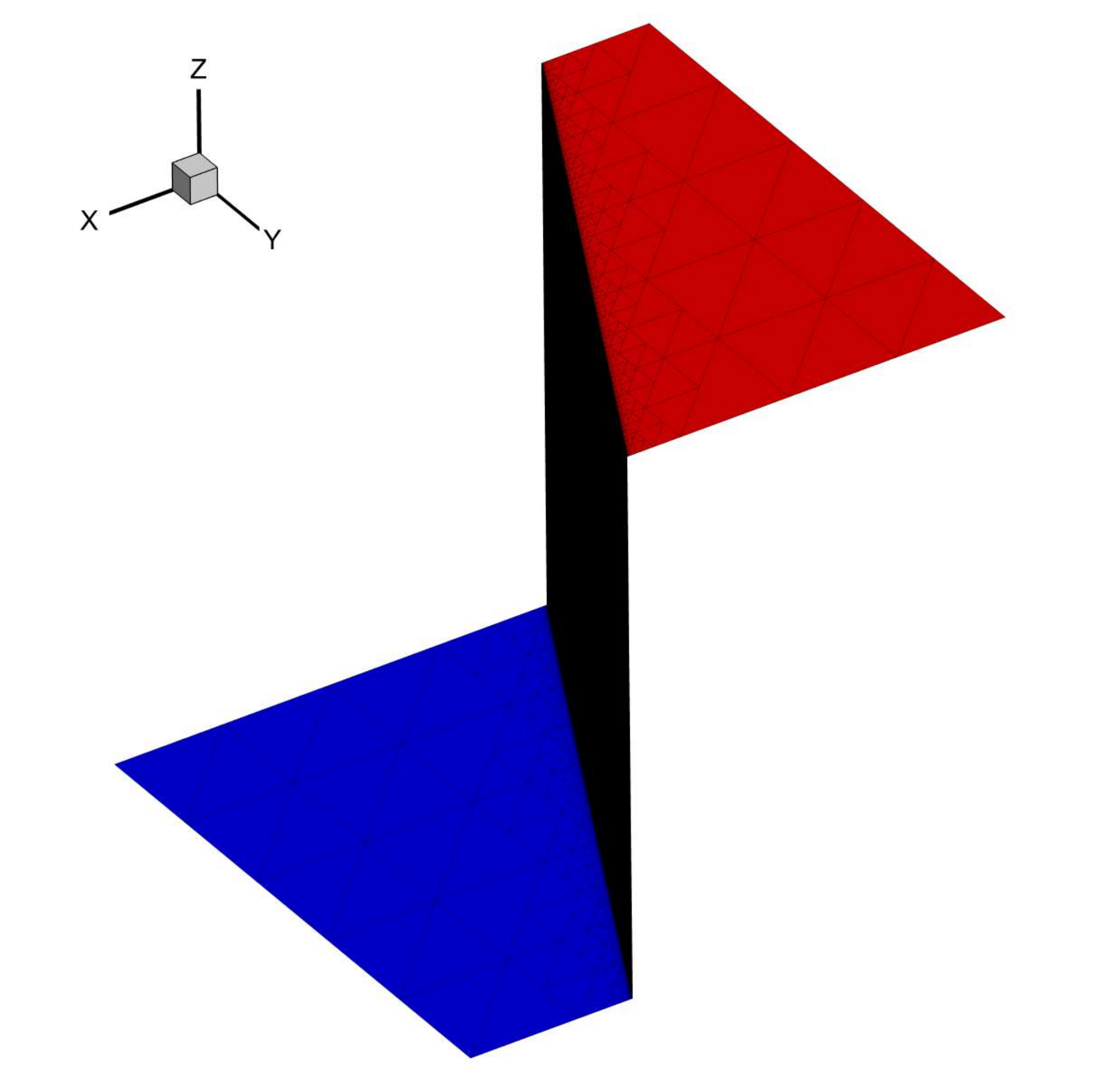} }}
    \caption{Convergence of 2D benchmark ($\Pe=10^{10}$) for uniform and local refinement with $m=2$ and $p=2$.}%
    \label{fig:convergence refine}%
\end{figure}

Local refinement is paramount in problems involving boundary and internal layers. Figure~\ref{fig:Mesh refine} depicts a sequence of meshes generated to capture an internal layer that arises in the advection dominated setting ($\Pe=10^{10}$). To reduce clutter, we only plot macro-elements. Here we employed an automatic gradient based refinement scheme with meshes of size $m=2$ and degree $p=2$. We illustrate the corresponding convergence curves for uniform and adaptive refinement and a solution plot for the finest level 9 in Figures \ref{fig:convergence refine}a and \ref{fig:convergence refine}b, respectively. We observe that local refinement greatly increases the efficacy of the approach.

\subsection{Domain decomposition and load balancing}
When computations need to be run on large distributed-memory systems, it is important to account for variable computational loads across processors and ensure that these loads are as equally sized as possible. Failure to account for load imbalances can severely impact performance and limit strong/weak scaling. Improved load balancing results in significant improvements in runtime and speed-up performance. 

To analyze the actual performance, we consider the ``load balancing performance factor'' (LBF) a measure of the load balancing efficiency \cite{zheng2005achieving}, which is calculated as the ratio of the average time spent among processors to the maximum time spent of all processors:

 \begin{align*}
    \text{LBF} \;=\; \frac{\text{Avg. workload of all proc's}}{\text{Max. workload among proc's}} \; .       
\end{align*}
A perfectly balanced scheme results in a factor of 1.0, with less optimal load balancing schemes having ratios smaller than 1.0. There are different definitions in the literature, but almost all are based on average and maximum time spent.

\begin{figure}[t]
    \centering
     \subfloat[$\text{LBF} = 0.97$ (with 8 proc's)]{{\includegraphics[width=0.33\textwidth]{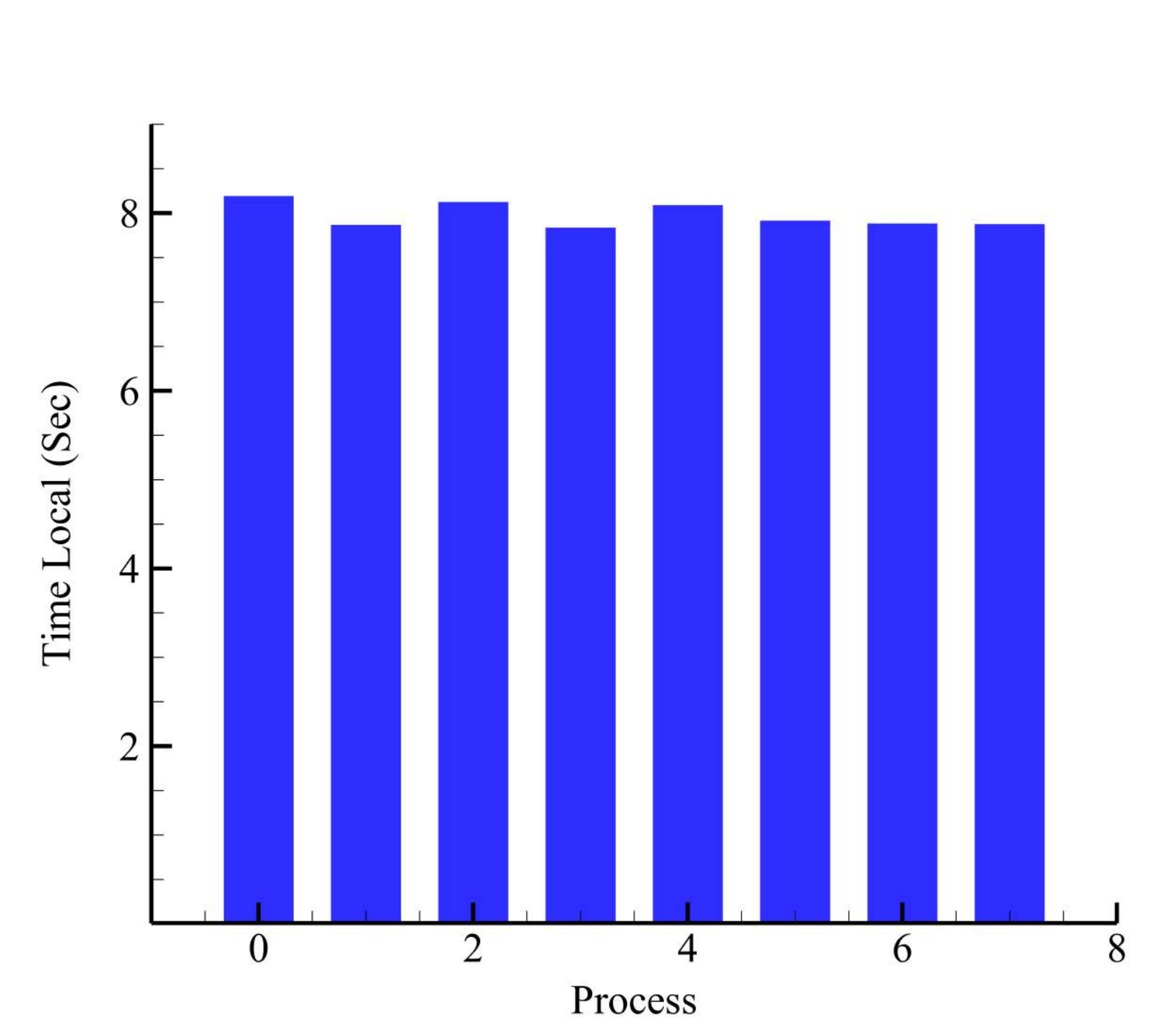} }}%
     \subfloat[$\text{LBF} = 0.98$ (with 16 proc's)]{{\includegraphics[width=0.33\textwidth]{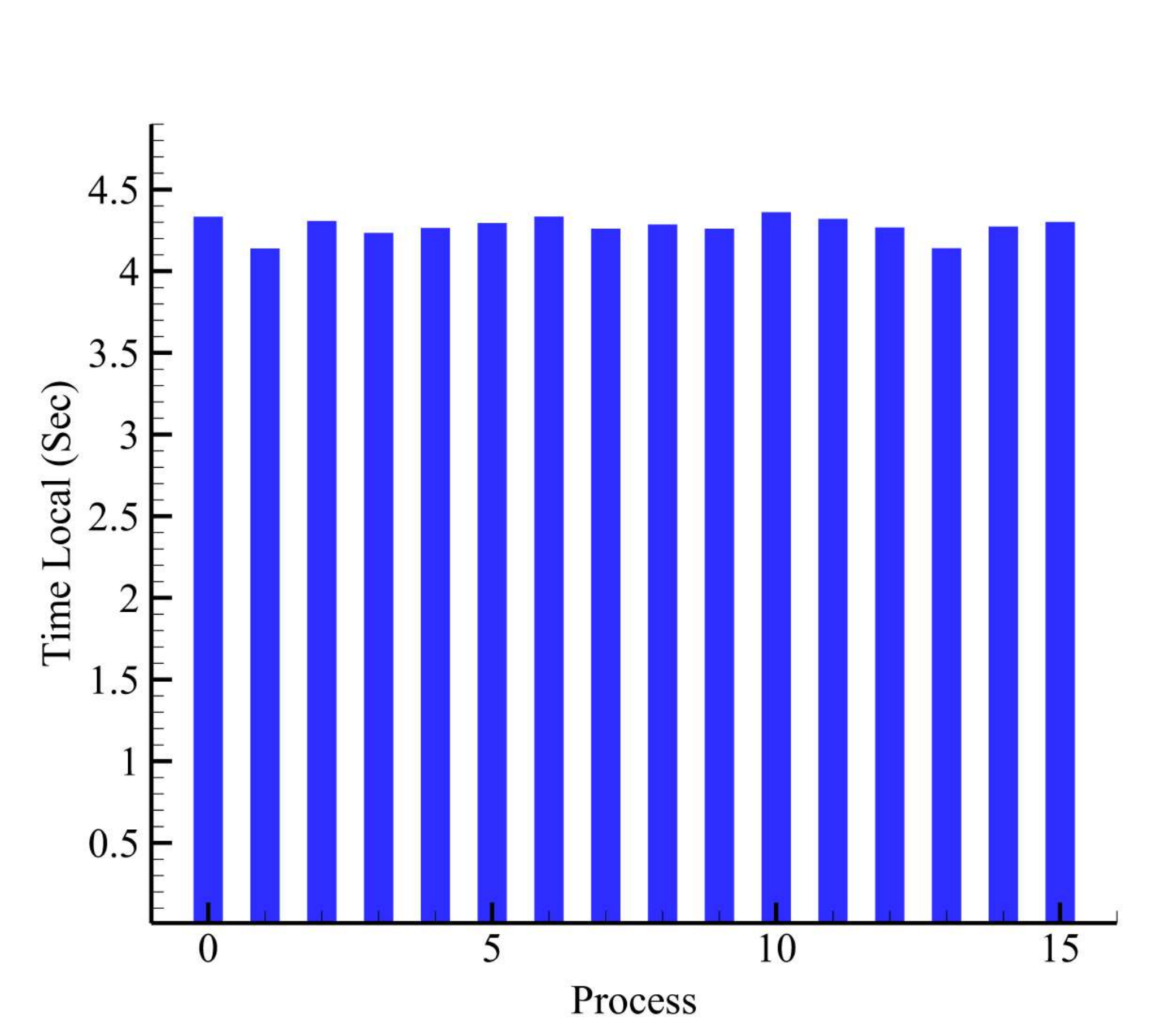} }}%
     \subfloat[$\text{LBF} = 0.97$ (with 32 proc's)]{{\includegraphics[width=0.33\textwidth]{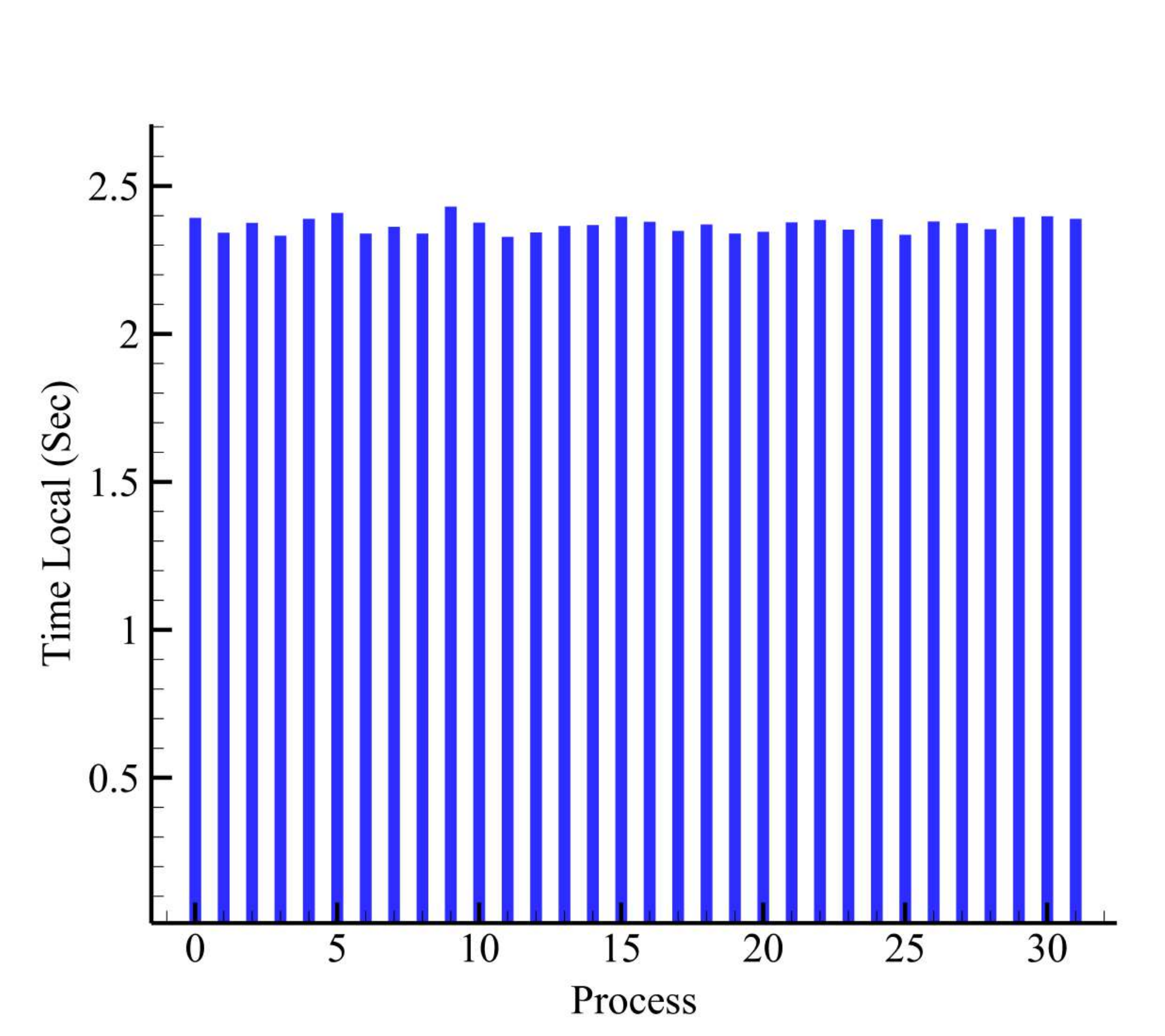} }}\\ 
     \subfloat[$\text{LBF} = 0.92$ (with 8 proc's)]{{\includegraphics[width=0.33\textwidth]{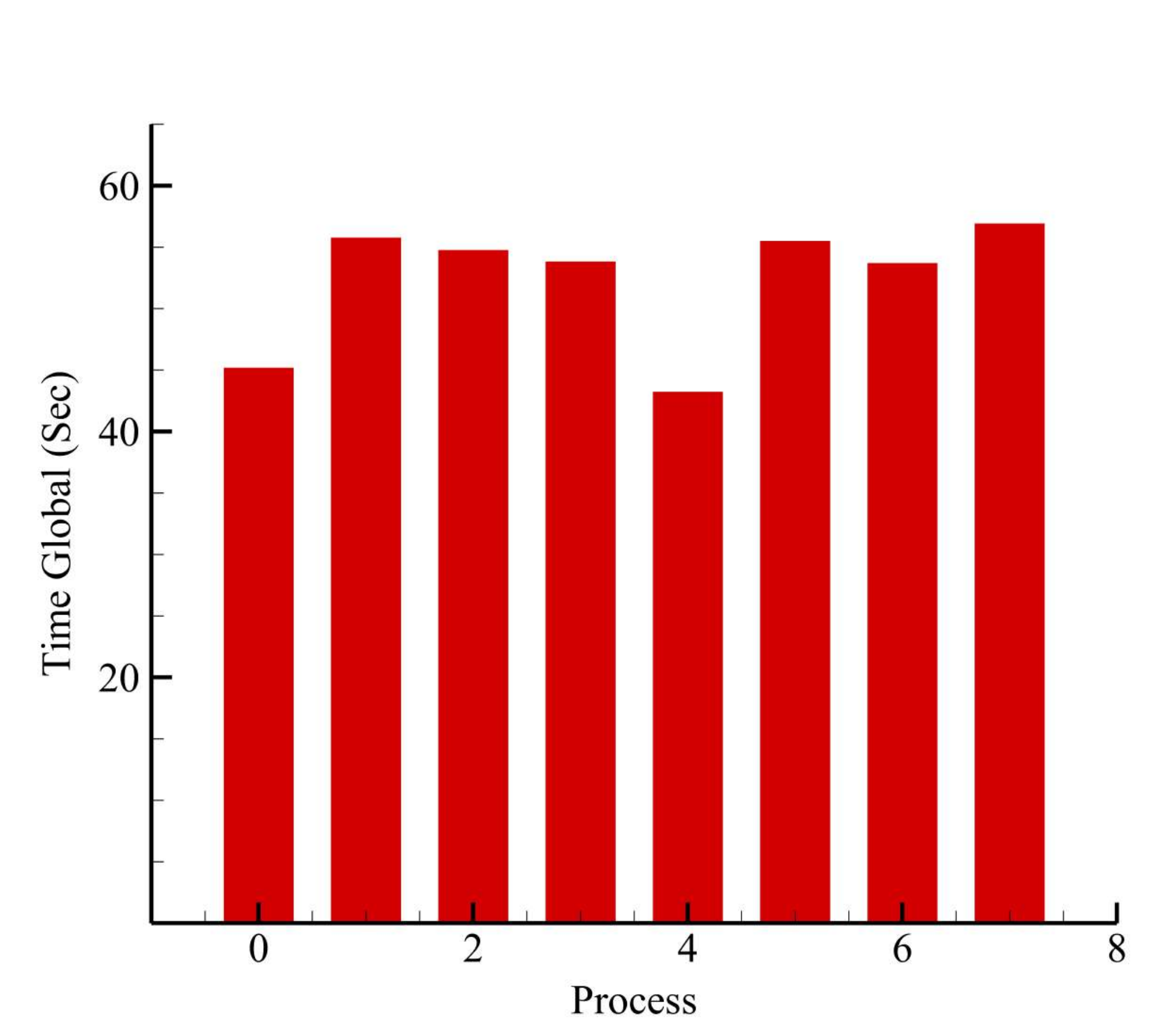} }}%
     \subfloat[$\text{LBF} = 0.95$ (with 16 proc's)]{{\includegraphics[width=0.33\textwidth]{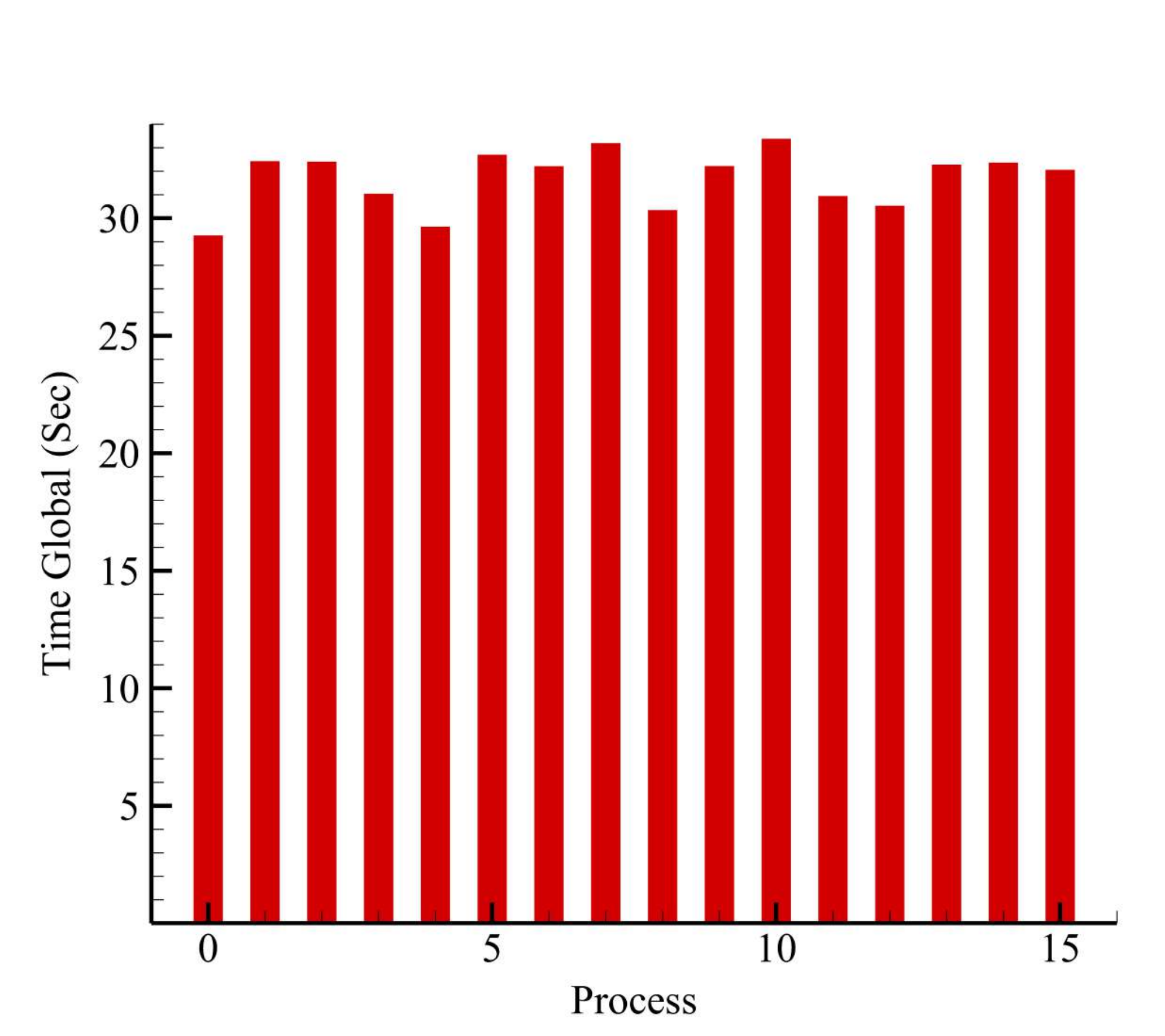} }}%
     \subfloat[$\text{LBF} = 0.97$ (with 32 proc's)]{{\includegraphics[width=0.33\textwidth]{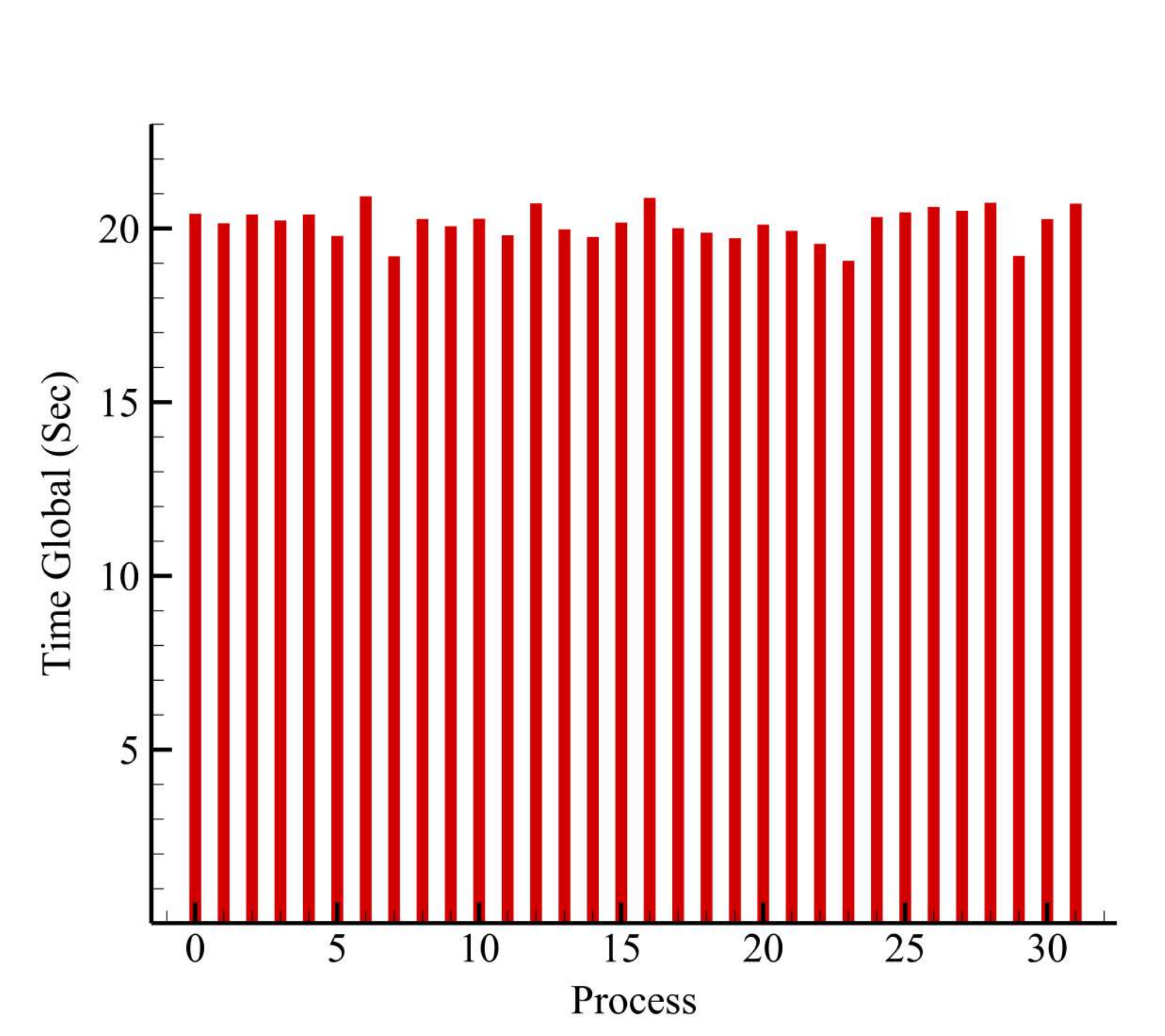} }} 
    \caption{Load balancing of macro-element HDG on a \textbf{single compute node} - time of local (blue) and global (red) solvers and load balancing factors (LBF) ($\text{dof}^{global} = 608,256$).}%
    \label{fig:BL coarse order}%
\end{figure}

\begin{figure}[t]
    \centering
     \subfloat[\centering $\text{LBF} = 0.91$ (on 1 node)]{{\includegraphics[width=0.33\textwidth]{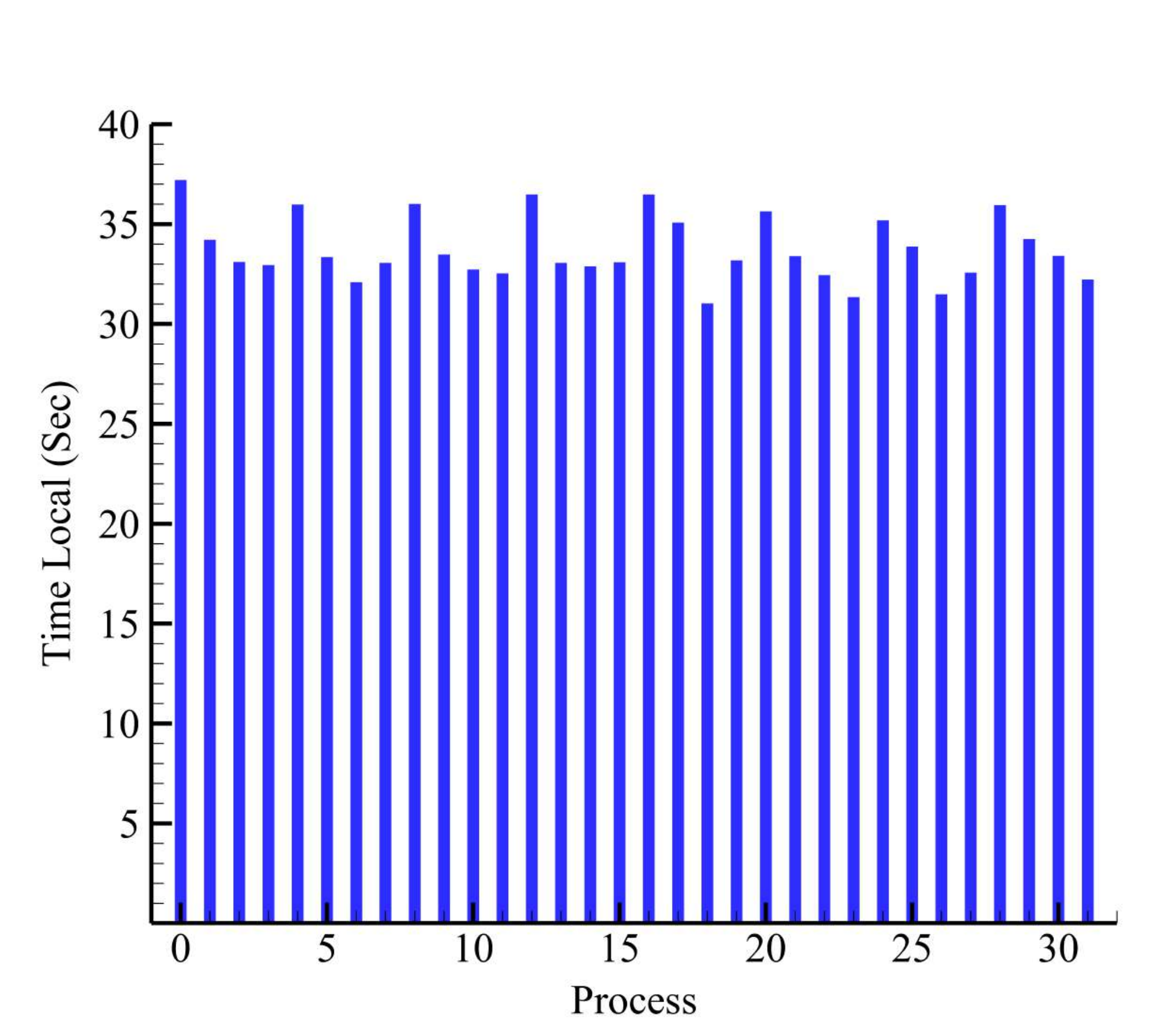} }}%
     \subfloat[\centering $\text{LBF} = 0.90$ (on 2 node)]{{\includegraphics[width=0.33\textwidth]{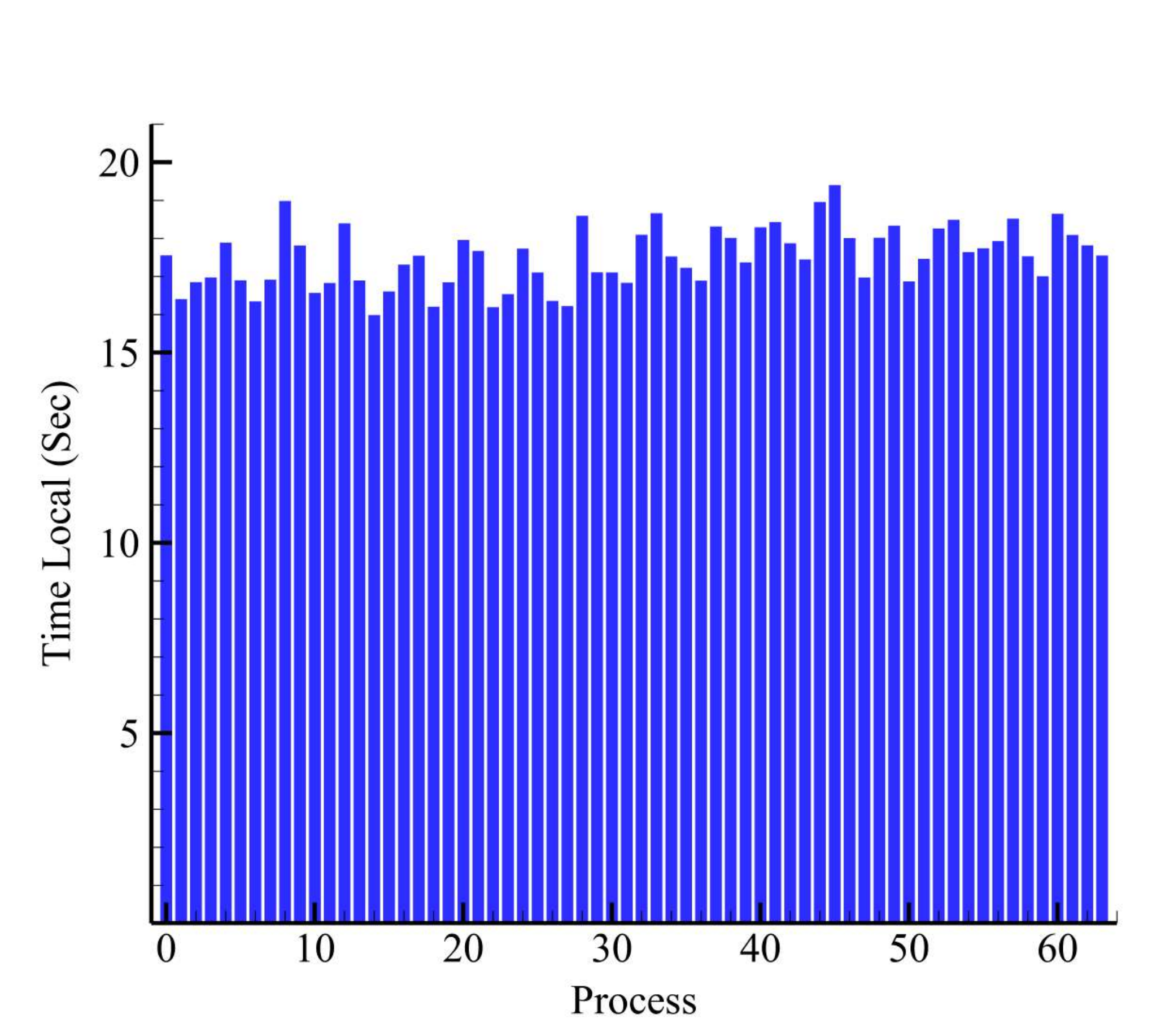} }}%
     \subfloat[\centering $\text{LBF} = 0.87$ (on 4 node)]{{\includegraphics[width=0.33\textwidth]{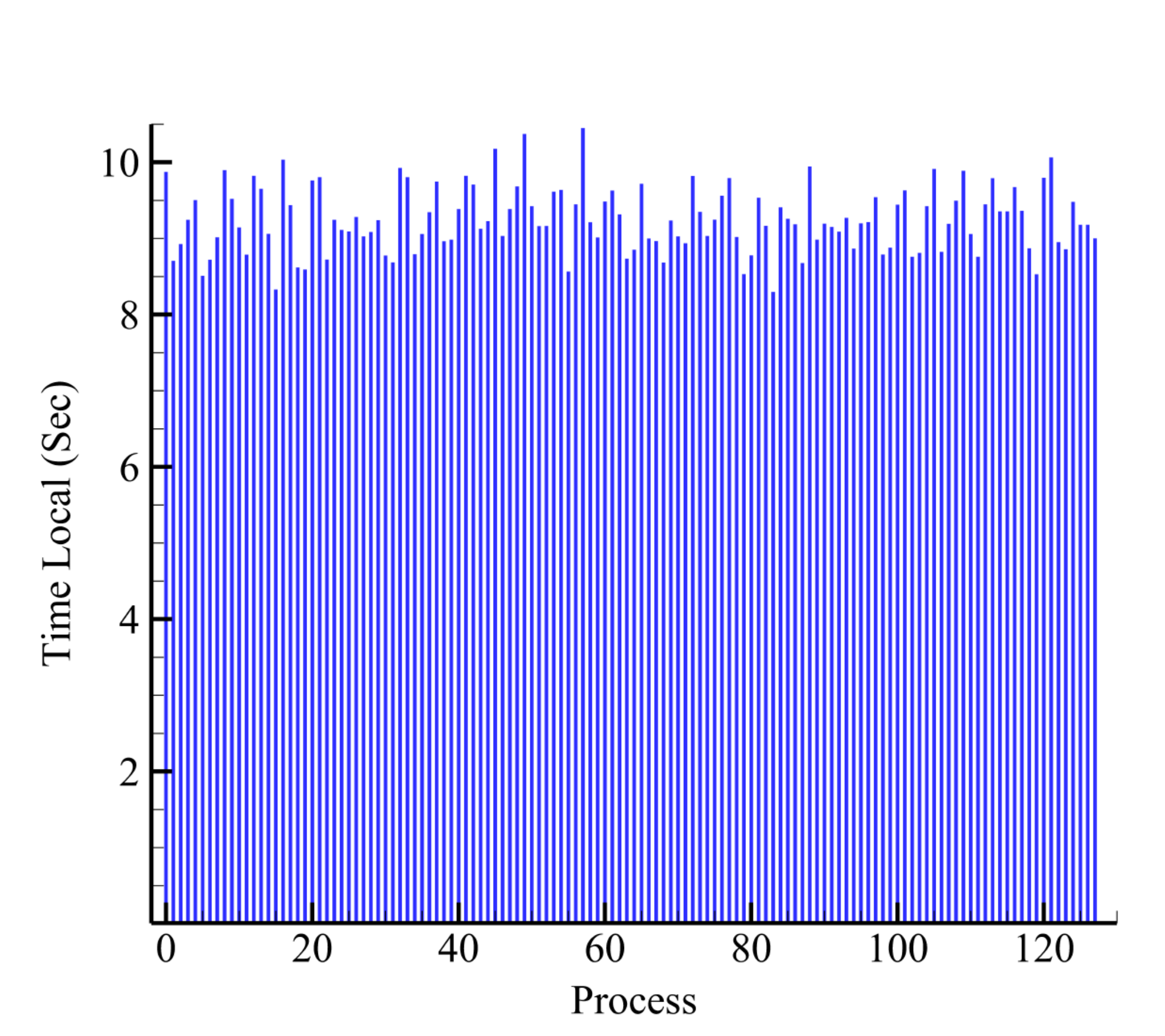} }}\\ 
     \subfloat[\centering $\text{LBF} = 0.87$ (on 1 node)]{{\includegraphics[width=0.33\textwidth]{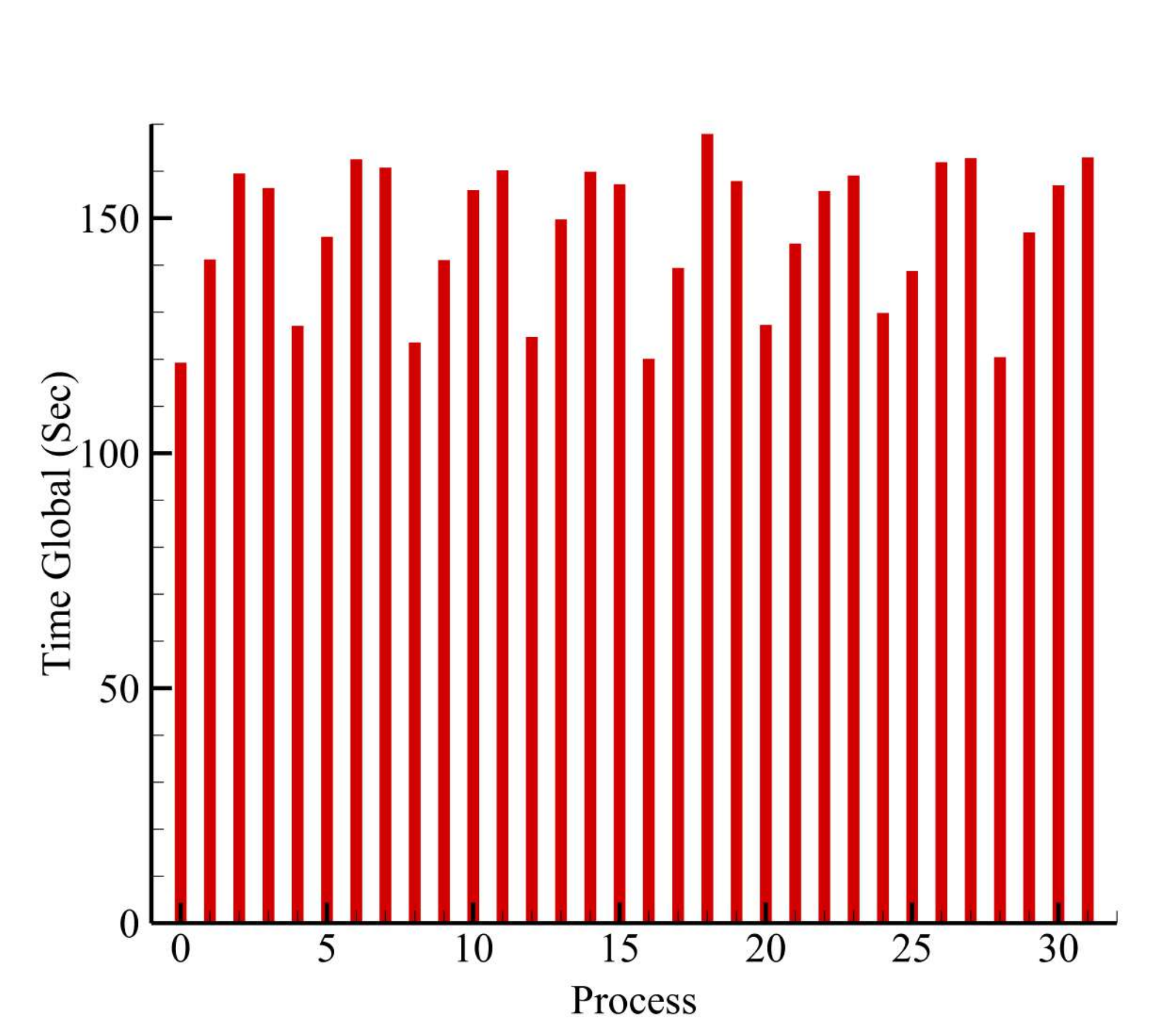} }}%
     \subfloat[\centering $\text{LBF} = 0.90$ (on 2 node)]{{\includegraphics[width=0.33\textwidth]{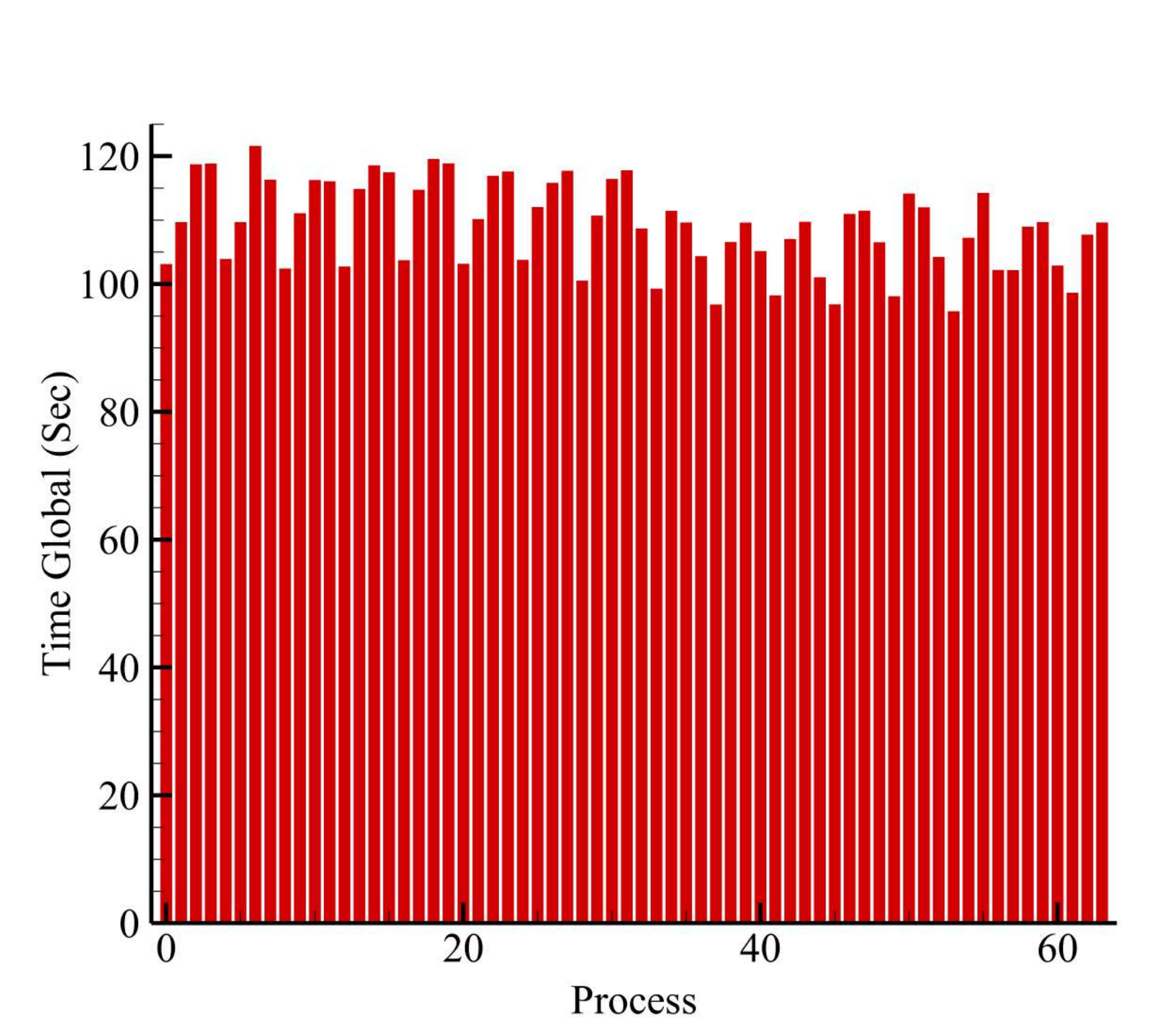} }}%
     \subfloat[\centering $\text{LBF} = 0.91$ (on 4 node)]{{\includegraphics[width=0.33\textwidth]{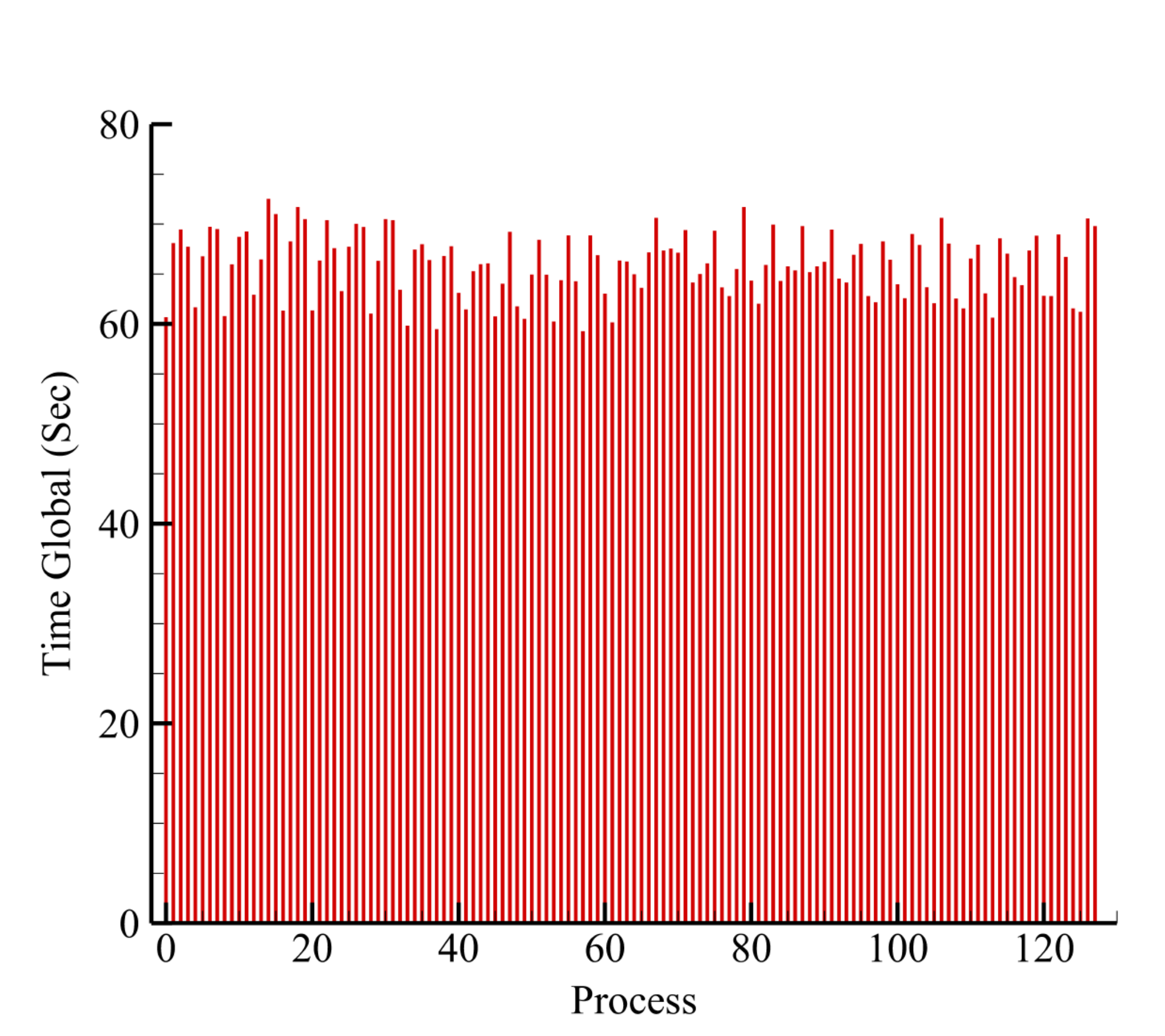} }} 
    \caption{Load balancing of macro-element HDG on \textbf{multiple compute nodes}, each with 32 processes - time of local (blue) and global (red) solvers and load balancing factors (LBF) ($\text{dof}^{local} = 3,440,640$, $\text{dof}^{global} = 1,216,512$).}%
    \label{fig:BL fine order}%
\end{figure}

Using the computing set-up and environment to be described in more detail in Section 6.1, we compute our 3D benchmark discretized with a total number of elements $N^{Elm} = 196,608$. We then collect LBFs for the our macro-element variant of the HDG method, where macro-elements contain two elements along a macro-element edge ($m = 2$). We first choose linear basis functions ($p=1$), where the global solver faces a system with $\text{dof}^{global} = 608,256$ degrees of freedom.
Figure \ref{fig:BL coarse order} depicts the breakdown of execution time for the case of parallel execution with 8, 16 and 32 processors (all on a single compute node) for the local and global solver. Based on these results, we see excellent load balancing in terms of local and global computations, with LBFs consistently above 0.9. We also examine how the workload of each processor itself scales with the number of compute nodes, where we use all 32 processors on each node. To generate sufficient load, we increase the polynomial degree to $p=2$, where the local and global solvers face systems with $\text{dof}^{local} = 3,440,640$ and $\text{dof}^{global} = 1,216,512$ degrees of freedom, respectively. The results are shown in Figure \ref{fig:BL fine order}. We observe a slight effect of communication between nodes, but still obtain LBFs of around 0.9. For 4 nodes and 128 processors, we observe LBFs of 0.87 and 0.91 for the local solver and the global solver, respectively.

\section{Computational efficiency: theoretical estimates}

The key question that remains is whether our macro-element variant of the HDG method is able to deliver competitive computational performance in terms of computing cost and scalability. In this section, we first investigate this question via theoretical estimates. 

We note that a recent preprint examined this question for a very similar method \cite{vymazal2018combined}, concluding that ``the cost of assembly and solution of the discrete transmission condition in three dimensions remains prohibitively expensive.'' A macro-element HDG approach was therefore discouraged as generally inefficient. We emphasize that our complexity analysis arrives at different results that also show limitations, but look much more encouraging than what was reported in \cite{vymazal2018combined}.

\subsection{Complexity analysis} 
Recall that $d\; (=2,3)$, denotes the spatial dimension. We assume a structured discretization of the unit cube $[0,1]^d$ into a set of $N = d! \, n^d$ simplicial macro-elements, where $n$ denotes the number of macro-elements in each spatial direction. Every macro-element is divided into a set of $m^{d}$ simplicial $C^0$-continuous finite elements, where $m$ denotes the number of elements per macro-element in each spatial direction. We further assume a polynomial basis on each simplicial element that is of total degree $p$. In Table \ref{tab:table1} and Table \ref{tab:table2}, we summarize the defining quantities and the dependent quantities that we use in our analysis.

 \begin{table}[h!]
  \begin{center}
    \caption{Defining quantities.}
    \label{tab:table1}
\begin{tabular}{| l | l |}
\hline
$d$ & spatial dimension ($=2, 3$) \\ 
\hline
$n$ &  number of macro-elements along edge of unit cube $[0,1]^d$ \\ 
\hline
$m$ &  number of elements along a macro-element edge  \\ 
\hline
$p$ &  element polynomial order \\
\hline
\end{tabular}
  \end{center}
\end{table}

 \begin{table}
  \begin{center}
    \caption{Dependent quantities.}
    \label{tab:table2}
\begin{tabular}{| m{0.44\textwidth} | m{0.50\textwidth} |}
\hline
$N = d! \, n^d$ &  total number of macro-elements\\
\hline
$M_{(d)} = m^{d}$ &  total number of elements per $d$-dimensional macro-element\\
\hline
$\begin{cases}Q_{(d)} & = \binom{m \, p + d}{d} \\
& = \frac{1}{d !} (m\cdot p + 1) \; \cdots  \;  (m\cdot p + d) \\
&  < \frac{1}{d!} (m \cdot p + d)^d \end{cases}$ &  total number of degrees of freedom per $d$-dimensional macro-element\\
\hline
$\begin{cases}D = A_d \, n^d + B_d \, d \, n^{d-1} (n-1)\\ \text{where } (A_2=1, B_2=1, A_3=6, B_3=2)
\end{cases}$ &  number of macro-element faces that make up the mesh skeleton (without boundaries). $A_d$ is the number of shared faces within a cube partitioned by simplices, and $B_d$ is the number of shared faces between two $d$-dimensional cubes.\\
\hline
$d+1$ &  number of faces of a $d$-dimensional simplex\\
\hline
$(2p+1)^d$ &  upper estimate of the number of non-empty intersections of $C^0$ test and trial functions within one macro-element.\\
\hline
$C^{\text{sparsity}}_{(d)} = \begin{cases}
    1 & \text{for dense arith.} \\
    (2p+1)^d / Q_{(d)}              & \text{for sparse arith.} \end{cases} $ &  upper estimate of sparsity of the matrices.\\
\hline
\end{tabular}
  \end{center}
\end{table}

We are interested in comparing the macro-element HDG method ($m > 1$) with the standard HDG method ($m=1$). We assume, and later verify, that a similar level of accuracy can be attained as long as the total number of elements stays equal, that is, $n \cdot m = \text{constant}$.

\subsubsection{Cost of the global solver}
The time complexity of the above ``global'' problem can be easily computed. The matrix-vector product can be localized to the macro-element level, that is, on macro-element $\element$ the matrix-vector product is,

\begin{enumerate}
	\item $\mat{x}_{(\element)} = \mat{B}_{(\element)} \mat{\hat{u}}_{(\element)},$
	\item $\mat{y}_{(\element)} = \mat{A}_{(\element)}^{-1} \mat{x}_{(\element)},$ 
	\item $\mat{\hat{v}}_{(\element)} = \mat{C}_{(\element)} \mat{y}_{(\element)}.$
\end{enumerate}
The matrix vector multiplications in step 1 and 3 require at most $\mathcal{O}\left( (d+1) Q_{(d-1)} \cdot Q_{(d)} \right)$ operations. An LU factorization for $\mat{A}_{(\element)}$ is computed once, which costs $\mathcal{O}\left( Q^3_{(d)} \right)$ operations. Application of the factorization in all subsequent steps costs $\mathcal{O}\left( Q^2_{(d)} \right)$ operations. This means that after the initialization process, one iteration of the solver costs $\mathcal{O}\left( Q^2_{(d)} \right)$ operations.

\begin{remark} The $N = d! \, n^d$ local problems can be run independently, in parallel!
\end{remark}

The final reduction step is performed for each sub-domain in the mesh skeleton. Suppose  $\mat{\hat{v}}_{(\face1 )}$ and $\mat{\hat{v}}_{(\face 2)}$  are the computed vectors associated with the two macro-elements adjacent to face $\face$ in the mesh skeleton. Then the reduction step on macro-element face is,
\begin{enumerate}
	\setcounter{enumi}{3}
	\item $\mat{\hat{w}}_{(\face)} = \mat{D}_{(\face)} \mat{\hat{u}}_{(\face)} - \mat{\hat{v}}_{(\face1)} -  \mat{\hat{v}}_{(\face2)}.$
\end{enumerate}
The matrix multiplication and subtraction require at most $\mathcal{O}\left( Q^2_{(d-1)} \right)$ operations.

\begin{remark} The $D$ local problems can be run independently, in parallel!
\end{remark}

\begin{remark} Communication happens when parts of $\mat{\hat{u}}$ are send as input to steps 1-3, and when the output of step 3 is channeled back to perform the reduction in step 4.
\end{remark}

\subsubsection{Cost of the local solver}
Once the global problem is solved, that is, the interface degrees of freedom $\mat{\hat{u}}$ are computed, the local solution can be reconstructed for each macro-element separately, by solving the following problem

\begin{align}
	\mat{A}_{(\element)} \mat{u}_{(\element)} = \mat{R}_{u(\element)} - \mat{B}_{(\element)} \mat{\hat{u}}_{(\element)}.
	\label{eq:local_matrix_problem}
\end{align}
Again, the computed LU factorization is used, which requires $\mathcal{O}\left( Q^2_{(d)} \right)$ operations. 

\subsection{Memory estimates}
We now summarize estimates of the required memory for the different sub-matrices per macro-element in the mesh and macro-element face in the mesh skeleton.

\begin{enumerate}
	\item $\mat{B}_{(\element)}$ and $\mat{C}_{(\element)}$:  $\text{const}^{\text{sparsity}}_{(d)} \cdot (d+1) \cdot Q_{(d-1)} \cdot Q_{(d)} \cdot 8$ bytes,  
	\item $\mat{A}_{(\element)} = \mat{L}_{(\element)} \mat{U}_{(\element)}$: $\text{const}^{\text{sparsity}}_{(d)} \cdot Q^2_{(d)} \cdot 8$ bytes,
	\item $\mat{D}_{(\face)}$:  $\text{const}^{\text{sparsity}}_{(d-1)} \cdot Q^2_{(d-1)} \cdot 8$ bytes.
\end{enumerate}
Here, we assume double precision arithmetic. The parameter $\text{const}^{\text{sparsity}}_{(d)} \in (0,1]$ denotes the fraction of non-zeros of the associated matrix. Furthermore, we assume that after $\mat{A}_{(\element)}$ is computed, its LU factorization is stored in-place, requiring no additional memory.

The matrices in 1. and 2. are stored for $N$ macro-elements in the mesh, possibly distributed among multiple processors. The matrix in 3. is stored for each of the $D$ macro-element faces in the mesh skeleton, also distributed among multiple processors.

\begin{remark}
For small values of $m$ it is more efficient to use dense linear algebra. However, for higher values of $m$, say $m>2$, sparse solvers may be more efficient. This will depend both on the sparsity of the matrix $\mat{A}$  (which depends on $m$ and $p$) and the size of the $L1-L2$ cache.
\end{remark}

\subsection{Comparison with standard HDG}
We are interested in the above estimates of operation counts for different $m,n$ such that $n \cdot m = const$. This case focuses on the setting where the total number of finite elements are equal, such that comparable accuracy may be expected between macro-element HDG and standard HDG. Let $\bar{n}, \bar{m}$ denote the dimensions of $n$ and $m$ for the macro-element HDG method. Then we can compare the number of operations in terms of these parameters with respect to the standard HDG method, $m = 1$, $n = \bar{m} \cdot \bar{n}$. The results are shown below in Table \ref{tab:comparison}. 
\begin{table}[h]
\centering
\begin{tabular}{| l | l | l |}
\hline
  							& macro-el.\ HDG 		& standard HDG \\
							\hline
initialization 	&   $\bar{n}^d (\bar{m} \cdot p + d)^{3d}$    				
							&   $\bar{m}^d \cdot \bar{n}^d (p+d)^{3d}$  						\\
step 1 				&   $\bar{n}^d (\bar{m} \cdot p + d)^{2d-1}$    				
							&   $\bar{m}^d \cdot \bar{n}^d (p + d)^{2d-1}$  					\\
step 2 				&   $\bar{n}^d (\bar{m} \cdot p + d)^{2d}$   				&    
								  $\bar{m}^d \cdot \bar{n}^d (p+d)^{2d}$						\\
step 3 				&   $\bar{n}^d (\bar{m} \cdot p + d)^{2d-1}$    				
							&   $\bar{m}^d \cdot \bar{n}^d (p + d)^{2d-1}$						\\
step 4 				&   $\bar{n}^d (\bar{m} \cdot p + d-1)^{2d-2}$    				
							&   $\bar{m}^d \cdot \bar{n}^d (p + d)^{2d-2}$ \\
\hline
\end{tabular}
\caption{Order of operations to initialize one iteration and then perform steps 1 through 4 in every iteration.}
\label{tab:comparison}
\end{table}

The initialization step, in which the LU factorizations are computed, is the most expensive, but is only performed once and is embarrassingly parallel. The other steps are performed at each iteration of the linear solver. Here step two is the leading term. 

Suppose now that the number of iterations is equal for the standard HDG method and the macro-element HDG method. In this case, it appears that it is not beneficial to have $\bar{m} > 1$. This means that from the perspective of the operation counts, the macro-element HDG method appears to be less efficient than the standard HDG method. There are, however, a number of parameters that are not taken into account in this analysis. The standard HDG method requires more communication between processors, because the number of local problems is far greater, and, typically, the number of iterations of the global problem is much lower for the macro-element variant. The remainder of this paper will study these aspects in the setting of a complete parallel implementation that is run on a modern heterogeneous compute cluster.

\begin{figure}
    \centering
     \subfloat[\centering Macro-element refinement of level 5 and solution.]{{\includegraphics[width=0.5\textwidth]{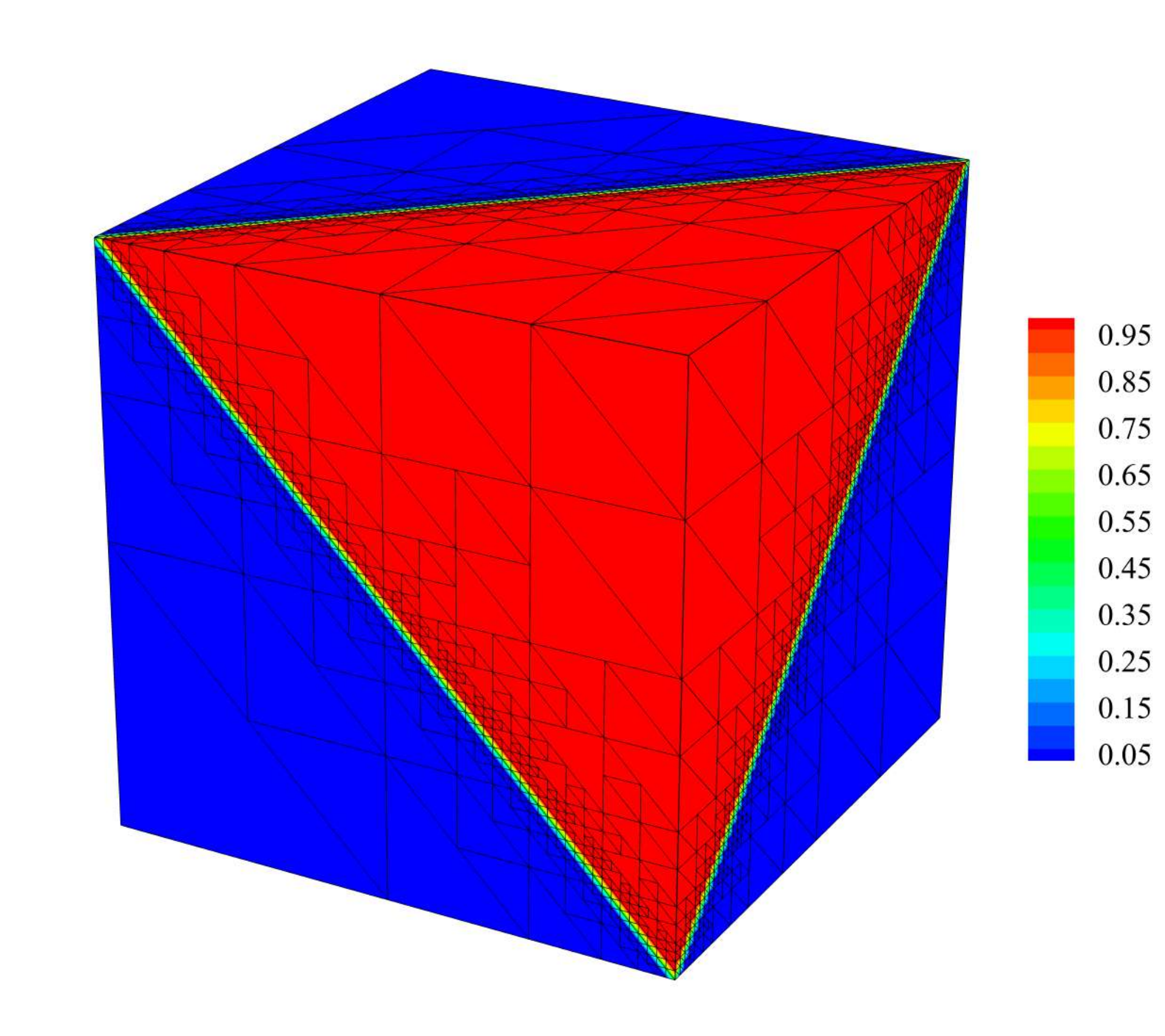} }}%
     \subfloat[\centering Volume slices.]{{\includegraphics[width=0.5\textwidth]{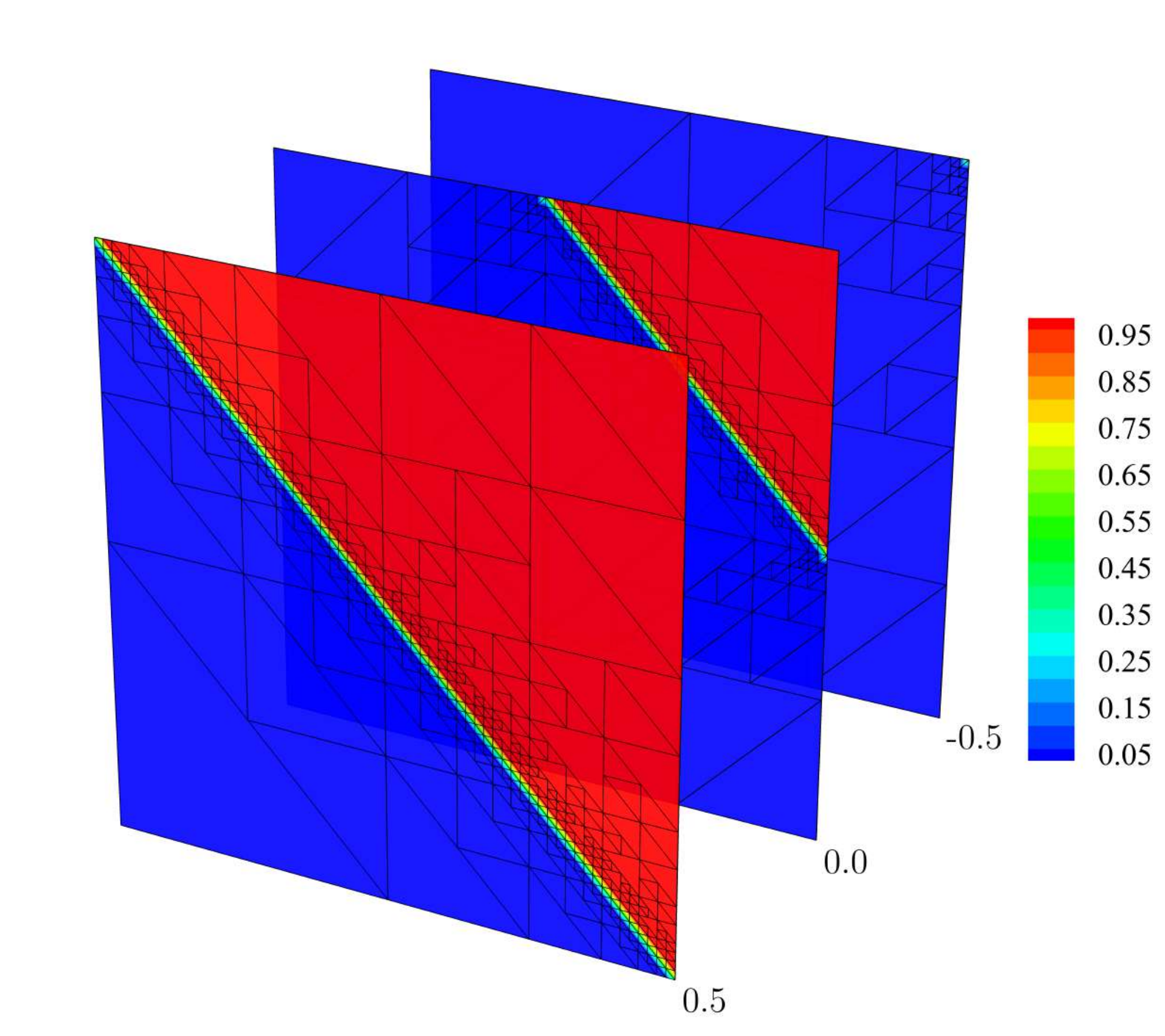} }}
    \caption{Macro-element refinement pattern and corresponding solution for the 3D benchmark.}%
    \label{fig:Test case Brick3D}%
\end{figure}

\section{Computational efficiency: numerical tests with a parallel implementation}

In the following, we will complement the theoretical estimates reported in the previous section by computing times measured from numerical tests with a parallel implementation. To this end, we reconsider the three-dimensional benchmark with an internal layer skew to a (structured) mesh that we introduced in Section 4.1. The geometry and a macro-element base mesh is shown in Figures \ref{fig:test case}b and \ref{fig:mesh}b. We choose the diffusion coefficient and the advective field as $\kappa = 10^{-8}$ and $\vect{a}=(1,1,1)$, respectively, arriving at a Peclet number of $\text{Pe} \gg 1$. In order to capture the internal layer accurately, we again employ a gradient based error estimator to generate an adaptive mesh with five levels of refinement. The macro-element refinement pattern is illustrated in Figure \ref{fig:Test case Brick3D} along with solution plots that demonstrate the accurate resolution of the internal layer.

\subsection{Computational setup and preconditioning methods}
Our implementation, written in Julia\footnote{The Julia Programming Language, \href{https://julialang.org/}{https://julialang.org/}}, includes the quadrature, element formation and global assembly routines.
For the local solver, we require the assembly of the local matrices, the computation of the local LU factorizations of $\mat{A}_{(\element)}$, and transferal of local values from the global solution. For $m=1$ and $m=2$, we use dense linear algebra provided by the package LAPACK\footnote{Linear Algebra PACKage, \href{https://netlib.org/lapack/}{https://netlib.org/lapack/}}, and for $m=4$ and $m=8$, we use sparse linear algebra, provided by the library UMFPACK \cite{davis2004algorithm}. 

For the global solver, we use GMRES iterative solvers from the open-source library PETSc\footnote{Portable, Extensible Toolkit for Scientific Computation, \href{https://petsc.org/}{https://petsc.org/}}. Here, we can employ two different solution variants. On the one hand, we can choose the matrix-free solution approach as detailed in Section 3.3. On the other hand, we can choose a matrix-based solution approach that operates on explicitly assembled and stored coefficient matrices.
In both approaches, we left-precondition equation \eqref{eq:global_problem} by the inverse of the block matrix $\mat{D}$. We emphasize that the matrix $\mat{D}^{-1}$ is never actually computed. The preconditioner is applied in step four of the matrix-free algorithm, which computes for each face, $\face$, in the mesh skeleton

\begin{align}
	\mat{\hat{w}}_{(\face)} = \mat{\hat{u}}_{(\face)} - \mat{D}_{(\face)}^{-1} \left( \mat{\hat{v}}_{(\face 1)} + \mat{\hat{v}}_{(\face 2)}\right) . 
\end{align}
Here, $\face1$ and $\face2$ denote the faces of the two adjacent macro-elements of face $\face$. Since all blocks $\mat{D}_{(\face)}$ are positive definite and symmetric, we compute and store its Cholesky factorization in-place. We note that the right-hand side is adapted in a similar manner. In the following, we call this variant the $\mat{D}^{-1}$ preconditioner for ease of notation.

We note that for the matrix-based approach, we tested several other preconditioning variants. In particular, we considered an algebraic additive Schwarz domain decomposition and one level of overlap, ASDD(1) \cite{{balay2019petsc}}, where each of the overlapped sub-problems is approximated with an incomplete LU factorization with zero fill-in, ILU(0). In our experience, the combination of ASDD(1) and ILU(0) provides a satisfactory balance between communication cost and the number of GMRES iterations for a variety of flow regimes, accuracy orders and number of processors \cite{roca2013scalable,diosady2011domain} for the matrix-based solution approach. We found, however, that no significant increase in efficiency could be achieved with respect to the $\mat{D}^{-1}$ preconditioner, which we hence used in all computations in this paper.

Our implementation is ported to the parallel computing environment Lichtenberg II (Phase 1) provided by the High-Performance Computing Center at the Technical University of Darmstadt, where it was compiled with GCC (version 9.2.0), Portable Hardware Locality (version 2.7.1) and OpenMPI (version 4.1.2). Our results are obtained on the cluster system, where we use several compute nodes. Each compute node has two Intel Xeon Platinum 9242 with 48 cores each at a base clock frequency of 2.3 GHz. Each compute node provides a main memory up to 384 GB. Further details on the available compute system can be found on the Lichtenberg webpage\footnote{https://www.hhlr.tu-darmstadt.de}.

\subsection{The matrix-based vs.\ the matrix-free approach for the global solver}

We first assess the efficiency of the global solver, comparing the matrix-based approach and the matrix-free approach, see \cite{franciolini2020efficient,kronbichler2012generic,kronbichler2019fast,kronbichler2022enhancing}. In the matrix-based approach, the global system  $\left( \mat{D} - \mat{C} \mat{A}^{-1} \mat{B} \right) \mat{\hat{u}} = \mat{f}$ must be explicitly formed at the macro-element level, assembled and stored at the global level, and then solved via an iterative solver. The macro-element HDG method inherits the advantages of the standard HDG method. Since $\mat{\hat{u}}$ is defined and single-valued along the element faces, the global system matrix  $\left( \mat{D} - \mat{C} \mat{A}^{-1} \mat{B} \right)$ is smaller than that of many other DG methods \cite{peraire2010hybridizable,cockburn2016static, nguyen2015class}. Moreover, it is compact in the sense that only the degrees of freedom between neighboring faces that share the same element are connected \cite{peraire2010hybridizable}.  

 \begin{table}
\caption{Number of degrees of freedom (dof) in the global problems referred to in Tables \ref{Tab:Tol1e-2} and \ref{Tab:Tol1e-6}.} 
\centering
\begin{tabular}{l|cc|cc|cc|cc}\toprule
& \multicolumn{2}{c|}{$p = 2$} & \multicolumn{2}{c|}{$p = 3$} & \multicolumn{2}{c|}{$p = 4$}& \multicolumn{2}{c}{$p = 5$}
\\\cmidrule(lr){2-3}\cmidrule(lr){4-5}\cmidrule(lr){6-7}\cmidrule(lr){8-9}
          & $m =1$  & $m =2$ 	& $m =1$   & $m =2$ & $m =1$   & $m =2$ & $m =1$   & $m =2$\\\midrule           
 $\text{dof}^{glob}$ 			& \footnotesize 1,517,376 	& \footnotesize 483,000 	& \footnotesize 2,528,960		& \footnotesize 901,600	  & \footnotesize 3,793,440		 & \footnotesize 1,449,000	& \footnotesize 5,310,816		& \footnotesize 2,125,200\\\bottomrule
\end{tabular}
\label{Tab:Order and DOfs size}%
\end{table}

We consider a discretization of the cube with 124,096 elements with polynomial degrees $p=2$ to $p=5$ that we compute on 2,848 processors. In Table \ref{Tab:Order and DOfs size}, we report the number of degrees of freedom for the global problems in the two HDG variants considered. In Table \ref{Tab:Tol1e-2}, we report the time and number of iterations of the GMRES linear solver, with a drop tolerance of 10e-2. We compare the standard HDG method versus the macro-element HDG method with eight elements per macro-element ($m=2$). In addition, we compare for each HDG variant a matrix-based (MB) approach versus a matrix-free (MF) approach, where both variants use the $\mat{D}^{-1}$ preconditioner as described above.

We see that in both methods, the computing time required for the global solver increases, when $p$ is increased, which is expected as the number of degrees of freedom increases and the structure of the global matrix changes. We observe that for a small number of iterations, the matrix-free approach is significantly more efficient than the matrix-based approach for both HDG variants. In addition, the increase in computing time is much larger for the matrix-based approach than for the matrix-free approach. We also see that the matrix-free approach is slightly faster for the macro-element HDG method, due to the reduction in degrees of freedom and communication across compute nodes compared to the standard HDG method.

 \begin{table}
\caption{\textbf{Global solver:} time and number of iterations of a \textbf{GMRES solver (with drop tolerance $\bm{{10}^{-2}}$)} and $\mat{D}^{-1}$ preconditioner. We compare standard HDG vs.\ macro-element HDG and matrix-based (MB) vs.\ matrix-free (MF) approaches for different $p$ ($N^{Elm} = 124,096$, $Proc's = 2,848$).} 
\centering
\begin{tabular}{l | cccc | cccc}\toprule
&\multicolumn{4}{c |}{Time global solver [sec]}  	&\multicolumn{4}{c}{Number of iterations} 
\\
&\multicolumn{2}{c}{Standard HDG}  	&\multicolumn{2}{c | }{Macro-el.\ HDG ($m=2$)}  	
&\multicolumn{2}{c}{Standard HDG}  	&\multicolumn{2}{c}{Macro-el.\ HDG ($m=2$)}  	
\\\cmidrule(lr){2-3}\cmidrule(lr){4-5}\cmidrule(lr){6-7}\cmidrule(lr){8-9}
            	& MB   	 & MF  & MB   & MF 
            	& MB 	 & MF  & MB   & MF  \\\midrule            
 $p = 2$  	&0.94		&0.05		&0.75		&0.01		&12	&12	&10	&10\\[5pt]
 $p = 3$   	&1.26		&0.10		&1.32		&0.03		&11	&11	&9	&9\\[5pt]
 $p = 4$  	&3.23		&0.18		&4.19		&0.08		&11	&11	&9	&9\\[5pt]
 $p = 5$	&4.31		&0.22		&13.00	&0.14		&10	&10	&9	&9\\\bottomrule
\end{tabular}
\label{Tab:Tol1e-2}%
\end{table}

 \begin{table}
\caption{\textbf{Global solver:} time and number of iterations of a \textbf{GMRES solver (with drop tolerance $\bm{{10}^{-6}}$)} and $\mat{D}^{-1}$ preconditioner. We compare standard HDG vs.\ macro-element HDG and matrix-based (MB) vs.\ matrix-free (MF) approaches for different $p$, ($N^{Elm} = 124,096$, $Proc's = 2,848$).} 
\centering
\begin{tabular}{l | cccc | cccc}\toprule
&\multicolumn{4}{c |}{Time global solver [sec]}  	&\multicolumn{4}{c}{Number of iterations} 
\\
&\multicolumn{2}{c}{Standard HDG}  	&\multicolumn{2}{c | }{Macro-el.\ HDG ($m=2$)}  	
&\multicolumn{2}{c}{Standard HDG}  	&\multicolumn{2}{c}{Macro-el.\ HDG ($m=2$)}  	
\\\cmidrule(lr){2-3}\cmidrule(lr){4-5}\cmidrule(lr){6-7}\cmidrule(lr){8-9}
            	& MB   	 & MF  & MB   & MF 
            	& MB 	 & MF  & MB   & MF  \\\midrule            
 $p = 2$  	&1.95		&3.04		&1.09		&0.92	&1259		&1259		&939		&939\\[5pt]
 $p = 3$   	&4.05		&8.23		&3.98		&3.26	&1735		&1741		&1304		&1304\\[5pt]
 $p = 4$  	&11.74	&15.26	&9.75		&8.98	&1829		&1830		&1259		&1258\\[5pt]
 $p = 5$	&40.53	&58.37	&24.73	&23.69	&4717		&4766		&1417		&1419\\\bottomrule
\end{tabular}
\label{Tab:Tol1e-6}%
\end{table}

We conduct the same numerical experiments, but with a drop tolerance of 10e-6, and report the results in Table \ref{Tab:Tol1e-6}. We observe that when we decrease the drop tolerance for the GMRES linear solver, that is, we increase the solution accuracy of the linear system, the matrix-based approach becomes competitive. The reason lies in the significant increase in iterations required for the smaller drop tolerance. While the matrix-based approach requires a large initial cost for building the matrix that does not exist in the matrix-free approach, each iteration of the matrix-free approach is more expensive due to the larger number of matrix-vector products. The more iterations we require, the less important the initial cost becomes. 
As we use the same preconditioner, the number of iterations in the matrix-based and matrix-free approaches are practically the same. 
Comparing the standard HDG and macro-element HDG variants, we observe that the advantage of the latter in terms of shorter computing time and fewer iterations (and hence improved conditioning) for the global solver becomes more pronounced with larger system sizes, due to the reduction in degrees of freedom and communication across compute nodes. 

Another important aspect to be observed is that the relative advantage of the macro-element variant versus the standard HDG method becomes particularly apparent for moderate polynomial degrees such as quadratics and cubics. We see in Table \ref{Tab:Tol1e-6} that for $p=2$, the matrix-free macro-element HDG method is more than three times faster than the matrix-free standard HDG method, while for $p=5$, this ratio is down to about two times, although for the latter, standard HDG requires more than three times the number of iterations in the global solver. It can therefore be expected that for larger system sizes that are discretized with elements of moderate polynomial degrees, the advantage of macro-element HDG over standard HDG will be largest. Based on this observation, we will focus on quadratic elements in the remainder of this study.

\subsection{Overall system size: computing time for global vs.\ local operations}

In the next step, we compare the runtime performance of the standard HDG method and macro-element HDG method with mesh refinement, based on the computing time for the local operations (local solver plus step 2 of the matrix-free global solver) and the time for the global operations (remaining steps of the matrix-free global solver), as well as the number of GMRES iterations. Based on our observation in the previous sub-section that the macro-element HDG method is efficient for moderate polynomial degrees, we fix the polynomial order at $p = 2$ and use a fixed number of 512 processors.

Table \ref{Tab:DOF Element ref with p2} reports the number of degrees of freedom for a sequence of meshes generated by globally refining an initial mesh with an adaptively refined structure according to Figure \ref{fig:Test case Brick3D}. Both HDG variants use the same meshes, where the standard HDG method assumes fully discontinuous elements  ($m=1$) and the macro-element HDG method combines groups of eight tetrahedral elements ($m=2$) or 64 tetrahedral elements ($m=4$) into one $C^0$ continuous macro-element. We observe that the macro-element HDG variant, with increasing $m$, significantly reduces the total amount of degrees of freedom in both the local and global problems. In particular, it reduces the number of degrees of freedom of the global system by about one order of magnitude.

\begin{table}
\caption{\textbf{Overall problem size:} we compare standard HDG ($m=1$) vs.\ macro-element HDG ($m=2,4$) in terms of degrees of freedom (dof) for different $N^{Elm}$ (all with $p=2$).} 
\centering
\begin{tabular}{l|cc|cc|cc}\toprule
& \multicolumn{2}{c|}{$m = 1$} & \multicolumn{2}{c|}{$m = 2$}& \multicolumn{2}{c}{$m = 4$}
\\\cmidrule(lr){2-3}\cmidrule(lr){4-5}\cmidrule(lr){6-7}
   & $\text{dof}^{local}$   & $\text{dof}^{global}$ 	& $\text{dof}^{local}$   & $\text{dof}^{global}$  & $\text{dof}^{local}$ & $\text{dof}^{global}$ \\\midrule           
 $N^{Elm} = 45,504$  		&1,820,160		&560,448		&796,320		&179,640		&469,260			&70,740\\[5pt]
 $N^{Elm} = 124,096$  		&4,963,840		&1,517,376		&2,171,680		&483,000		&1,279,740			&187,740\\[5pt]
 $N^{Elm} = 364,032$  		&14,561,280		&4,425,984		&6,370,560		&1,401,120		&3,754,080			&538,920\\[5pt]
 $N^{Elm} = 992,768$  		&39,710,720		&12,026,112		&17,373,440		&3,793,440		&10,237,920			&1,449,000\\\bottomrule
\end{tabular}
\label{Tab:DOF Element ref with p2}%
\end{table}

\begin{table}
\caption{ \textbf{Time for global vs.\ local operations:} we compare the time for the local solver and the local part of the matrix-free global solver (step 2) vs.\ the time for the remaining parts of the global solver (GMRES drop tolerance ${{10}^{-6}}$). We use standard HDG ($m=1$) vs.\ macro-element HDG $\left(m=2,4\right)$ (all with $p=2$, $\mat{D^{-1}}$ preconditioner, $Proc's = 512$).} 
\centering
\begin{tabular}{l|ccc|ccc|ccc}\toprule
& \multicolumn{3}{c|}{Time local op's [sec]} & \multicolumn{3}{c|}{Time global op's [sec]}& \multicolumn{3}{c}{Number of iterations}
\\\cmidrule(lr){2-4}\cmidrule(lr){5-7}\cmidrule(lr){8-10}
          &$m=1$   		& $m=2$ 	& $m=4$    		
          &$m=1$   		& $m=2$ 	& $m=4$  
          &$m=1$   		& $m=2$ 	& $m=4$ \\\midrule           
 $N^{Elm} = 45,504$  		&0.4	&0.1	&0.2	&4.8	&1.0	&0.8	&1970	&1066	&551\\[5pt]	
 $N^{Elm} = 124,096$  		&2.4	&0.5	&1.0	&23.4	&2.6	&2.2	&3118	&1231	&1048\\[5pt]
 $N^{Elm} = 364,544$		&11.1	&2.2	&3.9	&101.6	&10.0	&5.7	&5036	&1673	&1412\\[5pt]
 $N^{Elm} = 992,768$  		&46.2	&26.0	&24.3	&433.4	&120.5	&36.2	&8392	&7526	&3304\\\bottomrule
\end{tabular}
\label{Tab:matrixFree,Mesh, p2}
\end{table}

In Table \ref{Tab:matrixFree,Mesh, p2}, we can observe that the reduction of the size of the global solver translates into a significant reduction in computing time for a given mesh. An important component is also the improved conditioning of the smaller system, which results in a significant reduction in the required number of iterations of the GMRES solver. For the largest mesh with 992,768 elements, the time for the global operations in the macro-element HDG method with $m=4$ is about one order of magnitude smaller than the one of the standard HDG method, in part due to the reduction of the number of iterations from 8,392 to just 3,304. Moreover, we observe that for $m=4$, local and global operations require the same order of magnitude in terms of time. Hence, we conclude that the macro-element HDG method, due to its flexibility in changing the computational load required per macro-element, can achieve a balance between local and global operations.

\subsection{Macro-element size: computing time for global vs.\ local operations}

In the next step, we focus on the macro-element HDG, investigating the effect of different macro-element sizes and its ability to flexibly balance local vs.\ global operations. To this end, instead of keeping the number of processors fixed, we now switch to a variable number of processors, but keep the ratio of the number of macro-elements to the number of processors fixed at one (macro-elements / proc's = 1). Based on our observation that the macro-element HDG method is efficient for moderate polynomial degrees, we again fix the polynomial order at $p = 2$.
Table \ref{Tab:matrixFree,dof,m2,4,8, p2} reports the number of degrees of freedom for a sequence of meshes generated by globally refining the initial mesh with an adaptively refined structure according to Figure \ref{fig:Test case Brick3D}, where the macro-element HDG method combines patches of eight tetrahedral elements ($m=2$), 64 tetrahedral elements ($m=4$) and 512 tetrahedral elements ($m=8$) into one $C^0$ continuous macro-element. We observe again that with increasing $m$, the total amount of degrees of freedom is significantly reduced. The smaller system size of the global problem and its improved conditioning leads in turn to a significant reduction in the required number of GMRES iterations.

\begin{table}
\caption{\textbf{Macro-element size:}
Number of degrees of freedom (dof) of the global problem and number of iterations of the GMRES solver for increasing $m$. ‘*’ denotes that GMRES did not converge in the given number of iterations.} 
\centering
\begin{tabular}{l|rrr|rrr}\toprule
& \multicolumn{3}{c|}{$\text{dof}^{global}$} & \multicolumn{3}{c}{Number of iterations} 
\\\cmidrule(lr){2-4}\cmidrule(lr){5-7}
          	& $m=2$ 	& $m=4$   & $m=8$ 		
       		& $m=2$ 	& $m=4$	  & $m=8$\\\midrule    
 $N^{Elm} = 24,576$   		& 97,920 	    & 38,880       & 18,360 			& 999		    & 648			& 297	 \\[5pt]                
 $N^{Elm} = 196,608$  		& 760,320      & 293,760      & 132,192 	& 2,256		    & 1,101			& 764	\\[5pt]
 $N^{Elm} = 1,572,864$     & 5,990,400	& 2,280,960	& 998,784	& $^{*}3,000$	& 2,990	        & 1,885	 \\\bottomrule
\end{tabular}
\label{Tab:matrixFree,dof,m2,4,8, p2}%
\end{table}

  \begin{table}
\caption{\textbf{Time for global vs.\ local operations:}
we compare the time for the local solver and the local part of the matrix-free global solver (step 2) vs.\ the time for the remaining parts of the global solver (GMRES drop tolerance ${{10}^{-6}}$). We use macro-element HDG with different $m$ (with $p=2$, $\mat{D^{-1}}$ preconditioner, \textbf{macro-elements / proc's = 1}). ‘*’ denotes that GMRES did not converge in 3,000 iterations.} 

\centering
\begin{tabular}{l|ccc|ccc}\toprule
& \multicolumn{3}{c|}{Time local operations [sec]} & \multicolumn{3}{c}{Time global operations [sec]} 
\\\cmidrule(lr){2-4}\cmidrule(lr){5-7}
          		& $m=2$ 	& $m=4$   & $m=8$ 		
           		& $m=2$ 	& $m=4$	  & $m=8$\\\midrule    
 $N^{Elm} = 24,576$  			&0.9		&1.3	&2.7			&2.2		&1.2	&0.7	\\[5pt]                
 $N^{Elm} = 196,608$  		&2.1		&2.2	&5.5	&7.8		&2.6	&2.2	\\[5pt]
 $N^{Elm} = 1,572,864$  		&2.9		&6.6	&11.3				&$^{*}28.5$	&21.9	&14.8	\\\bottomrule
\end{tabular}
\label{Tab:matrixFree,patch,m2,4,8, p2}%
\end{table}

Table \ref{Tab:matrixFree,patch,m2,4,8, p2} reports the computing time for the local operations (local solver plus step 2 of the matrix-free global solver) and the time for the global operations (remaining steps of the matrix-free global solver) with increasing macro-element size $m$. We observe that the time for the local versus global operations can be flexibly tuned by adjusting $m$ and hence the computational load local to each macro-element. In our study, a macro-element size if $m=4$ achieves the best results for the two coarser meshes, while $m=8$ clearly wins for the finest mesh. This observation indicates that for very large systems, larger macro-elements are preferable to balance local and global operations and achieve the fastest computing times for the overall problem.

\subsection{Parallel scalability of the global and local solvers}

While the computing time and iteration counts reported so far provide an estimate of how well the macro-element HDG method performs with respect to the macro-element structure, the element size and the polynomial degree, it does not directly cover another decisive metric for evaluating the performance of a solver, namely parallel scalability \cite{fischer2020scalability}. 
We now examine the strong scalability of both the local solver and the global solver, where we use a direct linear solver based on LU factorization and a matrix-free approach with a GMRES linear solver and the $\mat{D^{-1}}$ preconditioner, respectively. We run our scalability study with the following number of processors - 48, 96, 192, 384, 768, 1,536 and 3,072. We use the same adaptive tetrahedral mesh with 637,888 elements. The macro-element HDG method uses macro-elements that are formed by grouping together eight and 64 elements ($m = 2, 4$), while the standard HDG method uses standard discontinuous elements ($m=1$). The resulting degrees of freedom amount to about 26, eleven and seven million local unknowns, and eight, two and 0.9 million global unknowns for $m= 1, 2, 4$, respectively. Regarding the local/global scalability, we report the speed-up ratio with respect to the base-line computing time that we obtained with 48 processors.

\begin{figure}[t]
    \centering
    \subfloat[\centering Scaling of  local solver.]{{\includegraphics[width=0.5\textwidth]{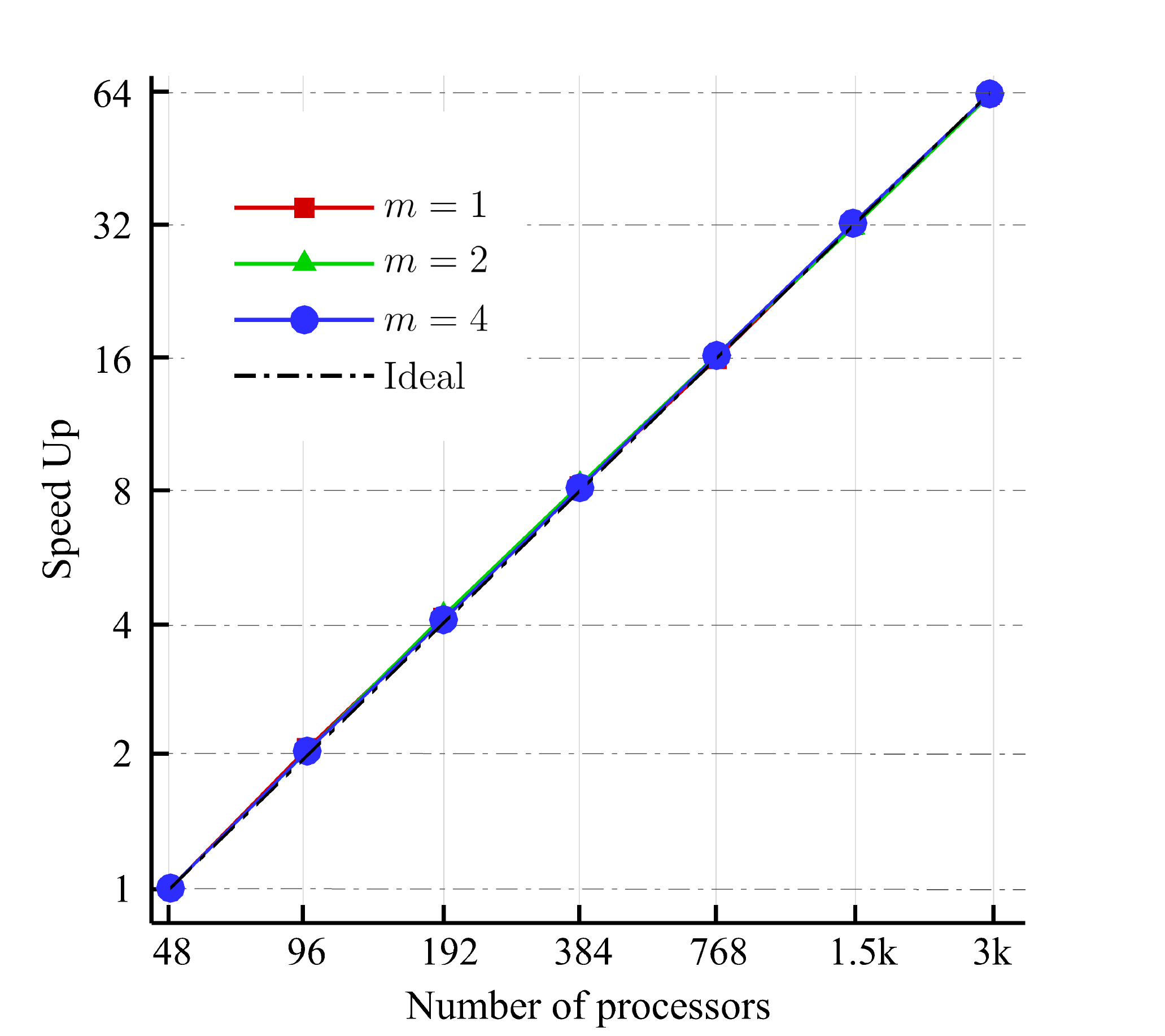} }}%
    \subfloat[\centering Scaling of global solver.]{{\includegraphics[width=0.5\textwidth]{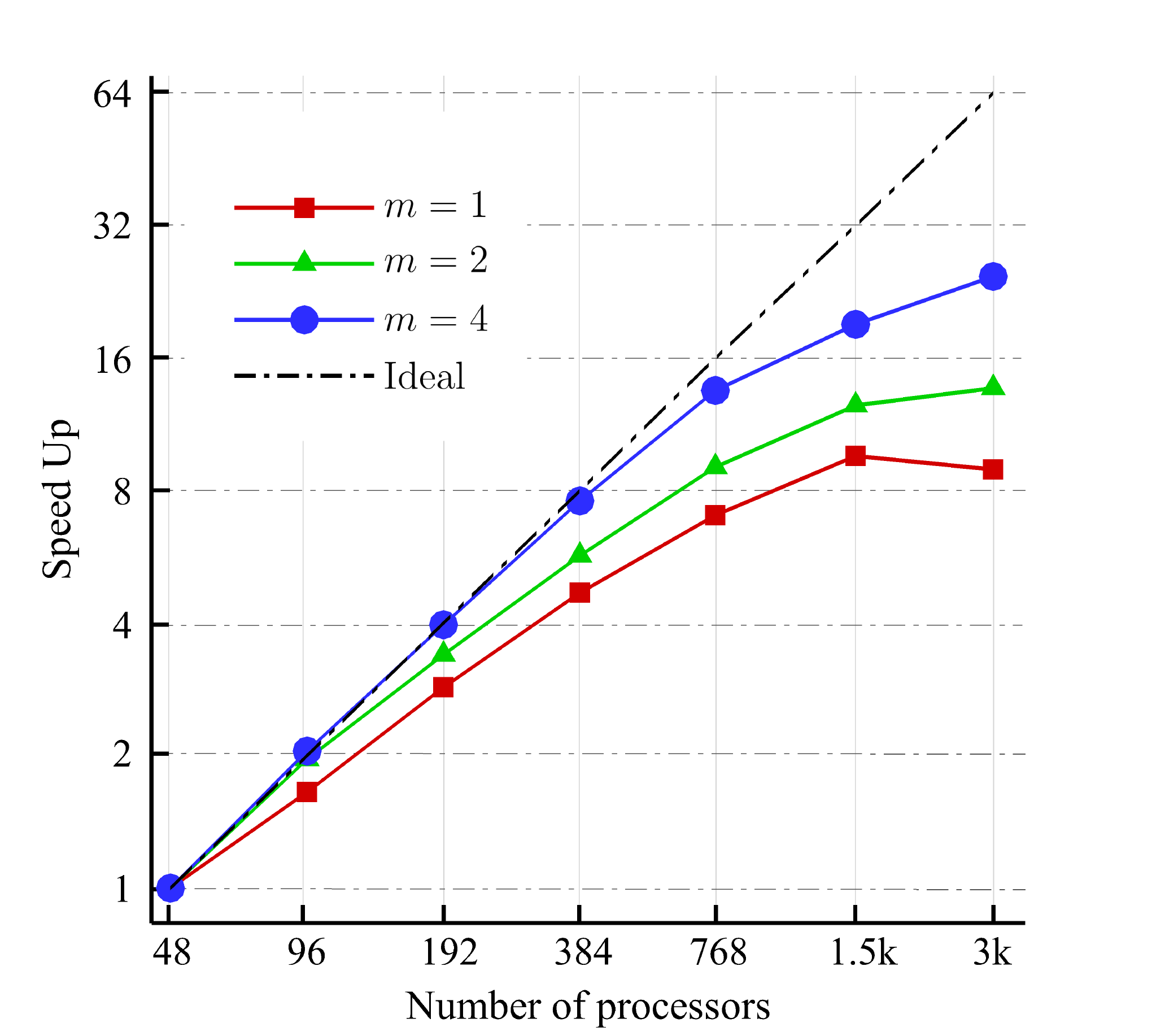} }}
    \caption{Strong scaling of the local solver and global the solver for standard HDG $\left( m=1\right) $ vs.\ macro-element HDG $\left( m=2,4\right) $ ($N^{Elm}=637,888$ with $p=2$).}%
    \label{fig:Scaling Element 3D}%
\end{figure}

Figure \ref{fig:Scaling Element 3D} plots the strong parallel scaling of the standard HDG method ($m=1$) and the macro-element HDG method with $m=2$ and $m=4$. We observe in Figure \ref{fig:Scaling Element 3D}a that the computing time of the local solver is perfectly scalable in a strong sense for both HDG variants. For the global solver, however, we see in Figure \ref{fig:Scaling Element 3D}b that the macro-element HDG method improves strong scalability compared to the standard HDG method.  While the parallel efficiency of the standard HDG method drops below 80\% on just 96 processors, the macro-element HDG method with $m=4$ maintains almost perfect parallel efficiency up to 763 processors. On the one hand, this can be attributed to the larger communication time in the global solver of the standard HDG method \cite{cockburn2009hybridizable, samii2016parallel}. On the other hand, the macro-element HDG method profits from a larger ratio of local operations (step 2) in the matrix-free global solver of the macro-element HDG method and a larger bandwidth of the local and global matrices.

\section{Summary and conclusions}

In this paper, we investigated a macro-element variant of the hybridized discontinuous Galerkin (HDG) method that combines elements of the continuous and hybridized discontinuous finite element discretization. We demonstrated a number of important practical advantages. First, due to the use of continuous elements within macro-elements, the macro-element HDG method mitigates the proliferation of degrees of freedom of standard HDG methods by significantly reducing the number of degrees of freedom with respect to a given mesh. Second, it preserves the unique domain decomposition mechanism of the HDG method that divides the overall problem into local problems per macro-element and a global problem. In particular, it offers additional flexibility in terms of tailoring the macro-element discretization and the associated local problem size to the available compute system and its parallel architecture. Third, it offers a direct approach to adaptive local refinement, as the macro-element mesh naturally accommodates hanging nodes, which allows uniform simplicial subdivision. Fourth, by employing the same continuous discretization for all macro-elements (possibly at different refinement levels), all local operations are embarrassingly parallel and automatically balanced, which makes re-course to load balancing procedures outside of the numerical method unnecessary. In this sense, the macro-element HDG method is particulary well suited for a matrix-free solution approach that consists to a large extent of macro-element local operations.

As for computational efficiency, we showed via theoretical estimates that - under the assumption that both variants operate on the same mesh and require a comparable number of iterations for the global solver - the macro-element HDG method appears to be computationally less efficient than the standard HDG method. Theoretical estimates, however, cannot take into account two important parameters, namely that the macro-element HDG variant requires less communication between processors and reduces the number of iterations of the global solver. 

To include these parameters, we compared the macro-element HDG method and the standard HDG method via a parallel implementation in Julia that we ported on a modern heterogeneous compute system (Lichtenberg II Phase 1 at the Technical University of Darmstadt, at position 187 in the TOP500 list 11/2022). We demonstrated that the matrix-free approach for the global solver is more efficient in the macro-element HDG method, benefitting from the reduction in degrees of freedom and communication across compute nodes. We also showed that unlike standard HDG, the macro-element HDG method is efficient for moderate polynomial degrees such as quadratics and cubics, as it is possible to increase the local computational load per macro-element irrespective to the polynomial degree. In terms of overall computing times for practical meshes with up to one million higher-order elements, we showed that due to the reduction in degrees of freedom, and in particular the reduction of the global problem size and the number of iterations for its solution, the macro-element HDG method can achieve a balance of local and global operations, and - in the situations considered here - deliver faster computing times than the standard HDG method, in the range of up to one order of magnitude. We also showed that the local solver in both variants scales almost perfectly. For the global solver, the macro-element HDG method shows clear advantages in terms of strong scalability, due to the reduction in communication and the favorable ratio in local versus global operations. 

It remains to be seen how well these properties demonstrated here for the advection-diffusion model transfer to more challenging problems. We plan to investigate this question in the future by applying the macro-element HDG method for the direct numerical simulation of turbulent flows modeled via the Navier-Stokes equations.

\section*{acknowledgements}
The authors gratefully acknowledge financial support from the German Research Foundation (Deutsche Forschungsgemeinschaft) through the DFG Emmy Noether Grant SCH 1249/2-1. The authors also gratefully acknowledge the computing time provided to them on the high-performance computer Lichtenberg at the NHR Centers NHR4CES at TU Darmstadt. This is funded by the Federal Ministry of Education and Research and the State of Hesse.

\bibliography{sections/references.bib}

\end{document}